\documentclass[11pt,preprint]{aastex}
\usepackage{longtable}

\usepackage{graphics}

 \usepackage{lscape}

\usepackage{aas_macros} %added later

\graphicspath{{./fig/}}

\usepackage{mediabb}
\usepackage{amsmath,amssymb}
\slugcomment{}
\shortauthors{Aizawa et al.}
\shorttitle{
}
\begin{document}
%%%%%%%%%%%%%%%%%%%%%%%%%%%%%%%%%%%%%%%%%%%%%%%%%%%%%%%%%%%%%%%%%%%%%%
%%%%%%%%%%%%%%%%%%%%%%%%%%%%%%
\title{Towards Detection of Exoplanetary Rings Via Transit Photometry: Methodology and a Possible Candidate}
%%%%%%%%%%%%%%%%%%%%%%%%%%%%%%%%%%%%%%%%%%%%%%%%%%%%%%%%%%%%%%%%%%%%%%
%%%%%%%%%%%%%%%%%%%%%%%%%%%%%%%%%%%%%%%%%%%%%%%%%%%%%%%%%%%%%%%%%%%%%%
\author{
Masataka \textsc{Aizawa},\altaffilmark{1}
Sho \textsc{Uehara},\altaffilmark{2}
Kento \textsc{Masuda},\altaffilmark{1,3,6}
Hajime \textsc{Kawahara},\altaffilmark{4,5} \\
and 
Yasushi \textsc{Suto}\altaffilmark{1,5}
} 
%%%%%%%%%%%%%%%%%%%%%%%%%%%%%%%%%%%%%%%%%%%%%%%%%%%%%%%%%%%%%%%%%%%%%%
\altaffiltext{1}{Department of Physics, The University of Tokyo, 
Tokyo, 113-0033, Japan}
\altaffiltext{2}{Department of Physics, Tokyo Metropolitan University, 
Tokyo 192-4397, Japan}
\altaffiltext{3}{Department of Astrophysical Sciences, Princeton University, Princeton, NJ 08544, USA}
\altaffiltext{4}{Department of Earth and Planetary Science, 
The University of Tokyo, Tokyo 113-0033, Japan}
\altaffiltext{5}{Research Center for the Early Universe, School of Science,
The University of Tokyo, Tokyo 113-0033, Japan}
\altaffiltext{6}{NASA Sagan Fellow}
\email{aizawa@utap.phys.s.u-tokyo.ac.jp}
%%%%%%%%%%%%%%%%%%%%%%%%%%%%%%%%%%%%%%%%%%%%%%%%%%%%%%%%%%%%%%%%%%%%%%
\parindent = 13pt
\begin{abstract}
Detection of a planetary ring of exoplanets remains as one of the most attractive but challenging goals in the field.  We present a methodology of a systematic search for exoplanetary rings via transit photometry of long-period planets. The methodology relies on a precise integration scheme we develop to compute a transit light curve of a ringed planet. We apply the methodology to 89 long-period planet candidates from the {\it Kepler} data so as to estimate, and/or set upper limits on, the parameters of possible rings. While a majority of our samples do not have a sufficiently good signal-to-noise ratio for meaningful constraints on ring parameters, we find that six systems with a higher signal-to-noise ratio are inconsistent with the presence of a ring larger than 1.5 times the planetary radius assuming a grazing orbit and a tilted ring.  Furthermore, we identify five preliminary candidate systems whose light curves exhibit ring-like features. After removing four false positives due to the contamination from nearby stars, we identify KIC 10403228 as a reasonable candidate for a ringed planet. A systematic parameter fit of its light curve with a ringed planet model indicates two possible solutions  corresponding to a Saturn-like planet with a tilted ring. There also remain other two possible scenarios accounting for the data; a circumstellar disk and a hierarchical triple. Due to large uncertain factors, 
we cannot choose one specific model among the three.  

%A systematic parameter fit of its light curve indicates two possible solutions with an orbital period $P=450$ years; one implies a planet of radius $ 0.88 R_J$ with a ring of $0.89 R_J<R<2.56R_J$ tilted by $59.4^{\circ}$ with respect to the orbital plane, while the other implies a planet of radius $1.44 R_J$ with a ring of $2.29 R_J<R<3.69R_J$ tilted by$12.3^{\circ}$.  We examine and compare various possibilities other than the ring hypothesis, and discuss future prospects.
\end{abstract}
%%%%%%%%%%%%%%%%%%%%%%%%%%%%%%%%%%%%%%%%%%%%%%%%%%%%%%%%%%%%%%%%%%%%%%
\keywords{methods: data analysis - planets and satellites: detections -
planets and satellites: rings - techniques: photometric}
%%%%%%%%%%%%%%%%%%%%%%%%%%%%%%%%%%%%%%%%%%%%%%%%%%%%%%%%%%%%%%%%%%%

\def\aj{AJ}%
         % Astronomical Journal
\def\actaa{Acta Astron.}%
         % Acta Astronomica
\def\araa{ARA\&A}%
         % Annual Review of Astron and Astrophys
\def\apj{ApJ}%
         % Astrophysical Journal
\def\apjl{ApJ}%
         % Astrophysical Journal, Letters
\def\apjs{ApJS}%
         % Astrophysical Journal, Supplement
\def\ao{Appl.~Opt.}%
         % Applied Optics
\def\apss{Ap\&SS}%
         % Astrophysics and Space Science
\def\aap{A\&A}%
         % Astronomy and Astrophysics
\def\aapr{A\&A~Rev.}%
         % Astronomy and Astrophysics Reviews
\def\aaps{A\&AS}%
         % Astronomy and Astrophysics, Supplement
\def\azh{AZh}%
         % Astronomicheskii Zhurnal
\def\baas{BAAS}%
         % Bulletin of the AAS
\def\bac{Bull. astr. Inst. Czechosl.}%
         % Bulletin of the Astronomical Institutes of Czechoslovakia 
\def\caa{Chinese Astron. Astrophys.}%
         % Chinese Astronomy and Astrophysics
\def\cjaa{Chinese J. Astron. Astrophys.}%
         % Chinese Journal of Astronomy and Astrophysics
\def\icarus{Icarus}%
         % Icarus
\def\jcap{J. Cosmology Astropart. Phys.}%
         % Journal of Cosmology and Astroparticle Physics
\def\jrasc{JRASC}%
         % Journal of the RAS of Canada
\def\mnras{MNRAS}%
         % Monthly Notices of the RAS
\def\memras{MmRAS}%
         % Memoirs of the RAS
\def\na{New A}%
         % New Astronomy
\def\nar{New A Rev.}%
         % New Astronomy Review
\def\pasa{PASA}%
         % Publications of the Astron. Soc. of Australia
\def\pra{Phys.~Rev.~A}%
         % Physical Review A: General Physics
\def\prb{Phys.~Rev.~B}%
         % Physical Review B: Solid State
\def\prc{Phys.~Rev.~C}%
         % Physical Review C
\def\prd{Phys.~Rev.~D}%
         % Physical Review D
\def\pre{Phys.~Rev.~E}%
         % Physical Review E
\def\prl{Phys.~Rev.~Lett.}%
         % Physical Review Letters
\def\pasp{PASP}%
         % Publications of the ASP
\def\pasj{PASJ}%
         % Publications of the ASJ
\def\qjras{QJRAS}%
         % Quarterly Journal of the RAS
\def\rmxaa{Rev. Mexicana Astron. Astrofis.}%
         % Revista Mexicana de Astronomia y Astrofisica
\def\skytel{S\&T}%
         % Sky and Telescope
\def\solphys{Sol.~Phys.}%
         % Solar Physics
\def\sovast{Soviet~Ast.}%
         % Soviet Astronomy
\def\ssr{Space~Sci.~Rev.}%
         % Space Science Reviews
\def\zap{ZAp}%
         % Zeitschrift fuer Astrophysik
\def\nat{Nature}%
         % Nature
\def\iaucirc{IAU~Circ.}%
         % IAU Cirulars
\def\aplett{Astrophys.~Lett.}%
         % Astrophysics Letters
\def\apspr{Astrophys.~Space~Phys.~Res.}%
         % Astrophysics Space Physics Research
\def\bain{Bull.~Astron.~Inst.~Netherlands}%
         % Bulletin Astronomical Institute of the Netherlands
\def\fcp{Fund.~Cosmic~Phys.}%
         % Fundamental Cosmic Physics
\def\gca{Geochim.~Cosmochim.~Acta}%
         % Geochimica Cosmochimica Acta
\def\grl{Geophys.~Res.~Lett.}%
         % Geophysics Research Letters
\def\jcp{J.~Chem.~Phys.}%
         % Journal of Chemical Physics
\def\jgr{J.~Geophys.~Res.}%
         % Journal of Geophysics Research
\def\jqsrt{J.~Quant.~Spec.~Radiat.~Transf.}%
         % Journal of Quantitiative Spectroscopy and Radiative Trasfer
\def\memsai{Mem.~Soc.~Astron.~Italiana}%
         % Mem. Societa Astronomica Italiana
\def\nphysa{Nucl.~Phys.~A}%
         % Nuclear Physics A
\def\physrep{Phys.~Rep.}%
         % Physics Reports
\def\physscr{Phys.~Scr}%
         % Physica Scripta
\def\planss{Planet.~Space~Sci.}%
         % Planetary Space Science
\def\procspie{Proc.~SPIE}%

\section{Introduction \label{s:sec1}}

As is the case of the Solar system, moons and planetary rings are
believed to exist in exoplanetary systems as well. Their detection,
however, has not yet been successful, and remains as one of the most
attractive, albeit challenging, goals in exoplanetary sciences. A
notable exception includes a system of giant circumplanetary rings of
J1407b \citep[e.g.][]{ 2015ApJ...800..126K}, but the inferred radius
$\sim 1$ AU implies that it is very different from Saturnian rings that
we focus on in the present paper.

In addition to the obvious importance of the ring discovery itself, its
detection offers an interesting method to determine the direction of the
planetary spin because the ring axis is supposed to be aligned with the
planetary spin as in the case of Saturn.  Thus the detection of ring
parameters yield a fairly complete set of dynamical architecture of
transiting planetary systems; the stellar spin via asteroseismology
\citep[e.g.][]{2013ApJ...767..127H, 2014PASJ...66...94B} and gravity
darkening \citep[e.g.][]{2011ApJS..197...10B, 2015ApJ...805...28M}, the
planetary orbit via transiting photometry and the Rossiter-McLaughlin
effect \citep[e.g.][]{2000A&A...359L..13Q,2005ApJ...622.1118O}, and the planetary spin
through the ring detection as discussed here.

The direct detection of planetary spin is very difficult, and so far
only four possible signals related to planetary
spins have been reported: the periodic flux variations of 2M1207b
\citep{2016ApJ...818..176Z}, and a rotational broadening and/or distortion
of the line profile of $\beta$ Pictoris b \citep{2014Natur.509...63S}, 
HD 189733b \citep{2016ApJ...817..106B}, and GQ Lupi b \citep{2016arXiv160700012S}. 
These interesting planets are very young and have sufficiently high
temperature ($>1600$ K) for their spin to be detected.  In contrast,
the same technique is not easily applicable for mature and cold planets
like Saturn. Thus the detection of the ring axis provides a
complementary methodology to determine the spin of more typical planets
with low temperature.

Since the total mass of planetary rings is small, they do not exhibit
any observable signature on the dynamics of the system.  Instead,
high-precision photometry and spectroscopy offer a promising approach
towards their detection, and, observations of reflected
light and transit are especially useful for this purpose.

 Possible signatures in reflected light due to the planetary rings
include the higher brightness, the characteristic phase function,
distinctive spectral variations, temporary extinction of the planet, and
discrepancy between reflection and thermal radiation intensities
\citep[e.g.][]{2004A&A...420.1153A, 2005ApJ...618..973D}.  For instance,
\cite{2015A&A...583A..50S} attempted to explain the line broadening of
the reflected light of 51 Peg b with a ringed planet model, and they
conclude that it is not due to the ring. This is because their solution
requires a non-coplanar configuration, which would be unlikely for
short-period planets.  Searches for rings through the light reflection
of the host star can be made for non-transiting planets even though
their signals are typically small. Therefore, while we focus on the
transit photometry in the rest of this paper, the reflected-light
method is indeed useful and complementary as well. 

\cite{1999CRASB.327..621S} is the first to propose the transit
photometry as a tool for the ring detection.  \cite{2001ApJ...552..699B}
derived the upper limit on the radius of a possible ring around HD
209458 b. \cite{2004ApJ...616.1193B} improved the model of 
\cite{2001ApJ...552..699B} by incorporating the influence of diffraction
on the light curves. They claimed that the Saturn-like ring system can
be detected with the photometric precision of the {\it Kepler} mission.
\cite{2009ApJ...690....1O} pointed out that the combination of the
transit photometry and the spectroscopic Rossiter-McLaughin effect
increases the detection efficiency and the credibility of the signal.
\cite{2015ApJ...803L..14Z} proposed that an anomalously large planet
radius indicated from transit photometry can be used to select
candidates for ringed planets. They also proposed that the 
anomalous stellar density estimated from the transit may be 
used as a probe of a ring. 

In addition to the above methodology papers, a systematic search for
ring systems using real data was conducted by
\cite{2015ApJ...814...81H}. They analyzed 21 short-period planets
($P\leq 50$ days) in the {\it Kepler} photometric data, and found no
appreciable signatures of rings around the systems. 
This is an interesting attempt, but their null detection is not surprising 
because the ring tends to be unstable as the planet gets closer to the central star. 
In addition, \cite{2011ApJ...734..117S} demonstrated that it is hard to 
detect the ring at below 0.1 AU in the case of solar-like stars. 

Instead, we attempt here a systematic search for rings around
long-period planet candidates that exhibit single or a few transit-like
signals in the {\it Kepler} photometric data. Since rings around those
planets, if exist, should be dynamically stable, even a null detection
would eventually put an interesting constraint on the formation
efficiency and properties of icy rings for those plantetary
environments.

The purposes of the paper are three-fold; to establish a methodology 
for the discovery of potential ringed planets, to apply the methodology to a catalog of
long-period planet candidates from {\it Kepler}, and to detect and/or
constrain the possible ringed planets. Section 2 presents our simple
model of a ringed planet, and describes the expected transit signal.  In
Section 3, we explain how to select target objects for our search, and
classify them into four groups according to the amplitude and nature of
the signal-to-noise ratio of their light curves relative to the expected
signature by possible ringed planets.  In Section 4, we place upper
limits on ring parameters for seven systems with a good signal-to-noise
ratio.  In Section 5, we select five tentative ringed-planet candidates
from the high signal candidates classified in Section 3. While four out
of the five are likely to be false positives, one system, KIC 10403228,
passes all the selection criteria that we impose.  Therefore we attempt
a systematic parameter survey for the possible ring around KIC 10403228 
in Section 6. Also we examine and discuss various other possibilities
that may explain the observed ring-like anomaly.  Final section is
devoted to conclusion and future prospects.
\section{A simple model for a ringed planet  \label{s:sec2}}

\subsection{Basic parameters that characterize a ringed planet system}

Our simple model of a ringed planet adopted in this paper basically
follows \cite{2009ApJ...690....1O}.  The ring is circular, and has a
constant optical depth $\tau$ everywhere between the inner and outer radii of
$R_{\rm in}$ and $R_{\rm out}$. We denote the radii of the star and
planet by $R_{\star}$ and $R_{\rm p}$.

\begin{figure}[htpb]
  \centering
  %\begin{tabular}[b]{@{}p{0.40\textwidth}@{}}
    \includegraphics[width=.49\linewidth]{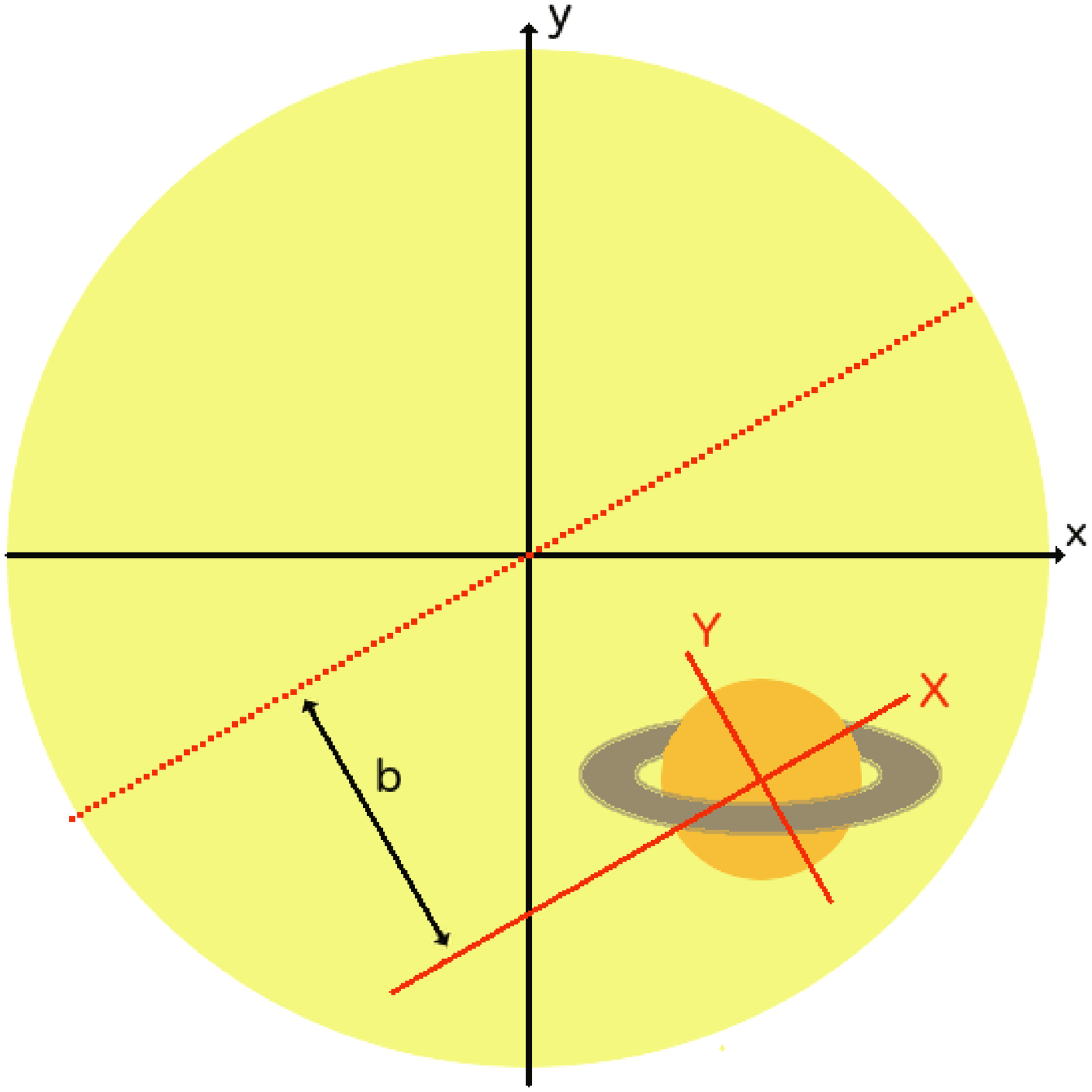} 
    %\centering\small (a)Transit of a ringed planet
  %\end{tabular}%
  %\quad
  %\begin{tabular}[b]{@{}p{0.40\textwidth}@{}}
   \includegraphics[width=.49\linewidth]{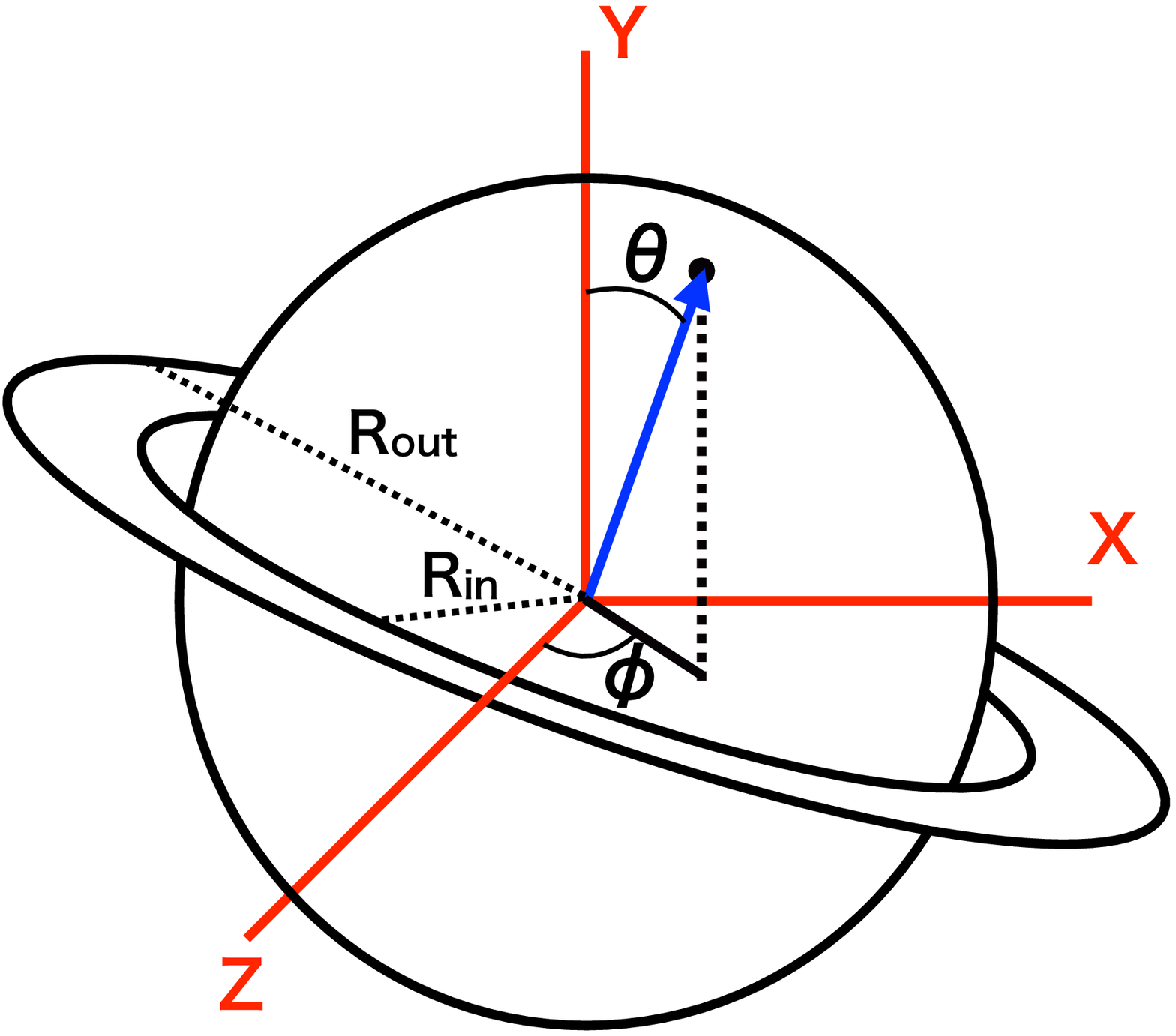} 
   % \centering\small (b)Enlarged view of a ringed planet
  %\end{tabular}
  \caption{\small (Left) Schematic illustration of a transit of a planet with a ring. The
    origin of $(X,Y)$ is shifted from the center of the planet to that
    of the star. The radius of the star is $R_{\star}$, the radius of
    the planet is $R_{\rm p}$ and the impact parameter of the planet
    is $b$. The $z$-axis and the $Z$-axis are 
    toward the observer.  (Right) Enlarged view of the planet with the
    ring. $R_{\rm in}$ and $R_{\rm out}$ are the inner and outer
    radii of the ring respectively. The obliquity angle $\theta$ and 
    azimuthal angle $\phi$ are defined with respect to $(X,Y,Z)$-coordinate. }
      \label{model}
\end{figure}

The configuration of the planet and ring during a transit is
illustrated in Figure \ref{model}. The $X$-axis is approximately aligned with 
the projected orbit of the planet on the stellar disk, and the 
$Z$-axis is towards the observer. This completes the $(X,Y,Z)$
coordinate frame centered at the origin of the ringed planet (left panel in Figure
\ref{model}). The normal vector of the ring plane is characterized by
the two angles $\theta$ and $\phi$ in a spherical coordinate (right panel in Figure
\ref{model}).

We also set up another coordinate system $(x,y,z)$ centered at the 
origin of the star in such a way that the major and minor axes of the
projected ring are defined to be parallel with $x$- and $y$-axes, respectively, 
with $z$-axis being towards the observer.

The ring is assumed to move along the planetary orbit with constant
obliquity angles $(\theta, \phi)$, and the planet is assumed to move
on a Keplerian orbit around the star.  The left panel in Figure \ref{model} 
illustrates the transit of the ringed planet, whose impact parameter
is $b$.

%%%%%%%%%%%%%%%%%%%%%%%%%%%%%%%%%%%%%%%%%%%%%%%%%%%%%%%%%%%
%%%%%%%%%%%%%%%%%%%%%%%%%%%%%%%%%%%%%%%%%%%%%%%%%%%%%%%%%%%
We assume a thin uniform ring with a constant optical depth $\tau$ for
the light from the direction normal to the ring plane.  Thus
the fraction of the background stellar light transmitted through the
inclined ring is given by $\exp(-\tau (\sin \theta \cos \phi)^{-1})$, and we define the shading parameter $T$ as $1-\exp(-\tau (\sin \theta \cos \phi)^{-1})$. In our simple ring model, 
the value of $T$, instead of $\tau$, fully 
specifies the effective optical transparency of the ring.

In summary, our simple ring model is characterized by five parameters; four 
($R_{\rm in}, R_{\rm out}, \theta, \phi$) specify the geometry of the ring, 
the other is a shading parameter $T$. Instead of $R_{\rm in}$ and 
$R_{\rm out}$, we use dimensionless parameters in fitting,:
%%%%%%%%%%%%%%%%%%%%%%%%%%%%%%%%%%%%%%%%%%%%%%%%%%%%%%%
\begin{equation}
r_{\rm in/p} \equiv \frac{R_{\rm in}}{R_{\rm p}},
\qquad
r_{\rm out/in} \equiv \frac{R_{\rm out}}{R_{\rm in}}.
\end{equation}
%%%%%%%%%%%%%%%%%%%%%%%%%%%%%%%%%%%%%%%%%%%%%%%%%%%%%%%

\subsection{Transit signal of a ringed planet}

The stellar intensity profile $I(x,y)$ under the assumption of the quadratic limb darkening law
is expressed in terms of two parameters $u_{1}$ and $u_{2}$:
%%%%%%%%%%%%%%%%%%%%%%%%%%%%%%%%%%%%%%%%%%%
\begin{equation}
\label{one}
\frac{I(x,y)}{I_{0}} =\left [ 1 - u_{1}( 1-\mu) - u_{2} ( 1-\mu ) ^{2} \right] 
\left( \text{ $\mu \equiv \sqrt{1-\cfrac{x^{2}+y^{2}}{R_{\star}^{2}}}$}\right), 
\end{equation}
%%%%%%%%%%%%%%%%%%%%%%%%%%%%%%%%%%%%%%%%%%%
where $I_{0}$ is the intensity at the center of the star. 
The physical conditions on the profile require the following 
complex constraints on $u_{1}$ and $u_{2}$:
%%%%%%%%%%%%%%%%%%%%%%%%%%%%%%%%%%%%%%%%%%%
\begin{equation}
\label{two}
u_{1} + u_{2} < 1,\; u_{1} > 0,\; u_{1} + 2 u_{2} > 0. 
\end{equation}
%%%%%%%%%%%%%%%%%%%%%%%%%%%%%%%%%%%%%%%%%%%
In this paper, we adopt $q_{1} = (u_{1} +
u_{2})^{2} $ and $q_{2} = u_{1}/(2(u_{1}+u_{2}))$ instead of ($u_{1}
$, $ u_{2}$) following \cite{2013MNRAS.435.2152K}. 
Then, Equations (\ref{one}) and (\ref{two}) are rewritten as
%%%%%%%%%%%%%%%%%%%%%%%%%%%%%%%%%%%%%%%%%%%
\begin{equation}
\frac{I(x,y)}{I_{0}} 
=\left [ 1 - 2 q_{2} \sqrt{q_{1}} ( 1-\mu) 
- \sqrt{q_{1}}(1-2q_{2}) ( 1-\mu ) ^{2} \right], 
\end{equation}
%%%%%%%%%%%%%%%%%%%%%%%%%%%%%%%%%%%%%%%%%%%
%%%%%%%%%%%%%%%%%%%%%%%%%%%%%%%%%%%%%%%%%%%
\begin{equation}
\label{limb}
\text{with}\;\;\;\;\;\;\; 0< q_{1} <1,\; 0 < q_{2} <1. 
\end{equation}
%%%%%%%%%%%%%%%%%%%%%%%%%%%%%%%%%%%%%%%%%%%
In this parametrization, $q_{1}$ and $q_{2}$ vary independently 
between 0 and 1. This is useful in finding best-fit parameters \citep{2013MNRAS.435.2152K}. 
For reference, the Sun has $q_{1} = 0.49$ and $ q_{2} = 0.34$ ($u_{1} = 0.47$ and $u_{2} = 0.23)$ \citep{2000asqu.book.....C}. 

Let $D(x,y,t)$ be the blocked fraction of light coming from the
location on the stellar disk $(x,y)$. Due to the motion of the planet
during a transit, $D(x,y,t)$ is time-dependent and given as
%%%%%%%%%%%%%%%%%%%%%%%%%%%%%%%%%%%%%%%%%%%
\begin{equation}
D(x,y,t) = \begin{cases}
    1 & :\text{if $(x,y)$ is within the planetary disk}  \\
    T & :\text{if $(x,y)$ is within the ring disk, 
but out of the planetary disk} \\
    0 & :\text{otherwise}.
  \end{cases} \label{discrete}
\end{equation}
%%%%%%%%%%%%%%%%%%%%%%%%%%%%%%%%%%%%%%%%%%%
Then the normalized flux from the the system is given by
%%%%%%%%%%%%%%%%%%%%%%%%%%%%%%%%%%%%%%%%%%%
\begin{equation}
\label{genkou}
F(t) = 1-  \frac{\int_{\text{stellar disk}}\, I(x,y) D(x,y,t) dx dy }{ I_{\rm all}},  
\end{equation}
%%%%%%%%%%%%%%%%%%%%%%%%%%%%%%%%%%%%%%%%%%%
where the second term indicates the fraction of light blocked by a
transiting ringed planet, and the total flux is
%%%%%%%%%%%%%%%%%%%%%%%%%%%%%%%%%%%%%%%%%%%
\begin{equation}
\label{eq:inte}
 I_{\rm all} = \int _{\text{stellar disk}} I(x,y) dx dy 
= \pi I_{0} R^{2} _{\star}\left [ 1 - \frac{2\sqrt{q_{1}} q_{2} }{3} 
- \frac{\sqrt{q_{1}}(1-2q_{2})}{6} \right]. 
\end{equation}
%%%%%%%%%%%%%%%%%%%%%%%%%%%%%%%%%%%%%%%%%%%

We develop a reliable numerical integration method that solves the boundary lines of
$D(x,y,t)$ as described in Appendix A. Our method achieves  
the numerical error less than $10^{-7}$ in relative flux, and this is much
smaller than a typical noise of the {\it Kepler} photometric data.

\subsection{Effects that are neglected in our model}

We briefly comment on three effects that we neglect in the analysis
below; finite binning during exposure time, planetary precession, and
forward-scattering of the ring. While all of them are negligible for the Saturnian ringed planet with a long period, they may become important in other situations. 

For the precise comparison of our light-curve predictions against
the {\it Kepler} long cadence, we may have to take account of the finite exposure time 
(29.4 min) properly. In fact, the binning effect is shown to
bias the transit parameter estimate in the case of short-period planets
\citep[]{2010MNRAS.408.1758K}.  For long-period planets that we focus on
here, however, the transit duration is sufficiently longer than the
finite exposure time. Thus the binning effect is not important.  In the 
case of the transit of Saturn in front of the Sun, for instance,
the fractional difference of the relative flux is typically an order of 
$10^{-5}$ between models with and without the binning effect. 
%as shown in Figure\ref{test_first} below. 
This value is an order-of-magnitude smaller than the
expected noise in the {\it Kepler} photometric data. 
Thus we can safely neglect the binning effect in the present analysis.

The precession of a planetary spin would generate observable seasonal
effects on the transit shape of a ringed planet 
\citep[][]{2010ApJ...716..850C, 2015ApJ...814...81H}.  Since our current
target systems are extracted from those with a single transit, however,
we can ignore the effect; the period of the precession is
proportional to the square of the orbital period, and thus the
precession effect during a transit is entirely negligible. Nevertheless, 
we note here that this could be an interesting probe of the dynamics of short-period
ringed planetary systems that exhibit multiple transits. 

In the present analysis, we consider the effect of 
light-blocking alone due to the ring during its transit.  In reality, forward
scattering (diffraction by the ring particles) may increase the flux of the
background light. Let us consider light from the star to the observer 
through the ring particle with diameter $d$. First, light is emitted from the disk of the star, 
and arrives at the ring particles. The angular radius of the star viewed from
the ring particles is about $R_{\star}/a$, where 
$a$ is the semi-major axis of the orbit, 
and $R_{\star}$ is the stellar radius. Next, the light is diffracted by 
the ring particles, and the extent of the diffraction is described 
by the phase function \citep{2004ApJ...616.1193B}; the rough diffraction 
angle can be estimated from the first zero of the phase function 
$\theta \simeq 0.61\lambda/d$, where $\lambda$ is wavelength of light. 
In particular, the effect of the diffraction becomes significant when 
the viewing angle $R_{\star}/a$ is comparable to the diffraction angle. Let us define the critical 
 particle size $d_{\rm crit}$ by equating $R_{\star}/a$ with  $\theta = 0.61\lambda / d$; 
%%%%%%%%%%%%%%%%%%%%%%%%%%%%%%%%%%%%%%%%%%%%%%%%%%%
\begin{equation}
d_{\rm crit} = 0.61\, \frac{a \lambda}{R_{\star}}  
= 0.63\, \text{mm}\left (\frac{a/R_{\star}}{2060} \right)  
\left( \frac{\lambda} {500 \text{nm}}  \right). \label{d_crit}
\end{equation} 
%%%%%%%%%%%%%%%%%%%%%%%%%%%%%%%%%%%%%%%%%%%%%%%%%%%
\cite{2004ApJ...616.1193B} discussed the effect of diffraction using $d_{\rm crit}$. 
When $ d \geq 10 d_{\rm crit}$, 
the diffraction angle is small, and light just behind the ring particles
is diffracted to the observer. In this case, the diffraction 
does not affect the direction of light, and  we may express the extinction due to 
absorption with a single parameter $T$.

When $d \leq d_{\rm crit}/10$, 
the diffraction angle is large, and the ring 
particles diffract light to wider directions. Then, the amount of light that 
reaches the observer significantly decreases, and we may model the extinction in terms of $T$. 

In both cases, $ d \geq 10 d_{\rm crit}$ and $ d \leq d_{\rm crit}/10$, the extinction can be
modeled with a single parameter $T$. In the case of Saturn with the typical
particle diameter $d =1$ cm, for instance, $d_{\rm crit}\simeq 0.63$ mm 
from Equation (\ref{d_crit}) satisfies $d>10 d_{\rm crit}$, so our model can be used to calculate the light curves of Saturn observed far from the Solar System. 

We should note that when the typical size of particles satisfies $d_{\rm crit}/10 \le d
\le 10 d_{\rm crit}$, the forward scattering induces the rise in the light curve 
before the ingress and after the egress, and this effect can become the key to
identify the signatures of the rings out of other physical signals. 
Incorporating the diffraction into the model, however, requires intensive computation, 
and this is beyond the scope of this paper. 

\section{Classification \label{s:sec3}}
\begin{figure}[htpb]
\begin{center}
\includegraphics[width=0.97\linewidth]{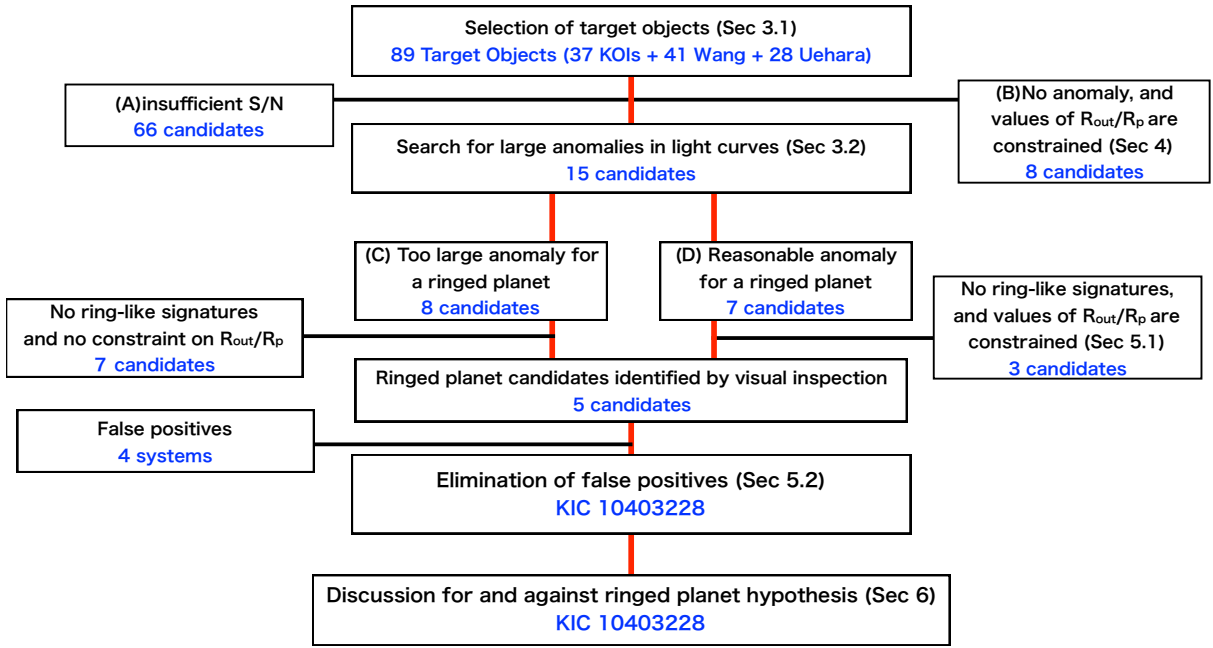}
\caption{\small Flow chart of our strategy of 
ringed-planet search. Procedure and the number of remaining 
candidates are described in each step. 
The details of each procedure are described in the corresponding 
sections. For classification into (A)$\sim$(D), model I is adopted in Table \ref{sec3_ring}. }
\label{flow}

\end{center}
\end{figure}

In what follows, we present our methodology to search for 
planetary rings in the real data. Figure \ref{flow} shows the 
flow chart of the analysis procedure and its application. 
Methods in each step of the chart are described along 
with the results of analysis in the following sections. 
\setlength{\unitlength}{2em} % for the picture environment

In this section, we first choose target objects found in the {\it Kepler} field. 
Then, we classify them into four categories depending on 
the observed anomalies in the light curves. 
The details of classification procedure may be found in Appendices B and C. 

\subsection{Target Selection}
The {\it Kepler} mission monitored more than 150,000 
stars over four years, and identified about 8,000 planet 
candidates as {\it Kepler} Objects of Interest (KOIs). 
In this paper, we focus on long-period planet candidates 
because icy ring particles as observed around Saturn are supposed to 
survive only at locations far from the host star.
Considering that the temperature of the snow line
is $170$ K \citep{1981PThPS..70...35H}, 
we choose 37 KOIs whose equilibrium 
temperatures are less than 200 K. In addition, we selected planet candidates reported 
by recent transit surveys; 41 candidates from a search by \cite{2015ApJ...815..127W} 
and 28 candidates from \cite{2016ApJ...822....2U}. 
In Table \ref{table_num}, the numbers of planetary candidates in three groups 
are listed with the number of transits.

\begin{figure*}[htbp]
\begin{center}
\includegraphics[width=0.5 \linewidth]{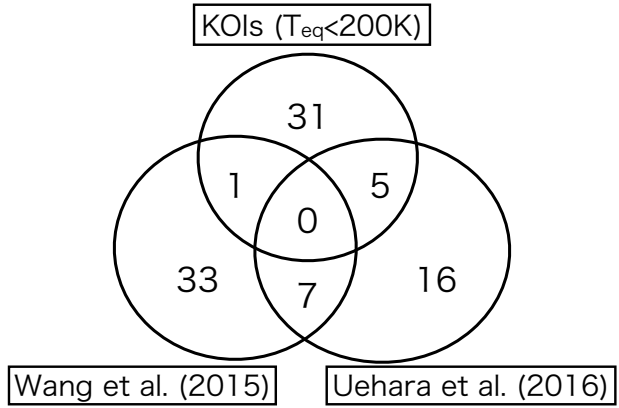}
\end{center}
{\caption{Venn diagram for target objects corresponding to Table \ref{table_num}.  } \label{benzu}}
\end{figure*}

We exclude several systems, which are not suited for our search. 
For KOI-5574.01 in KOIs and KIC 2158850 in \cite{2015ApJ...815..127W}, 
we cannot find the transit signal among the noisy light curves. 
For KOI-959.01 in KOIs with $P=10$ days and KIC 8540376 
in \cite{2015ApJ...815..127W} with $P= 31.8$ days, 
we cannot neglect the binning effect due to the short transit duration. 
After removing these systems, 89 planet candidates are left in total for our search. 
Tables \ref{table_num} summarizes the number of targets, and Figure \ref{benzu} 
shows the overlapped objects among KOIs, \cite{2015ApJ...815..127W}, and 
 \cite{2016ApJ...822....2U}

  \begin{table}[htpb]
\caption{\small Number of planet candidates in three groups. 
}
\begin{center}
\scalebox{0.8}{
\begin{tabular}{l c c c c  }
\hline
Group & \multicolumn{3}{c}{Number of systems}& Total number of candidates \\\hline \hline
 & One transit & Two transits& more than two transits&  \\ 
\hline\hline
KOIs ($T_{\rm eq} \leq 200$ K)&5 & 2 & 30 &  37  \\
\cite{2015ApJ...815..127W} & 17 & 14 & 10& 41\\
\cite{2016ApJ...822....2U}& 28&0&0&28  \\
\hline \hline
\end{tabular} 
  \label{table_num}
}
\end{center}

 \end{table}  %% 
 
\subsection{Classification of target objects} 
Inevitably a signature of a possible ring around a planet is very
tiny. Long-period planet candidates exhibit a small number of transits 
(Table \ref{table_num}), and the precision of the transit light curves 
is not improved so much by folding the multiple events. Therefore the search 
for a possible ring signature crucially relies on the quality of the 
few transiting light curves for individual systems. 

According to the automated procedures described in Appendices B and  C, we
classify the long-period planet candidates into the following four
categories. 
\parindent = 0pt

(A) insufficient $S/N$ to constrain ring parameters:\\ 
\ \ Since the anomalous feature due to the ring is very subtle,
one cannot constrain the ring parameters at all if the intrinsic
light-curve variation of the host  is too large to 
be explained by any ring model. Thus we exclude 
those systems that exhibit a noisy light curve out-of-transit. 
The exclusion criteria depend on the adopted ring model to some
extent, but are determined largely by the threshold signal-to-noise
ratio $(S/N)$ that we set as $S/N=10$. For definiteness,  we consider
4 different ring models (Table \ref{sec3_ring}), and the details of the
procedure are described in Appendices B and C.

(B) sufficient $S/N$ and no significant anomaly:\\ 
\ \  A fraction of the systems has a sufficiently good $S/N$ and exhibits no significant anomaly.
In such a case, we can put physically meaningful constraints on the possible ring parameters (Section 4).

(C) too large anomaly for a ringed planet:\\
\ \ In contrast to (B), some systems exhibit a large anomaly in the
transiting light curve that exceeds the prediction in the
adopted ring models. Nevertheless, different ring models may be able to explain 
the anomaly, and we still continue to search for ringed planets in this category (Section 5). 

(D) reasonable anomaly for a ringed planet\\
\ \ Finally a small number of systems with a good $S/N$ indeed exhibit 
a possible signature that could be explained by the ring model. 
We perform additional analysis to test the validity of the ring 
hypothesis in a more quantitative fashion (Section 5 and 6).
\parindent = 13pt

The above classification is done on the basis of observed anomalies, which 
are derived by fitting a planet model to light curves. 
The data are taken form the Mikulski Archive for Space Telescopes (MAST), and we 
use the Simple Aperture Photometry (SAP) data taken in the 
long-cadence mode (29.4 min). 
In fitting, we use only the first transit in the light curve for 
each candidate in deriving the observed anomaly for simplicity. 
After fitting the planet model to data, the long-period planet candidates are automatically 
classified into the above categories (A)$\sim$(D). Table \ref{sec3_ring} summarizes the results of classification for four models. In a later section, we use the classification 
according to model I, which contains more candidates 
in categories (B)$\sim$(D) than the other three. In fact, the choice of model I is partly reasonable because the distant planets potentially have tilted rings like Saturn because of low tidal force. 
%Table \ref{table_num} shows the large 
%difference among models, This implies that even though the ringed planets are detectable in some configurations, they cannot be detected in other configurations. 

As candidates in (A) have insufficient $S/N$ for further analysis, 
we do not consider them in the following analysis. 
In section 4, we obtain upper limits on $R_{\rm out}/ R_{\rm p}$ for candidates in 
(B). In section 5, we first search for the ringed planets 
in categories (C) and (D) by visual inspection, 
and later examine the reliability of transits more quantitatively. 
In section 6, we interpret the possible ringed planet candidate. 

  \begin{table}[htpb]

\caption{\small Parameters and classification. }

\centering
\begin{minipage}{\textwidth}

\scalebox{0.68}{
\begin{tabular}{c c c c c c}
\hline \hline
Parameters & Meaning &&& &  \\ 
\hline\hline
%Parameters in $p_{\rm dur} $& & & & &  \\
%\hline \hline
$P$(day) &Period  & $10759$  &&&  \\
$a/R_{\star} $&Scaled semi-major axis  &$2059.67$   &&&\\ 
%\hline \hline
%Parameters in $p_{\rm flux}$ &&&&&   \\
%\hline \hline
%\hline\hline
%Fixed parameters in $p_{\rm other} $& & & & &  \\
%\hline \hline
$q_{1} $ &Limb darkening parameter & $0.49$&   && \\
$q_{2}$ &Limb darkening parameter& $0.34$  &&&  \\
$t_{0}$(day) &Time of a transit center of a planet &$0$  &&&\\ 
$T$ &Shading coefficient &$1.0$   &  && \\ 
$r_{\rm in/p} $ &Ratio of $R_{\rm in}$ to $R_{\rm p}$& $1.0$  && &   \\ 
$r_{\rm out/in}$ & Ratio of $R_{\rm out}$ to $R_{\rm in}$ & \nodata  & & &  \\ 
$R_{\rm p}/R_{\star} $ & Planet to star radius ratio & \nodata   &&& \\ 
\hline \hline\\

& &{\bf model I (Fiducial)} &model I\hspace{-.1em}I  & model I\hspace{-.1em}I\hspace{-.1em}I & model I\hspace{-.1em}V   \\
\hline \hline
$b$ &Impact parameter & 0.8 & 0.5& 0.8 & 0.5   \\ 
$\theta$(deg)&Angle between $Y$-axis and axis of the ring &45& 45&$\arcsin(0.1)$&$\arcsin(0.1)$ \\  \hline
$\phi$(deg)&Angle between $Z$-axis and &45 &45& 0&0 \\ 
&   ring-axis projected onto $(Z,X)$-plane& & & &   \\ \hline
$r_{\rm eq}$ &Boundary value, above which the sky-projected 
&2.0 & 2.0 & 10.0 & 10.0  \\
&ring is larger than the planetary disk & & & &   \\
&(see Section B.3 for detailed explanation) & & & &   \\

%\hline \hline
%Category & $model(1)$ & $model(2)$ & $model(3)$ & $model(4)$ \\
 \hline\\

& Classification&\multicolumn{4}{c}{Number of classified systems }   \\\hline \hline
&(A) insufficient $S/N$  &66 &82 & 80&82  \\
&(B) sufficient S/N to $R_{\rm out}/R_{\rm p}$ &8 &1 &2 &1  \\
&(C) too strong anomaly  &8 & 3& 4&4  \\
&(D) possible candidate  &7 &3 &3 &2 \\
\hline \hline

\end{tabular} 
}
\end{minipage}

  \label{sec3_ring}
\end{table}

\section{Upper limits of $R_{\rm out}/R_{\rm p}$ for candidates in (B)} 
\begin{figure*}[htbp]
\begin{center}
\includegraphics[width=0.32 \linewidth]{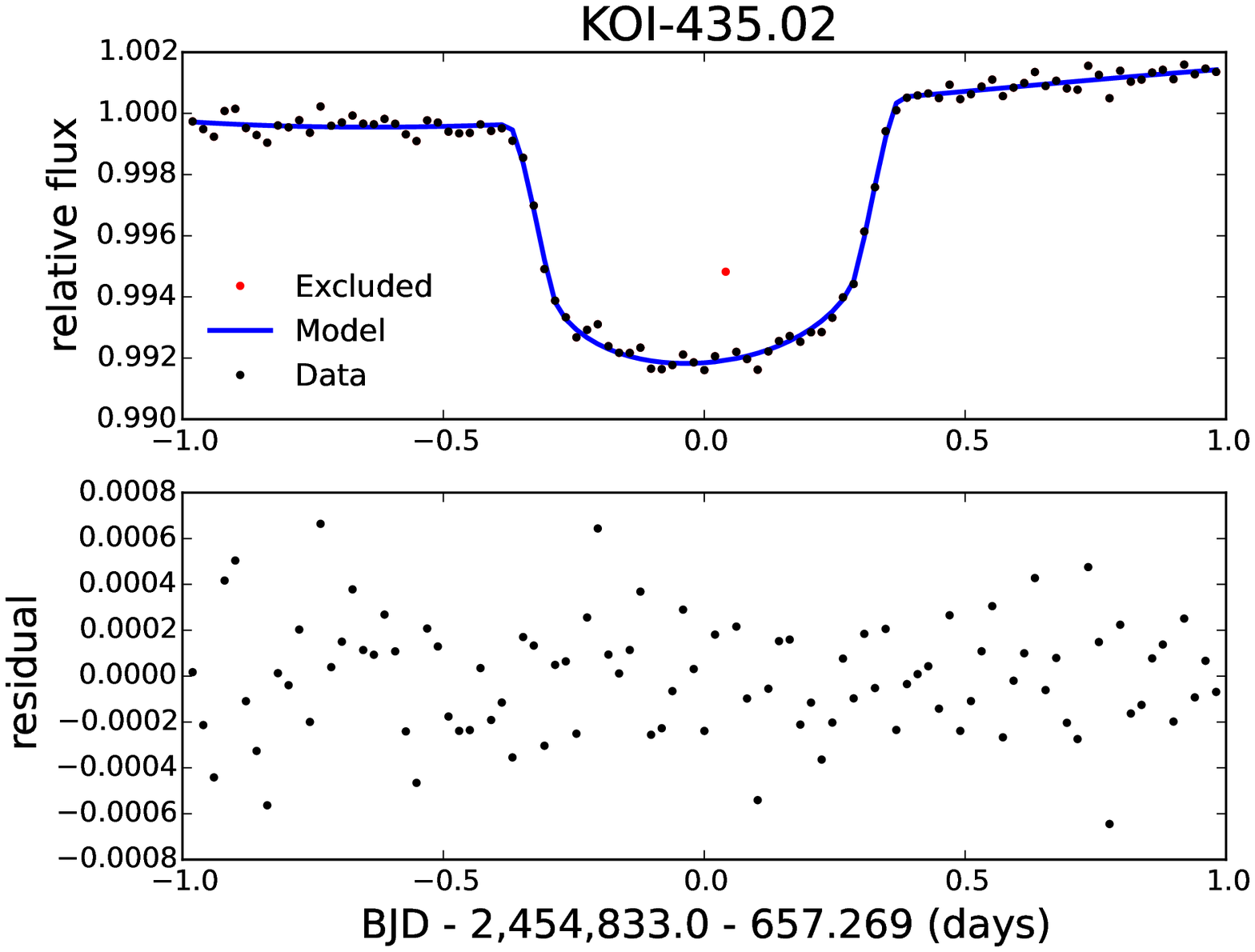}\hfill%
\includegraphics[width=0.32 \linewidth]{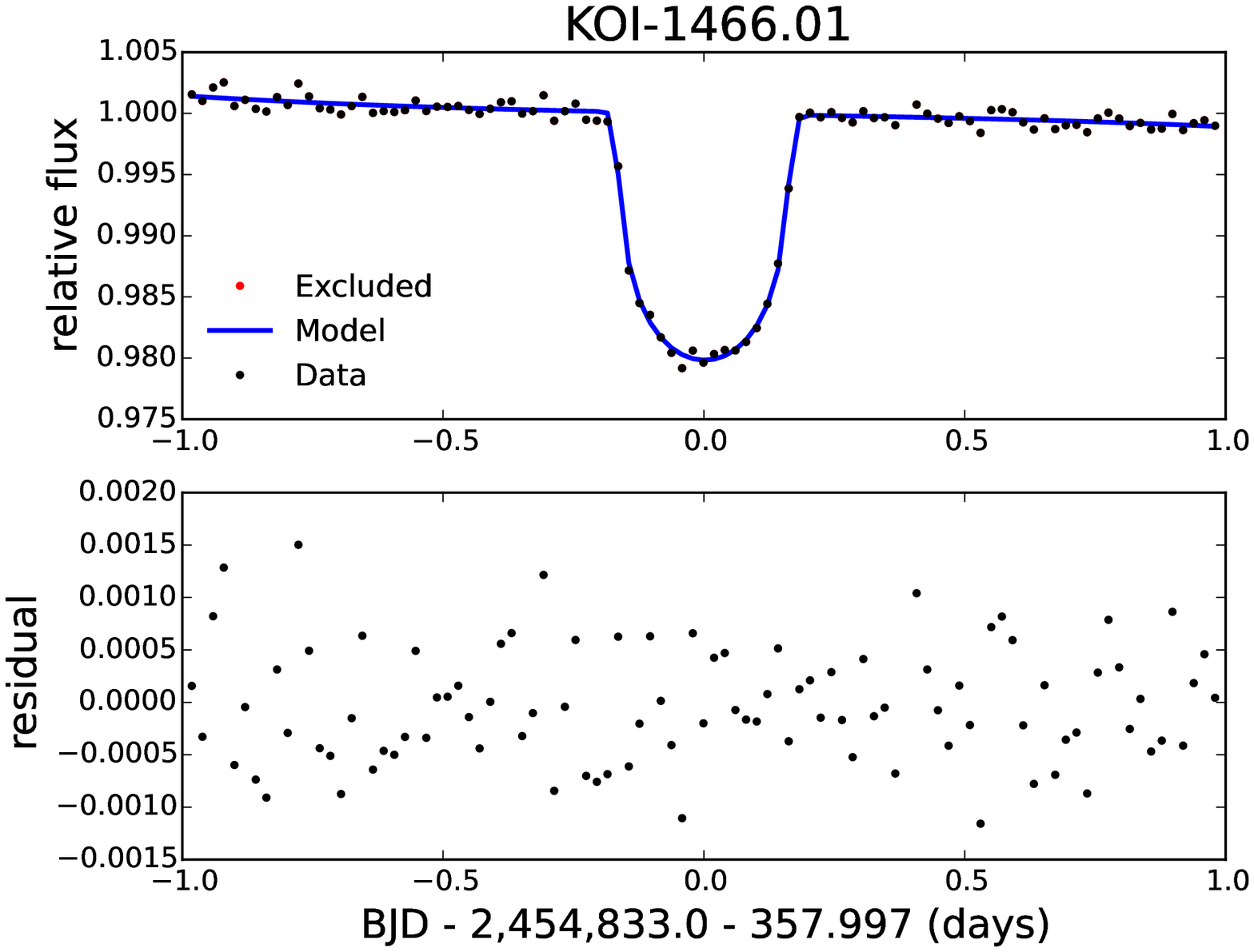}\hfill%
\includegraphics[width=0.32 \linewidth]{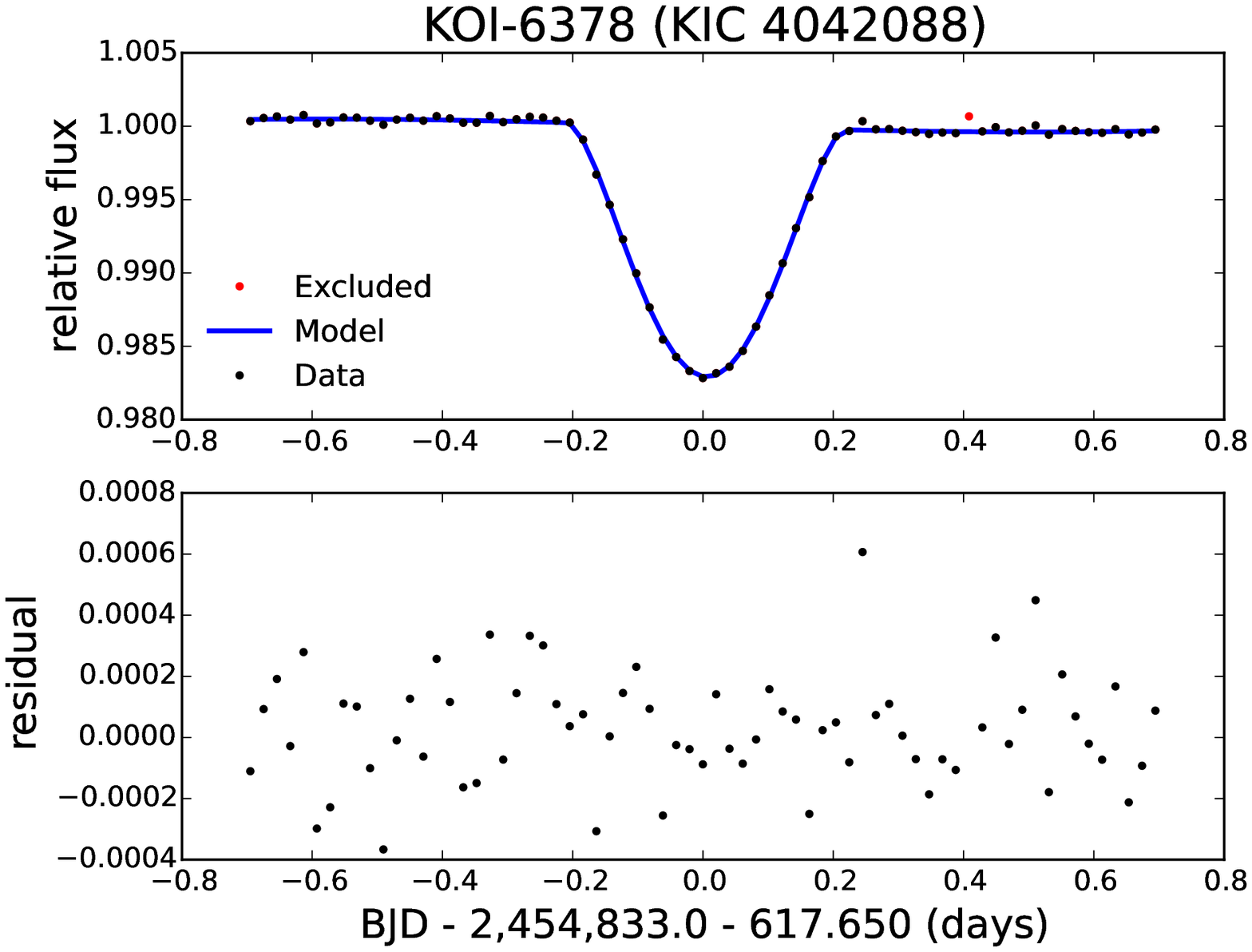}
\vspace{4pt}

\includegraphics[width=0.32 \linewidth]{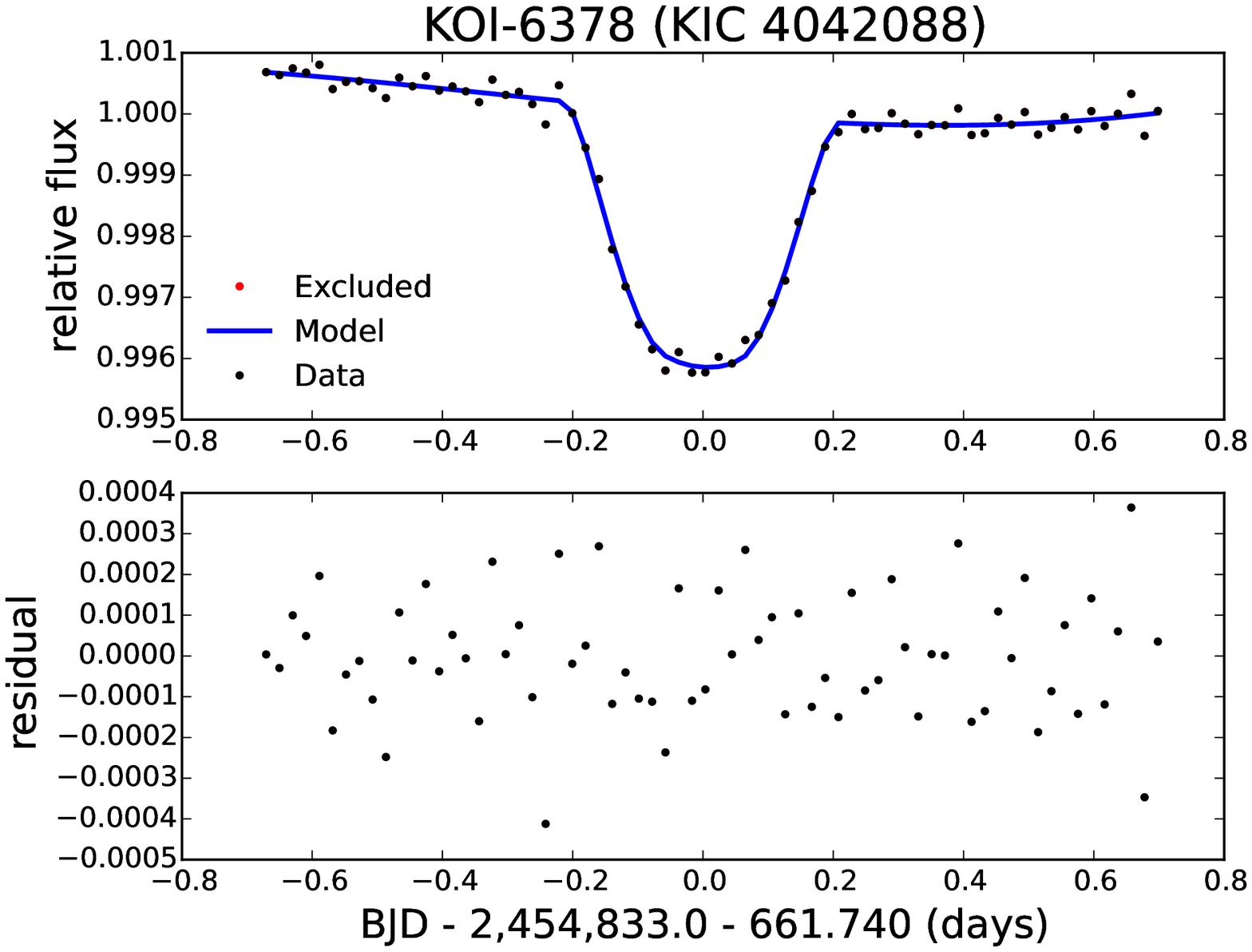}\hfill%
\includegraphics[width=0.32 \linewidth]{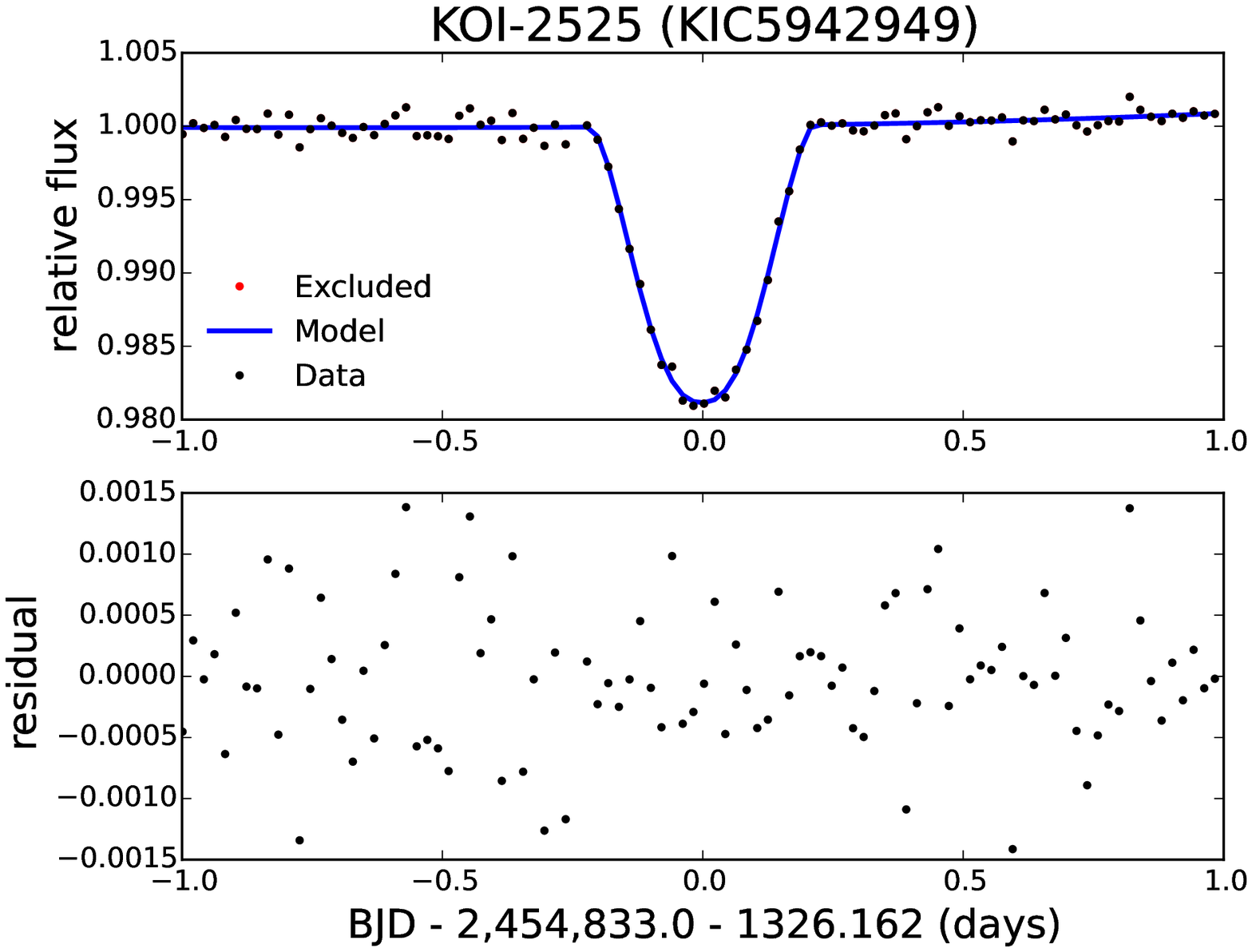}\hfill%
\includegraphics[width=0.32 \linewidth]{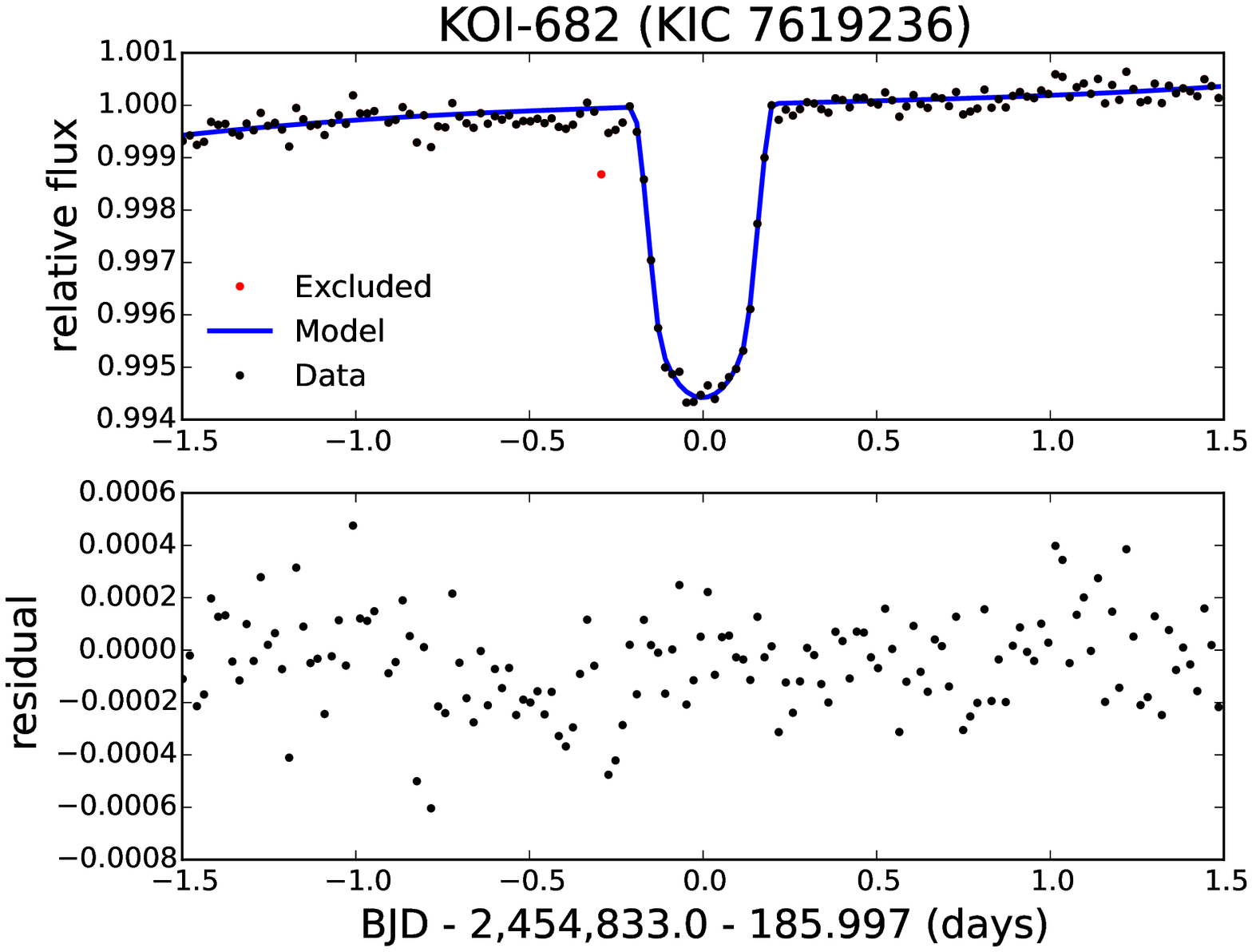}
\vspace{4pt}

%\centering
\includegraphics[width=0.32 \linewidth]{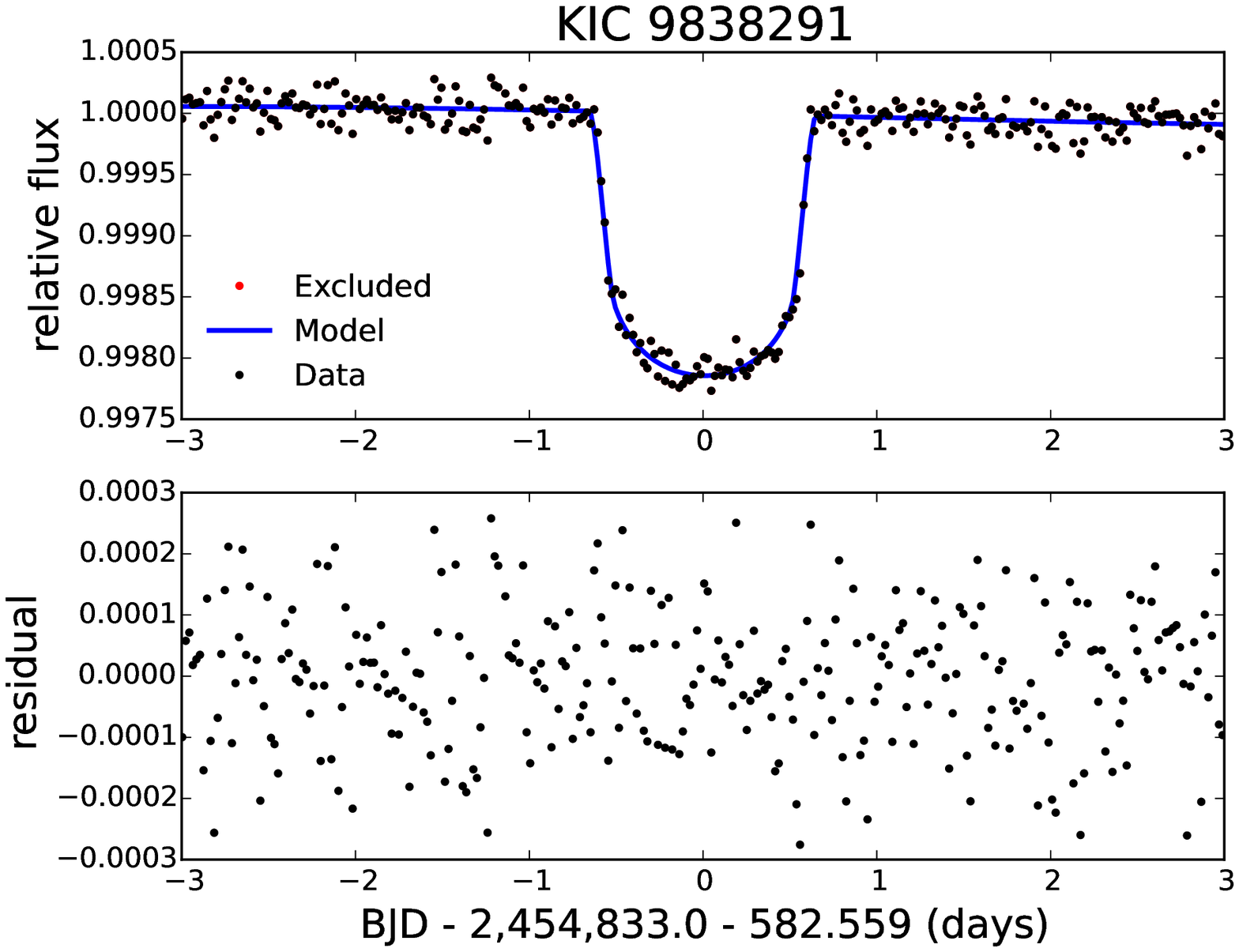} 
\hspace{0.8pt}
\includegraphics[width=0.32 \linewidth]{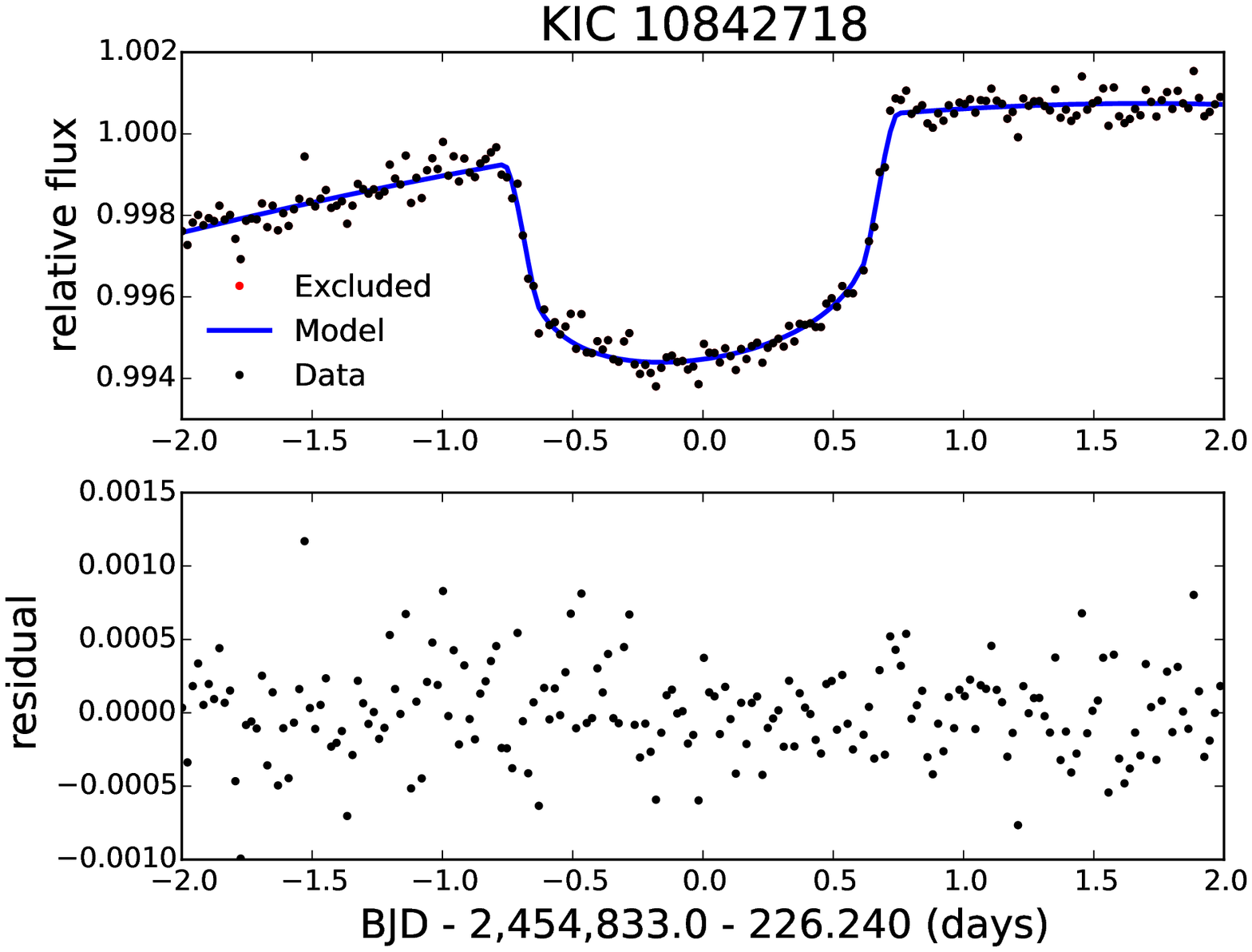} 
\hspace{-0.4pt}

\end{center}
{\caption{ Candidates in (B) with sufficient $S/N$ to constrain $R_{\rm out}/R_{\rm p}$. 
The light curves (black circles) are shown with the 
best-fit planet models (blue lines). The red points are excluded in fitting. 
The horizontal axis shows time in BKJD = BJD - 2454833.0 
(the Barycentric Kepler Julian Date) offset by the 
central time of the transit. 
} 
\label{class_B}}
\end{figure*}

Upper limits on $R_{\rm out}/R_{\rm p}$ are given for candidates in (B) 
as a result of classification.  
Figure \ref{class_B} shows the light curves and fitted curves 
of eight candidates classified to (B) in model I.
They show no appreciable anomalies 
in the residual relative to the single planet model. 
For these candidates, we could detect the ring signature if exists. 
Thus in turn, we can derive the upper limits on $R_{\rm out}/R_{\rm p}$. 
This is done by simply comparing the expected anomaly in model I 
and the observed anomaly in the light curve. 
The details of the method to place upper limits on 
$R_{\rm out}/R_{\rm p}$ are described in Appendix B and C, 
and the results are summarized in Table \ref{table_D}. 

 \begin{table}[htpb]
\caption{\small Upper limits of radii of outer rings.} 
\centering
\scalebox{0.8}{
\begin{tabular}{c c c c c}
 \hline 
Name & Upper limits of $R_{\rm out} /R_{\rm p}$ in model I 
& $R_{\rm p}(R_J)$ &$ a $(AU) & Transit Epoch (BKJD) \\
\hline Candidates in (B)& & &\\ \hline
KOI-435.02  &1.5 & 0.66 & 1.28 & 657.269  \\
KOI-1466.01 &1.5 &1.13 & 1.14 & 357.997\\
KIC 4042088  &1.2 & 2.94 & 0.78&617.65\\
KIC 4042088  &1.95 & 0.85 & 1.41&661.74 \\
KIC 5942949  &1.5&1.18 & 1.13 &1326.162 \\
KIC 7619236 &1.7 & 0.71&1.35 & 185.997 \\
KIC 9838291 & 1.9 & 0.42 & 14.3 &582.559  \\
KIC 10842718 &1.6& 0.75 & 7.60&226.300 \\
\hline Candidates in (D) & & &&\\ \hline
KOI-490.02 & 1.2 & 1.16 & 2.53& 492.772\\
KOI-868.01 & 1.2 & 0.76 & 0.74 & 208.401\\
KIC 8012732 &1.8 & 0.67 & 0.68 & 391.807\\
KIC 8410697 & 1.8 & 0.77 & 3.19 & 542.122\\
\hline
\end{tabular} 
}
\label{table_D}
\end{table}

\section{Search for ringed planets}
In this section, we search for ringed planets in categories (C) and (D), extract 
the tentative ringed planet candidates, and examine whether 
the transits are not  false positive. 

\subsection{Tentative selection of possible ringed planets}
\begin{figure*}[htbp]
\begin{center}
\includegraphics[width=0.32 \linewidth]{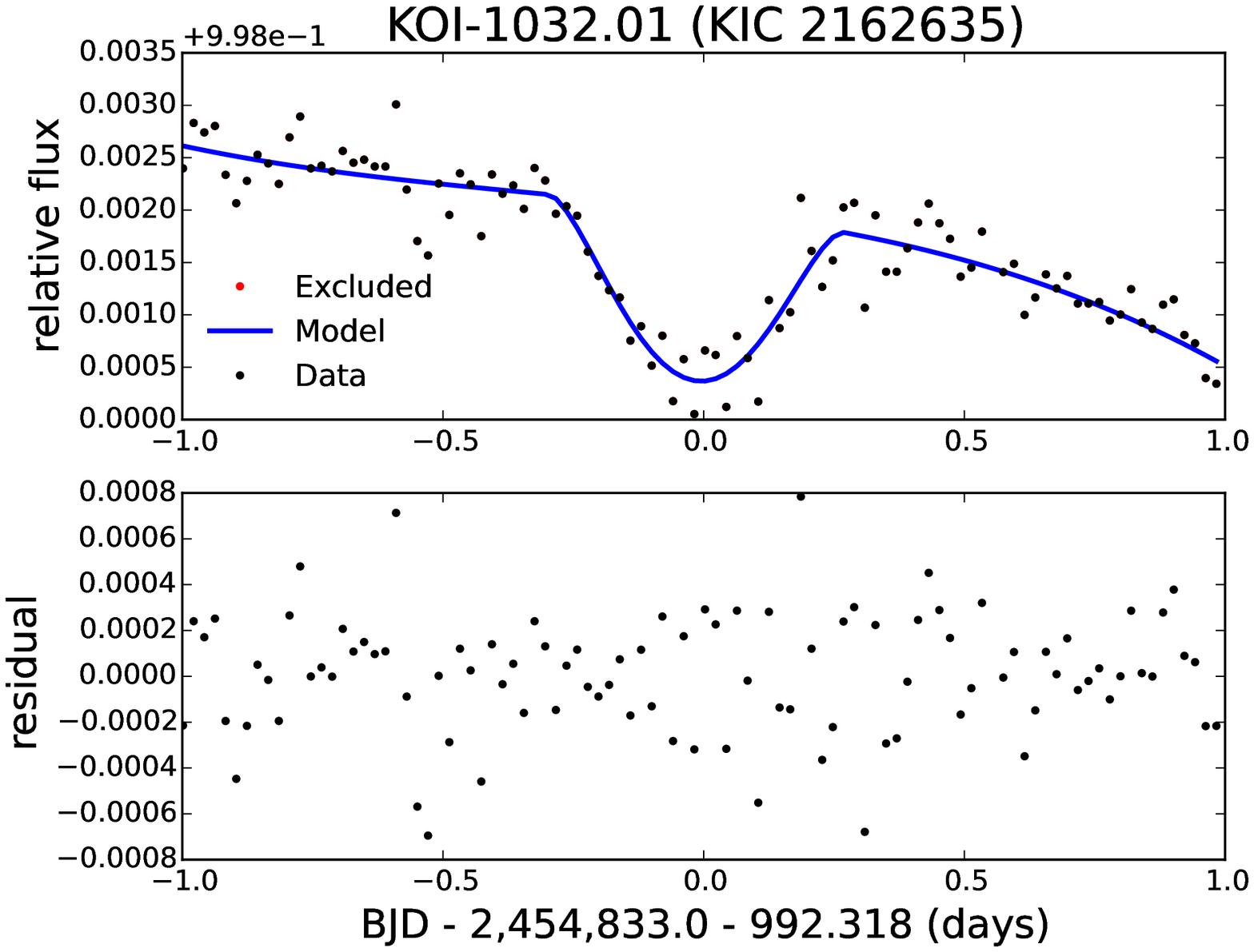}\hfill%
\includegraphics[width=0.32 \linewidth]{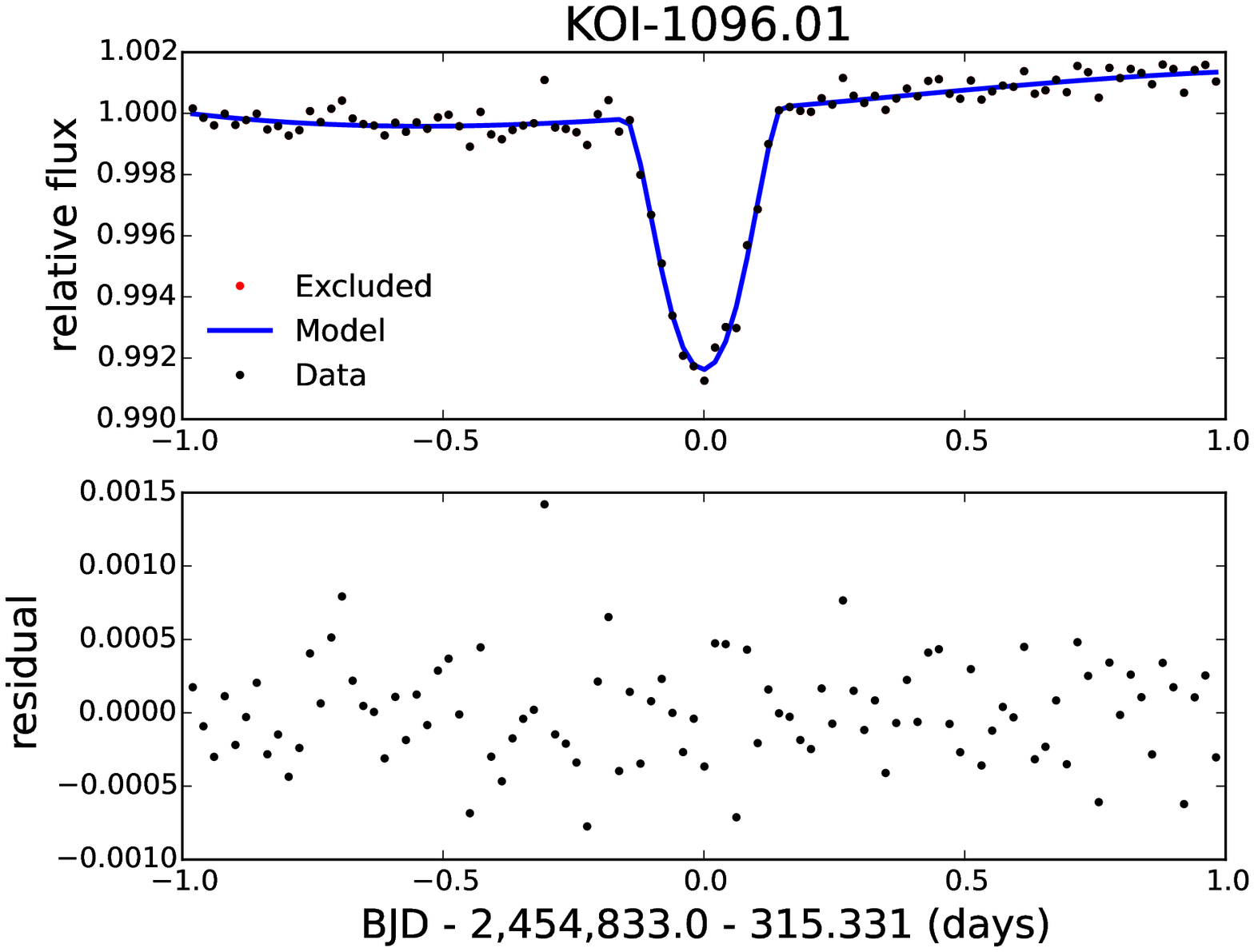}\hfill%
\includegraphics[width=0.32 \linewidth]{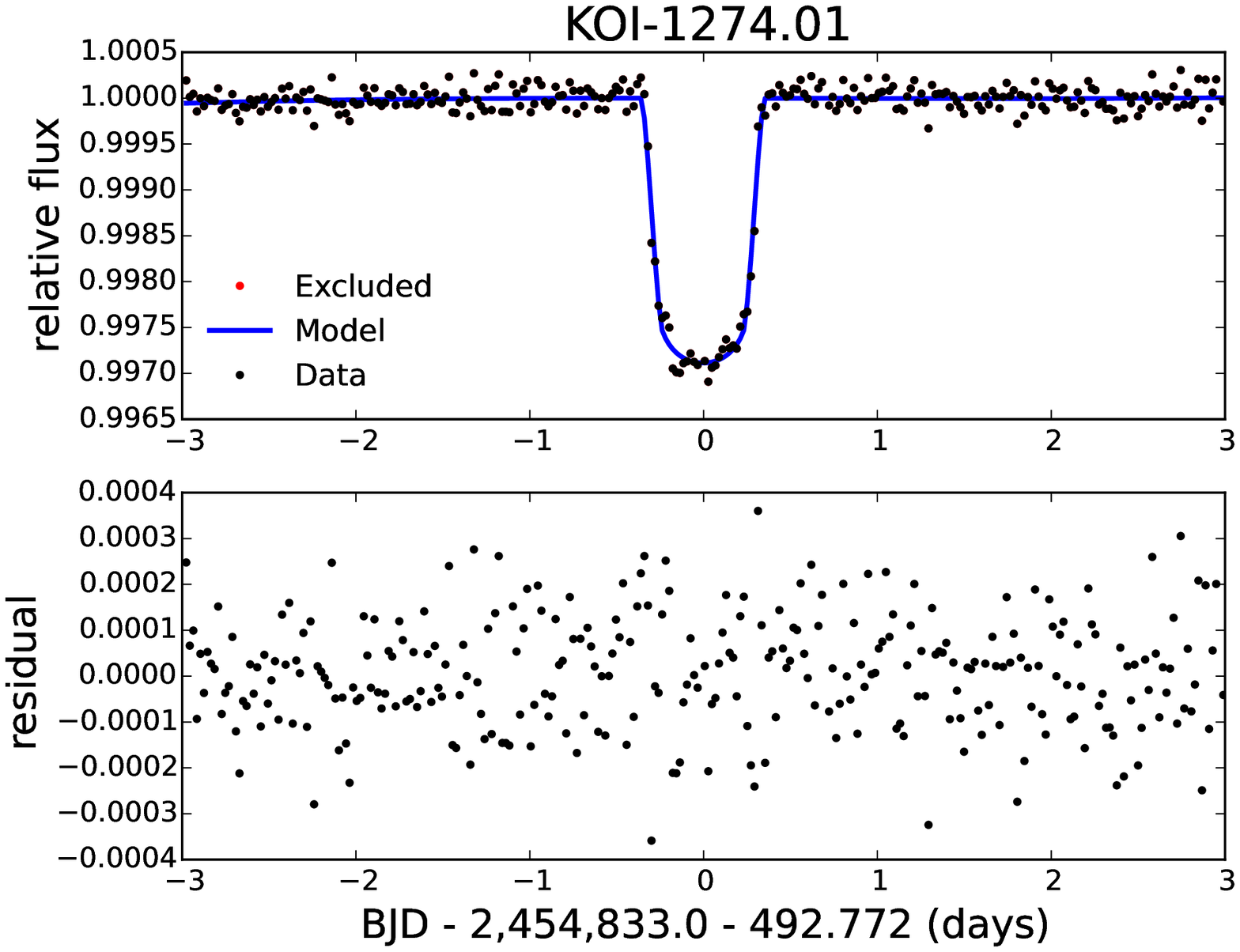}
\vspace{4pt}

\includegraphics[width=0.32 \linewidth]{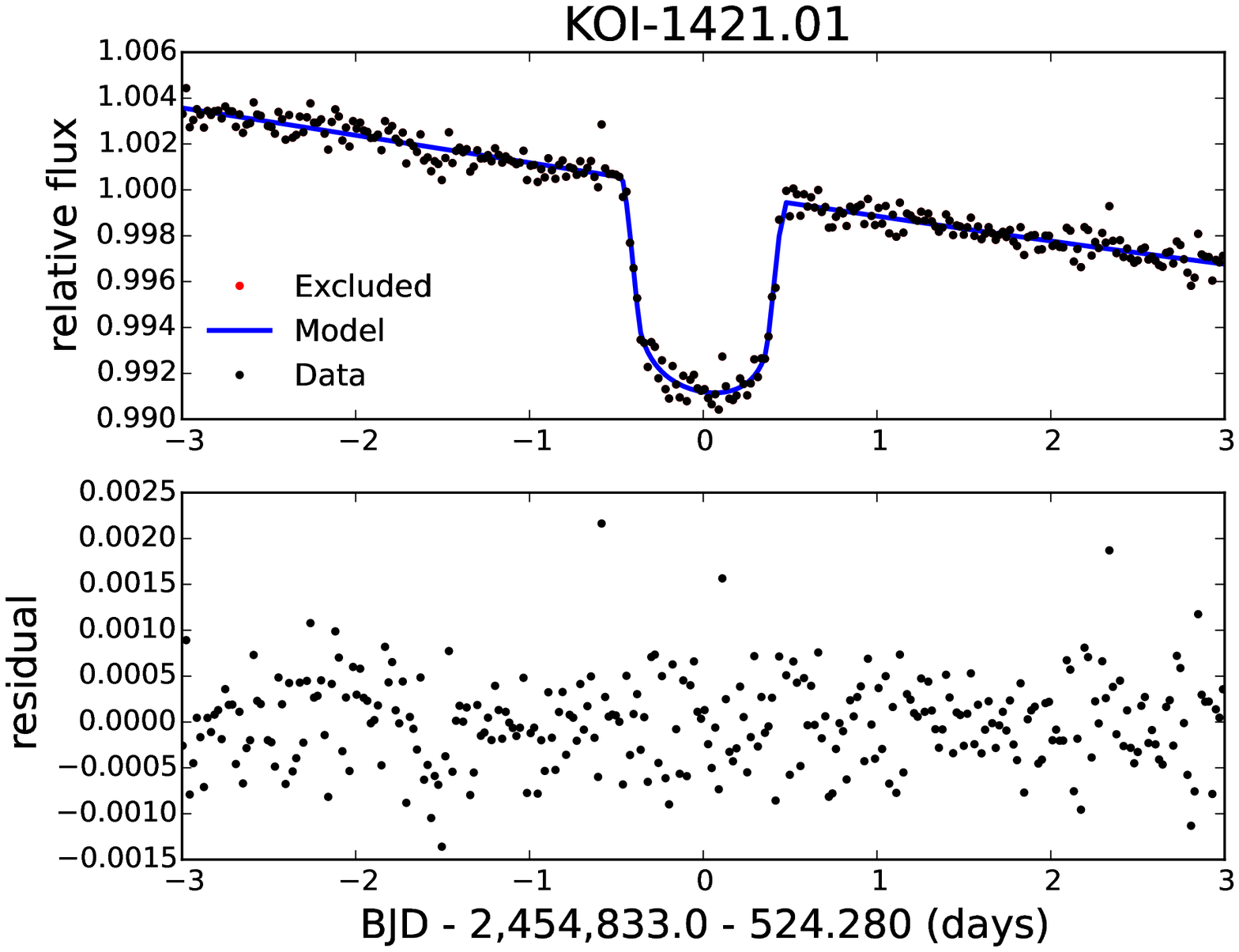}\hfill%
\includegraphics[width=0.32 \linewidth ]{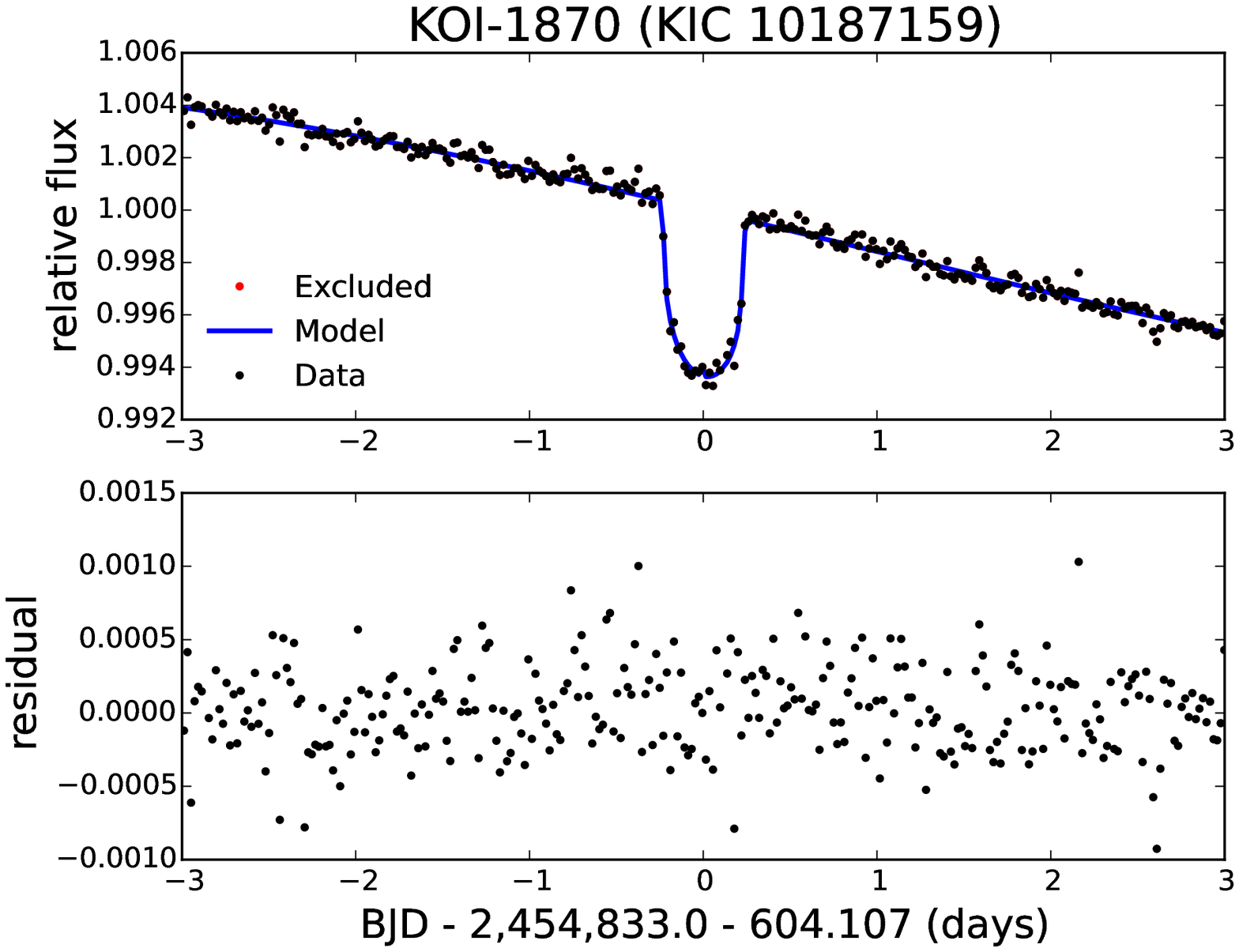}\hfill%
\includegraphics[width=0.32 \linewidth]{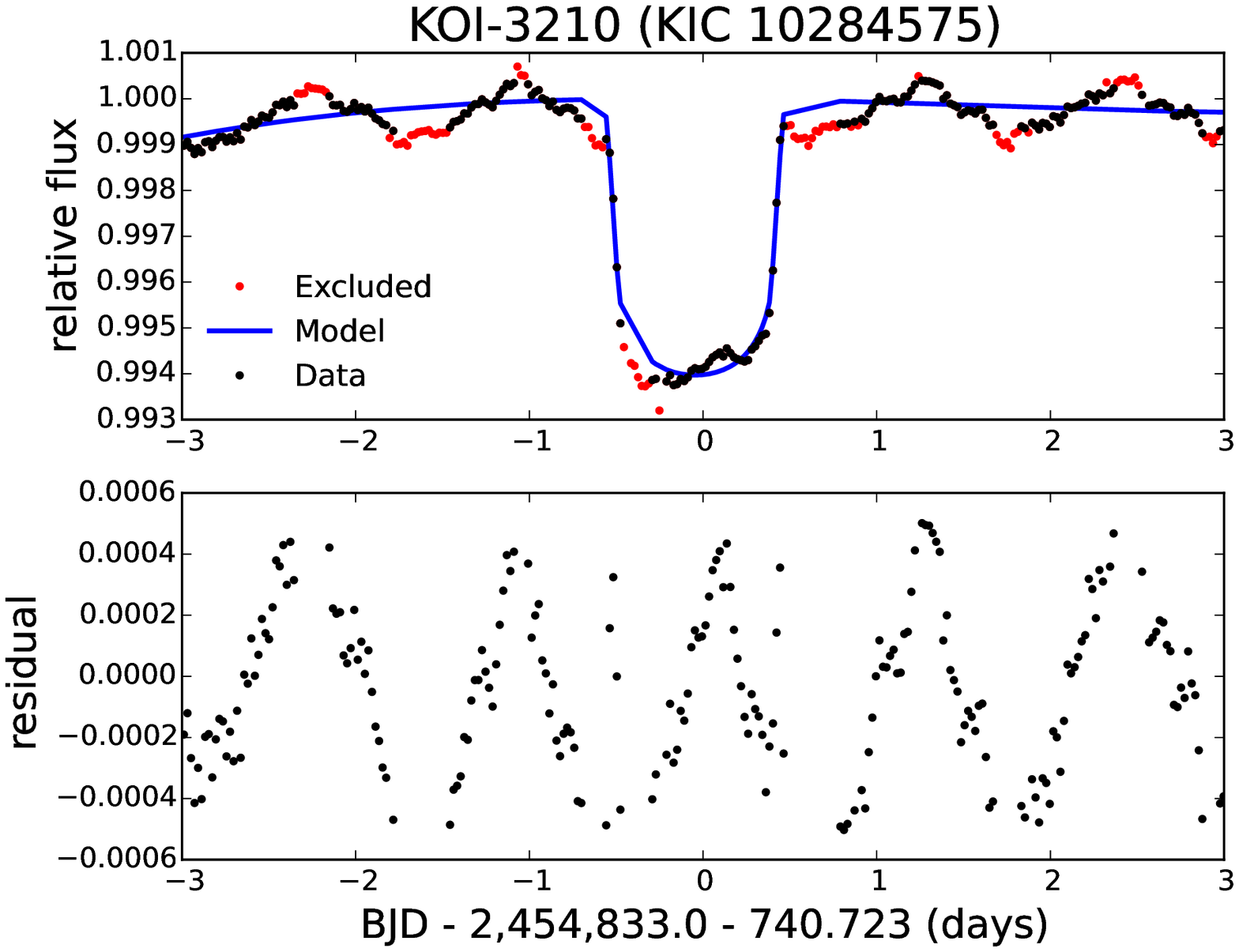}
\vspace{4pt}

\includegraphics[width=0.32 \linewidth]{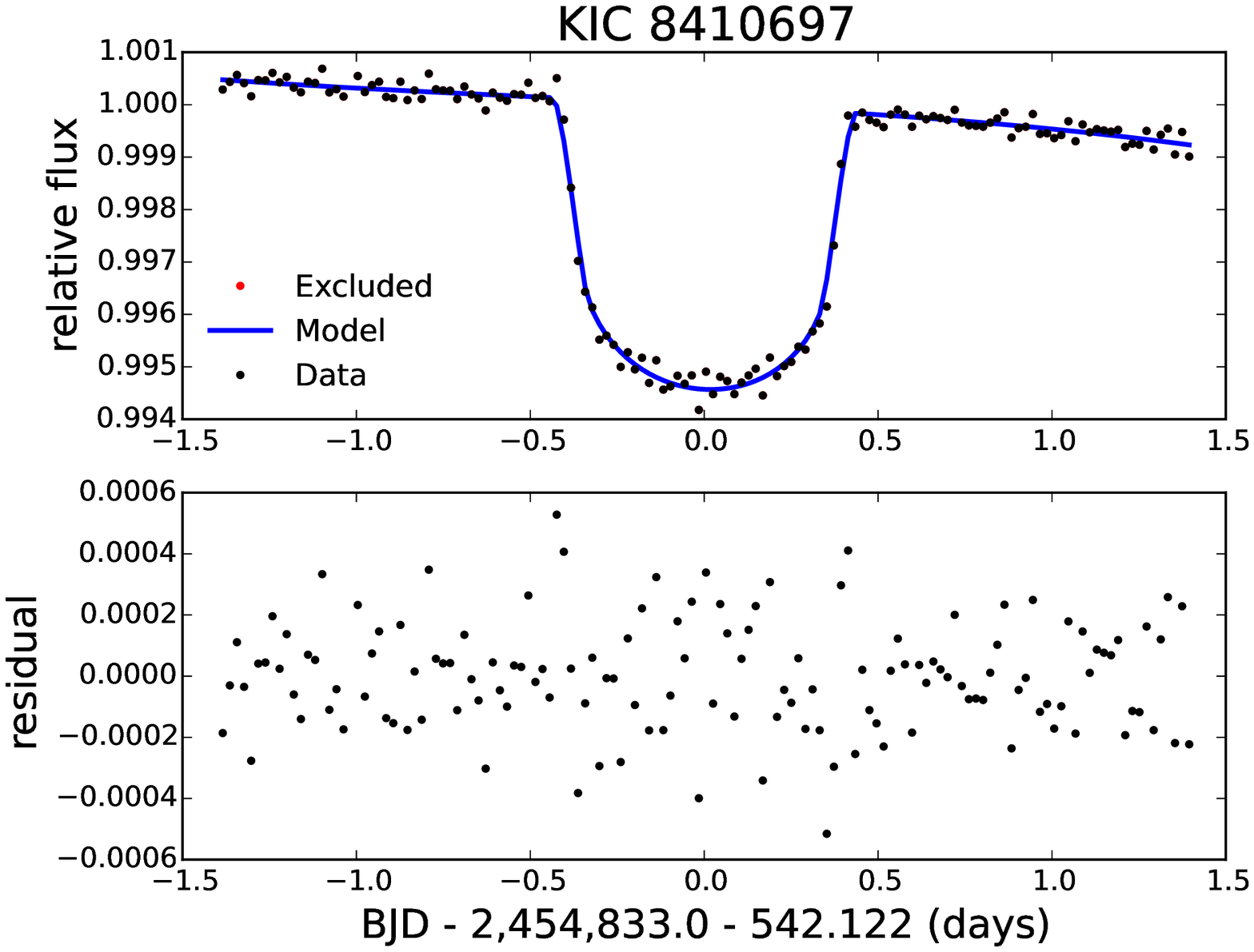}
\hspace{0.8pt}
\includegraphics[width=0.32 \linewidth]{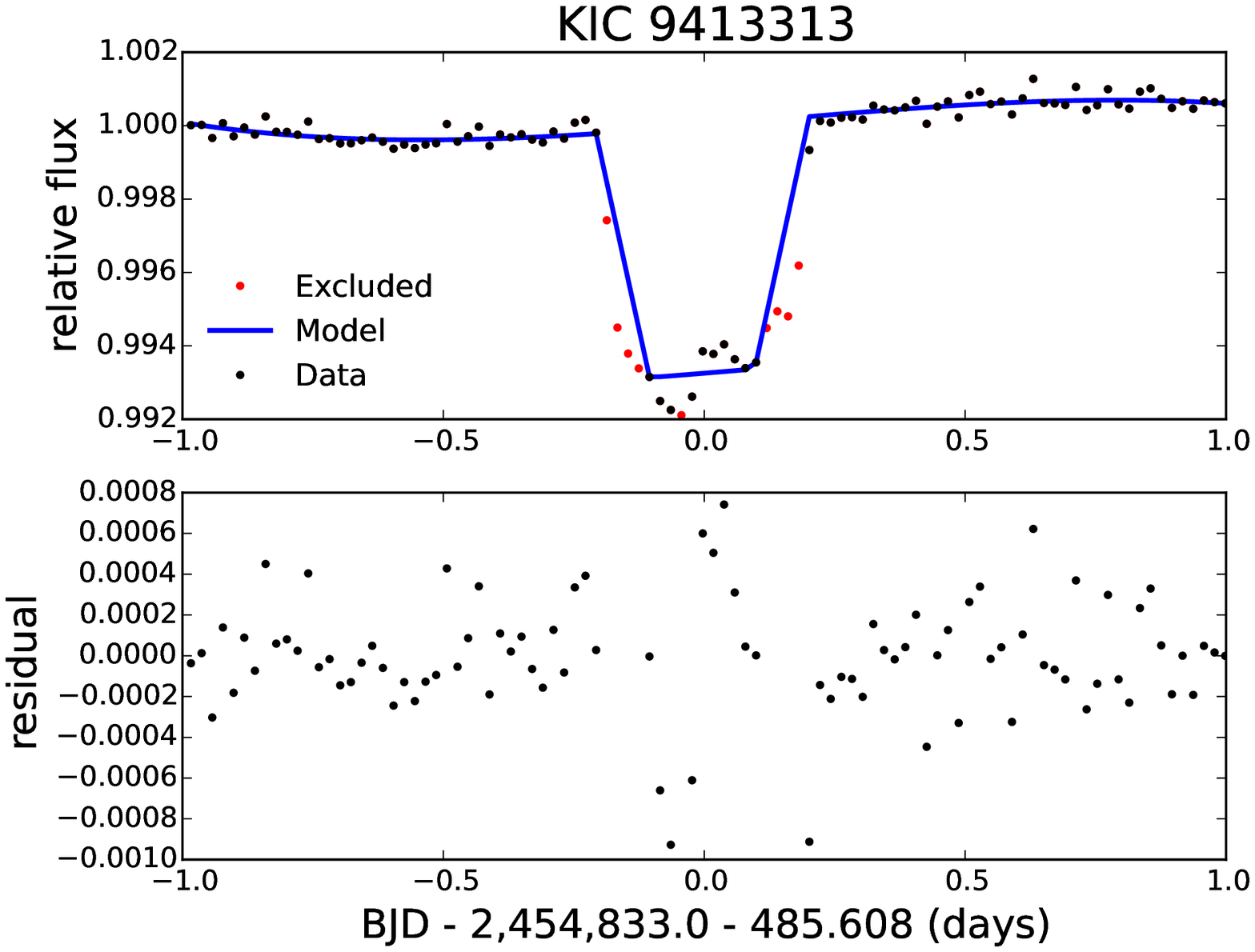}
\hspace{-0.4pt}

\caption{\small Candidates in (C) with too large anomaly for a 
ringed planet. The format of the figure is the same as Figure \ref{class_B}.
\label{class_C}}
\end{center}
\end{figure*}

\begin{figure*}[htbp]
\begin{center}
\includegraphics[width=0.32 \linewidth]{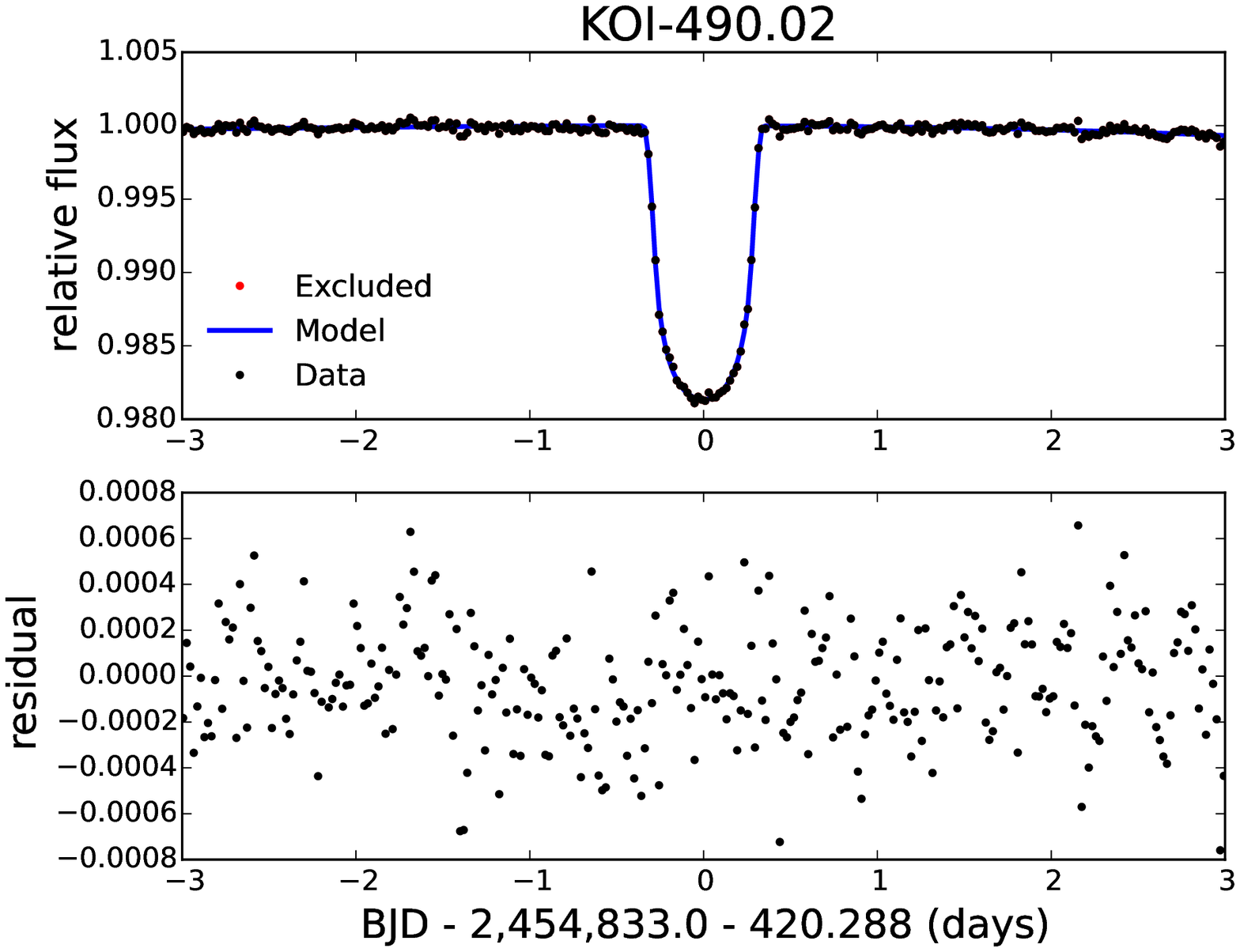}\hfill%
\includegraphics[width=0.32 \linewidth]{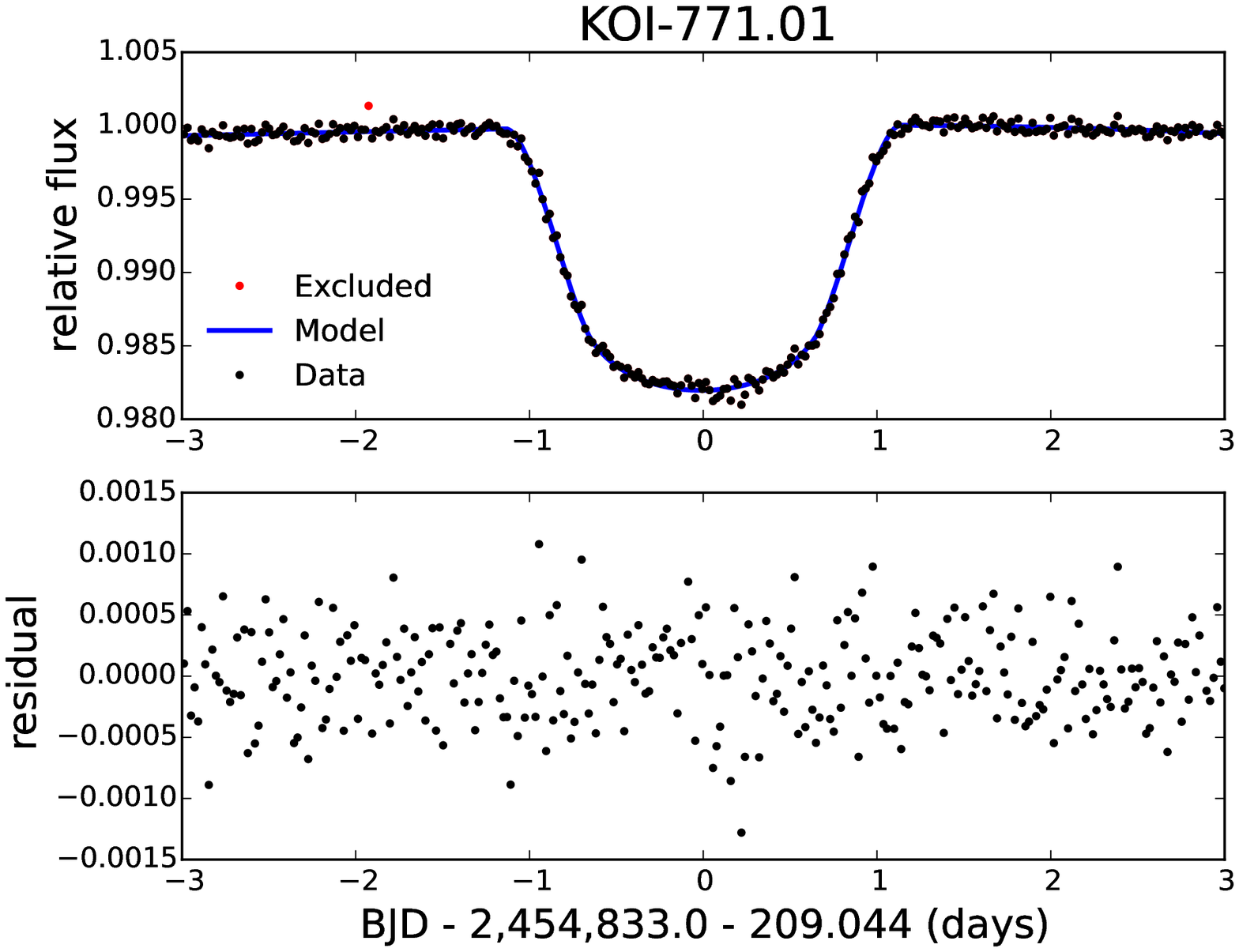}\hfill%
\includegraphics[width=0.32 \linewidth]{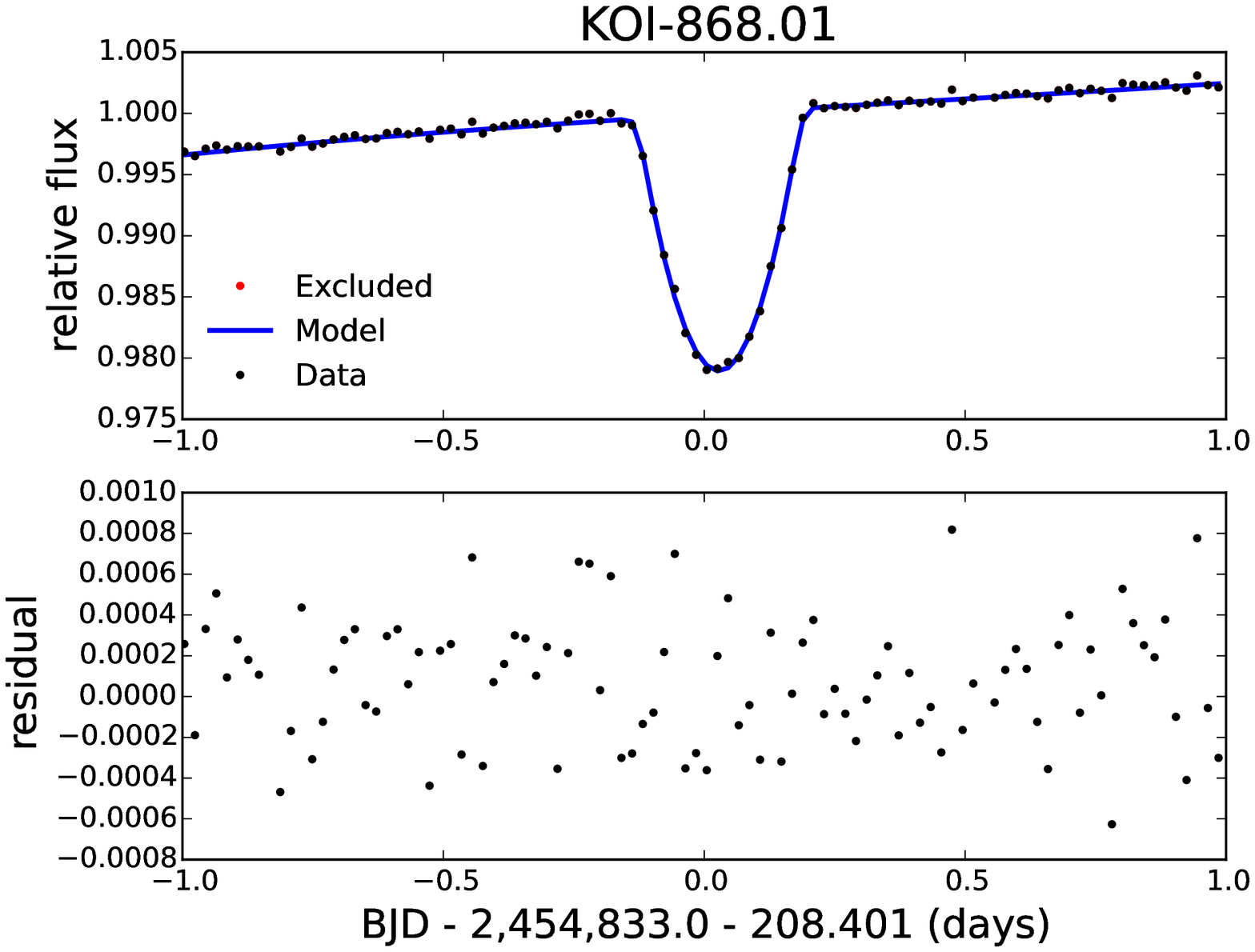}
\vspace{4pt}

\includegraphics[width=0.32 \linewidth]{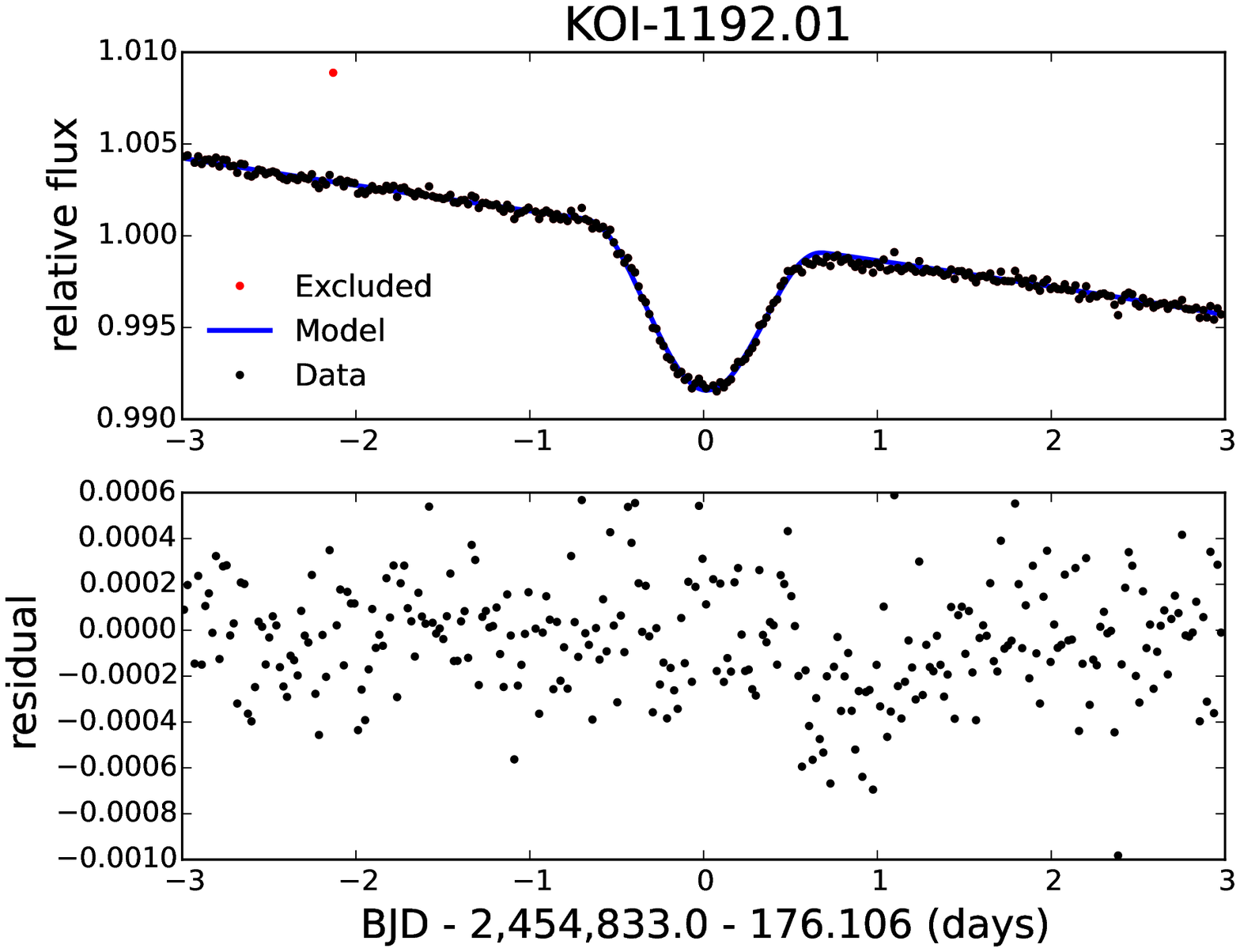}\hfill%
\includegraphics[width=0.32 \linewidth]{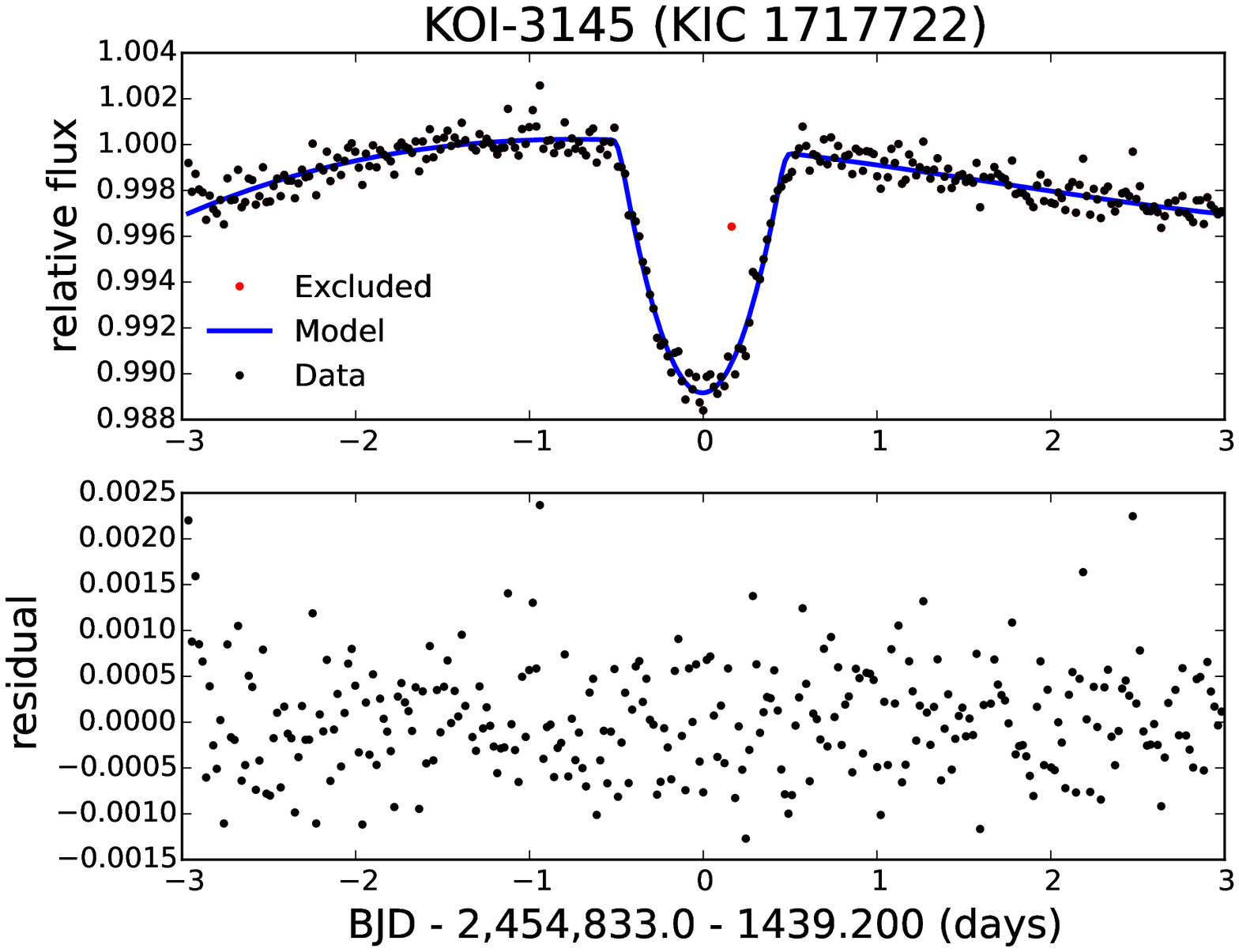}\hfill%
\includegraphics[width=0.32 \linewidth]{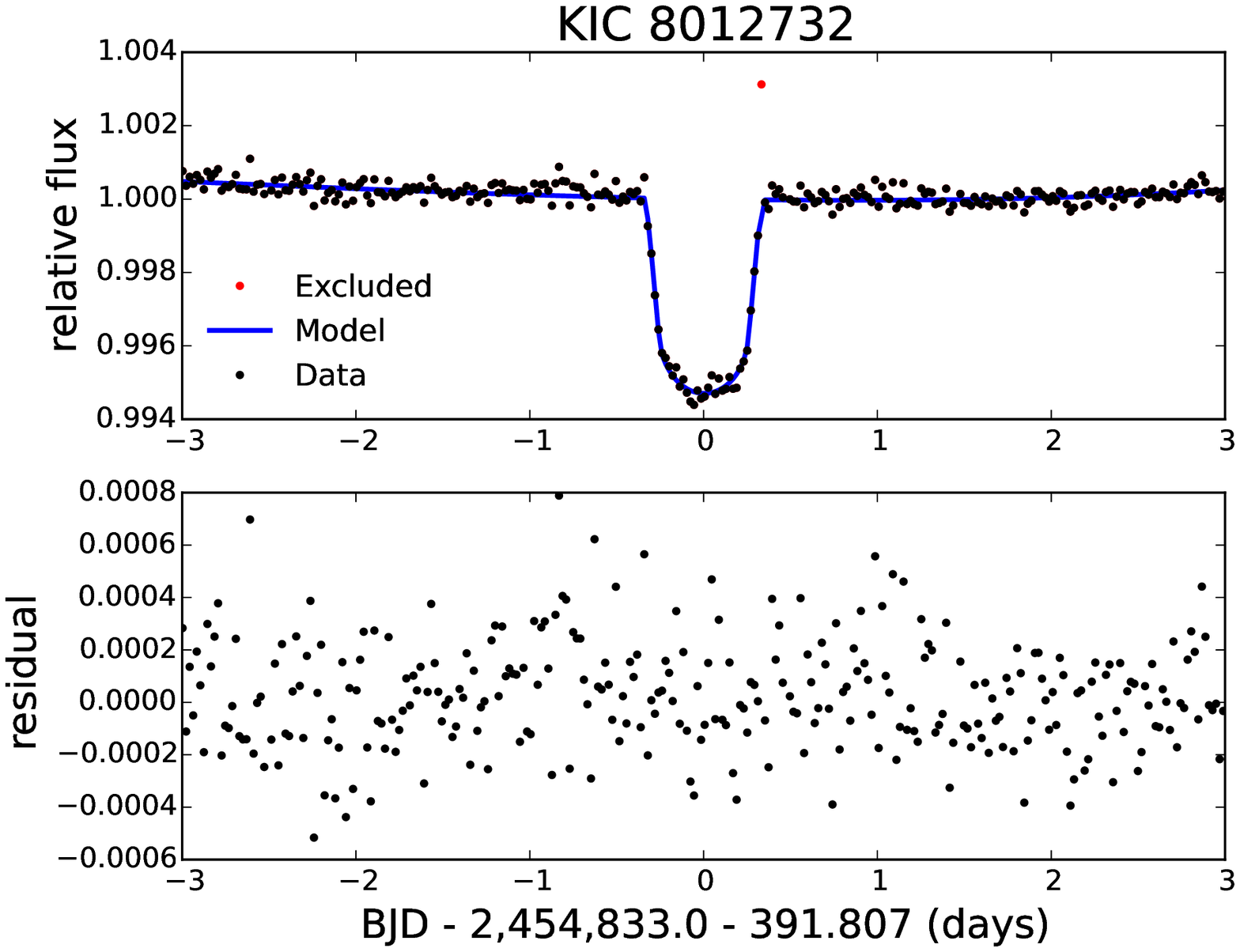}
\vspace{4pt}

\includegraphics[width=0.31 \linewidth]{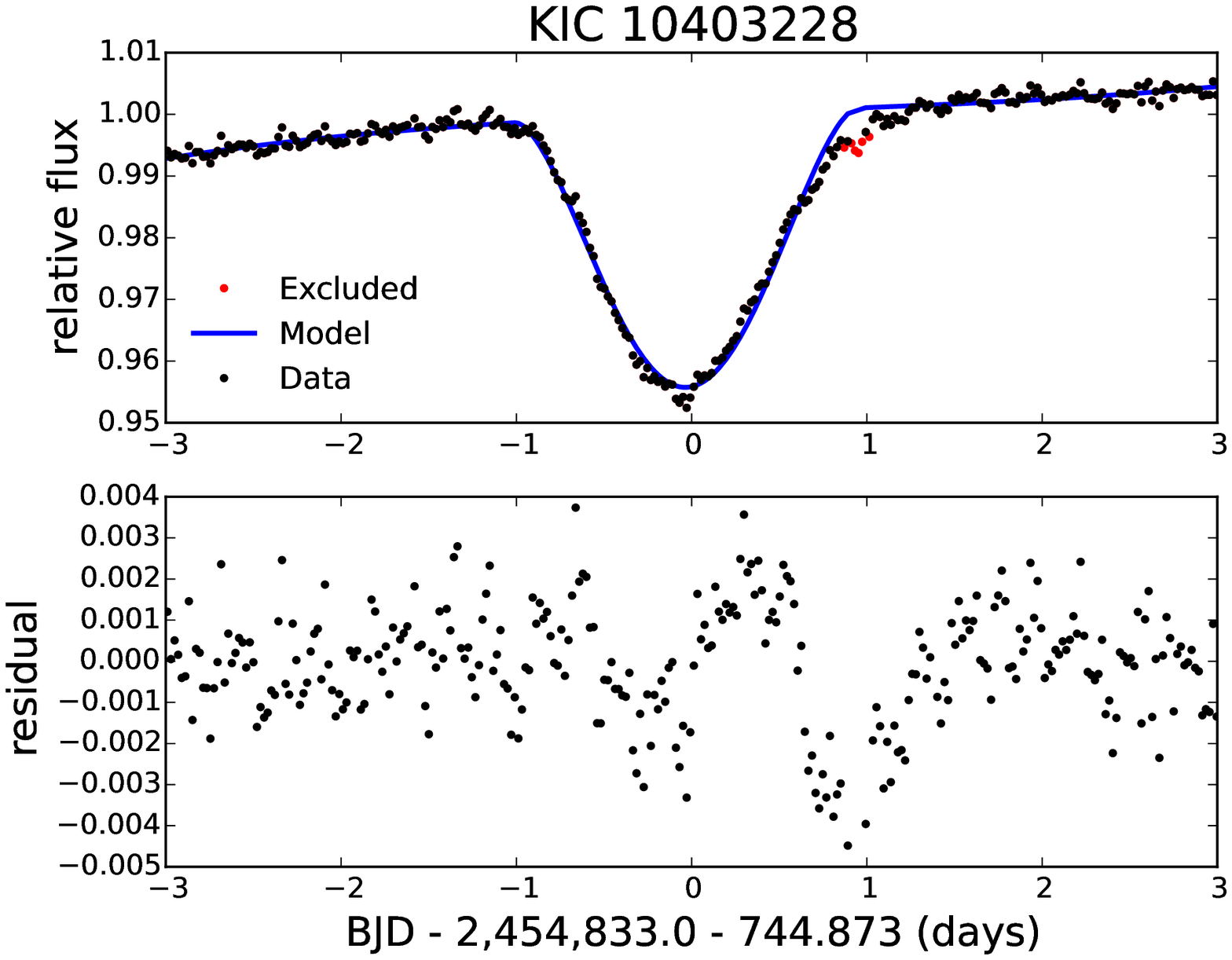}

\caption{\small Candidates in (D) with moderate anomaly for a ringed planet.
The format of the figure is the same as Figure \ref{class_B}. \label{class_D}}
\end{center}
\end{figure*}

Figures \ref{class_C} and \ref{class_D} show the 
light curves of candidates in categories (C) and (D), 
respectively. Candidates in (C), where 
the observed anomaly exceeds the prediction of 
model I, may be consistent with other ringed planets 
in different configurations. 
Thus, we search for ringed planets not only in (D) but also (C).  

We extract ringed planet candidates by visual inspection of their 
light curves on the basis of following properties expected for ringed planets: 
\begin{itemize}
\item Duration of ingress and/or egress is long.
\item Transit shape is asymmetric due to the non-zero $\phi$. 
\end{itemize} 
As a result, we identify five systems 
KOI-771(D), KOI-1032(C), KOI-1192(D), KOI-3145(D), and KIC 10403228(D) 
as tentative ringed planets. For the other four candidates in (D), which 
show no visible ring-like feature in the light curves, 
we obtain the upper limits on $R_{\rm out}/R_{\rm p}$ in the same 
method as in the previous section (Table \ref{table_D}). 
In total, we obtain the upper limits on 
$R_{\rm out}/R_{\rm p}$ for 12 candidates, and 
the six of them have $R_{\rm out}/R_{\rm p}\leq 1.5$. 

For six candidates in (C) with no ring-like features, 
we cannot set the upper limits of ring parameters, and 
we conclude that the signals are not due to rings, 
but are due to the temporal stellar activities. 

\subsection{Elimination of false positives} 
\begin{figure}[htpb]
  \centering
  %\begin{tabular}[b]{@{}p{0.45\textwidth}@{}}
  \includegraphics[width=.6\linewidth]{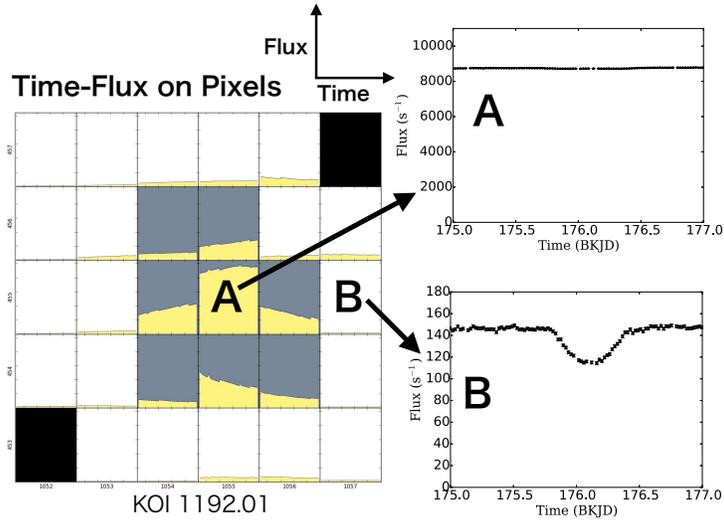}
  \caption{\small Indication that KOI-1192.01 is a false positive. 
  The light curves of the pixels labeled as A and B are shown 
  in the right panels. 
  The pixel located at A receives the largest amount of light among 
  pixels of the target star, and the pixel located at B receives the small amount of light. 
  The very different relative depths in the 
  two pixels indicate that the signal does 
  not originate from the target star at the center. }
  \label{FP_1}
\end{figure}

We examine the reliability of transit signals 
for the five preliminary candidates. 
As a result, we find that four are false positives, 
and KIC 10403228 still passes all criteria. 
More specifically, we regard a target as a 
false positive if one of the following criteria is satisfied \citep{2016ApJS..224...12C}. 
\begin{enumerate}
\item[Criterion 1:] The target object exhibits a significant secondary eclipse, 
which is expected for an eclipsing binary. 
\item[{\it - Results}:]
None of our candidates exhibits the secondary eclipse.
\item[Criterion 2:] The signal originates from the other nearby stars or 
instrumental noise. 
\item[{\it - Results}:] Inspecting Target Pixel Files, 
we found that the dips in the light curves of 
KOI-1032.01, KOI-1192.01, and KOI-3145 do not come from the target stars. 
Figure \ref{FP_1} shows an example of KOI-1192.01. 
Community Follow-up Observing Program (CFOP) 
classifies KOI-1032.01 as a false positive \citep{2016ApJ...822....2U}. 
\cite{2015ApJ...815..127W} and \cite{2016ApJ...822....2U} 
also indicate that KOI-1192.01 and KOI-3145 are false positivities. 
Moreover, we find that the transit depths in the light curves of KOI 771.01 
differ in many pixels, and the contaminations from the non-target stars are 
very strong. \cite{2015ApJ...815..127W} also pointed out that 
this system is false positive. 
For KIC 10403228, the transit depths differ in only two pixels, while it 
is constant in the other pixels, so we conclude that the signal 
is originated from the target star. The more detailed discussion 
of KIC 10403228 is presented in a later section. 

\item[Criterion 3:] The transit simultaneously occurs at different stars 
in different pixels. This indicates that the signal does not originate from 
the target but from the instrumental noise. 
\item[{\it - Results}:] The transit events of KOI-1032.01 
and KOI-1192.01 are located at the same time. This result is 
consistent with that of the Criterion 2.  
\item[Criterion 4:] The shape of the light curve is inconsistent with that of 
a transiting object. 
\item[{\it - Results}:] From Figures \ref{class_C} and \ref{class_D}, 
all signals fit well to transit-like features. 
\end{enumerate}
KIC 10403228 is the single system that passes all the criteria. 
Thus, we move on to the detailed pixel-based analysis next.

\subsection{Detailed pixel analysis on KIC 10403228}
KIC 10403228 is considered to be an M dwarf and has 
a nearby star separated by about 3 arcsec \citep{2014ApJ...788..114R}. 
According to the data taken by
United Kingdom Infrared Telescope (UKIRT),
the nearby star is located at (RA, Dec) = $(19^{\rm h}\, 24^{\rm m}\, 54.25^{\rm s},\; +47^{\circ}\,32'\, 57.5'')$ and its J-band flux is about $1/5$ of KIC 10403228. 
Here we examine the possibility that the transit 
is associated with this nearby star rather than KIC 10403228.

\begin{figure}[htpb]
  \centering
  %\begin{tabular}[b]{@{}p{0.45\textwidth}@{}}
  \includegraphics[width=.7\linewidth]{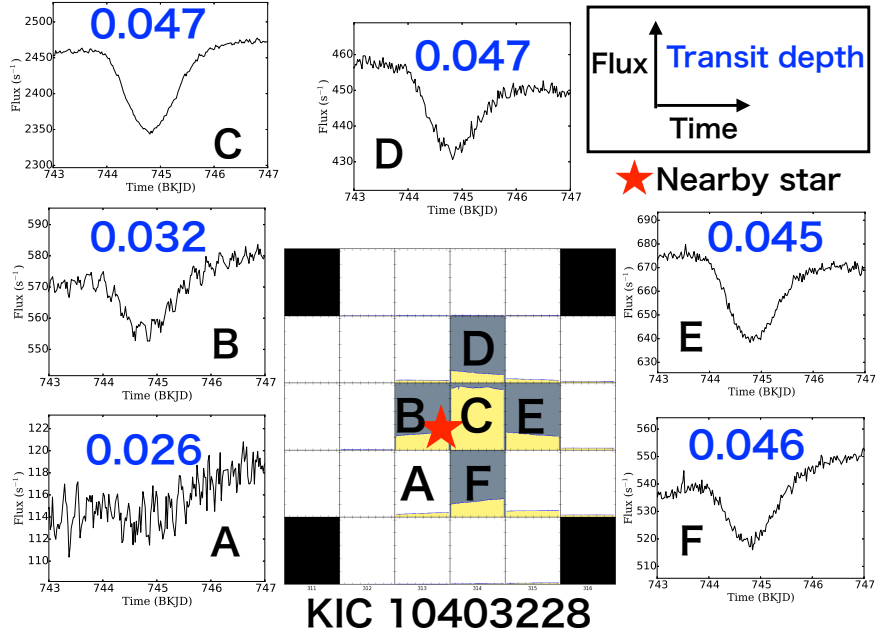}
  \caption{\small Light curves on pixels are shown along with the 
  approximate depth of the transit. A red star expresses the position of 
  the nearby star. The depth is shallow in the left side of the region, where the contamination 
  from the nearby star is large.
   }
  \label{True}
\end{figure}

Figure \ref{True} shows the light curve and fractional depth of the transit event 
in each of the pixels around KIC 10403228. The small transit depths in pixels A and B suggest that the source of the transit is not the nearby star shown by a red filled star, because otherwise the transit depths should be larger in those pixels close to the nearby star. To confirm this fact in a more quantitative way, 
we also calculate the centroid offset using the pixel-level light curves. 
As a result, we find that the flux centroid moves towards the nearby star during the transit
and that the displacement is comparable to the value expected from the observed transit depth ($5\%$) and the flux ratio in J-band ($5:1$). 
The variation of the transit depth and the centroid displacement 
consistently indicate that the transit is not due to the nearby star. 
While we may be able to evaluate the contamination on the light curve 
from this nearby star more quantitatively, it does not change our conclusion in any case, and we do not perform the detailed analysis for simplicity. 

We note that the transit signal contains a clear 
short-period modulation (panel B in Figure \ref{True}). 
Since the modulation is not visible at panel C, it is most likely 
due to the nearby star.  
Actually, there is another long-period modulation with 
$P\simeq 35$ days in the light curve, which may 
come from the target star. If these periods are related to the stellar spins, 
the nearby star is a fast rotating star, and the target star is a 
slow rotator. Thus, we may ignore the effect of 
gravity darkening of the target star. 

\section{Detailed analysis of a possible ringed planet KIC 10403228}
%Although the candidate KIC 10403228 is identified from model I, we survey a wider parameter space for the possible ring beyond model I. 
For the further study of KIC 10403228,  we present and discuss three possible models accounting for the data: ``planetary ring scenario", ``circumstellar disk scenario", and ``hierarchical triple scenario". We also discuss other possibilities than the above three models. 

%the a ringed planet model to the current transit. In section 6.2, we explore another possible explanation for the the ring-like anomaly %with a circumstellar disk. In section 6.3, we 

%Then, we discuss the reliability of the ring hypothesis. 

%\subsection{Fit of a ringed planet model to the observed light curve} 

\subsection{Interpretation with a ringed planet} 
We fit various models with and without the ring to the light curve of KIC 10403228 by minimizing 
the value of $\chi^{2}$ defined in Equation (\ref{chi_def}). 
In practice, we use $\pm 3.09$ days-time window to trim 300 data points centered around $T_{0} = 744.773$ day (BKJD [= BJD$-2454833$ day]). 
To remove the long-term flux variations in the light curve, 
we adopt the model in Equation (\ref{model_fit}) that is 
composed of a fourth-order polynomial 
and the transit model $F(t)$ in Equation (\ref{genkou}).
The standard deviation $\sigma$ is estimated to be $9.17 \times 10^{-4}$ from the 
out-of-transit data. This value is  about 1.3 times larger than the error recorded 
in the SAP data. 

As the transit of KIC 10403228 is observed just once, 
we cannot infer the orbital period from the 
timing of the transit. However, we can infer it from Kepler's law. 
%To achieve the grazing and large transit of KIC 10403228, 
%the geometrical width of the planetary path 
%during the transit should be comparable to the $2R_{\star}$. 
%So the transit duration $T_{\rm tra}$ is approximately given by 
The depth and V-shape of the observed transit imply that  
the transiting object is relatively large and grazing. Thus, 
we approximate the total transit duration $T_{\rm tra}$ as 
\begin{equation}
T_{\rm tra} \simeq  P \left(\frac{ 2R_{\star}}{2\pi  a}\right)\left(\frac{\sqrt{1-e^{2}}}{1 + e \sin \omega}\right), \label{duration_app}
\end{equation}
where the last factor is a correction term due to an eccentricity $e$ with 
$\omega $ being the argument of periapse. 
From Kepler's law and Eq (\ref{duration_app}), one obtains
\begin{equation}
 P \simeq450 \text{ years} 
\left(\frac{ \rho_{\star}}{\text{ 12.6 g cm}^{-3}} \right) \left ( \frac{T_{\rm tra}}{\text{2 days}}\right)^{3}\left(\frac{1 + e \sin \omega}{\sqrt{1-e^{2}}}\right)^{3} . \label{linear}
\end{equation}
 We obtain $P \simeq 450$ years if we adopt $e=0$ and $T_{\rm tra} = 2$ days for the transit duration of KIC 10403228, and the stellar density $\rho_{\star}=12.6 \pm 6.0 $ g cm$^{-3}$ from \cite{2015ApJ...815..127W}. 
The stellar density in \cite{2015ApJ...815..127W} is adopted from 
\cite{dressing2013occurrence}, who estimated the stellar properties 
by comparing the observed colors taken in 2MASS and SDSS with the Dartmouth model \citep{dotter2008dartmouth}. 

Before fitting, we simply examine how often we expect to see a transit of a planet 
with $P \simeq 450$ years. Assuming that all the stars host 
planets with $P \simeq 450$ years, the expected number of 
transit detections is given by
\begin{equation}
\label{eq:n_tra}
n_{\rm tra} = 0.045\left(\frac{N_{\rm target}}{150,000} \right) \left(\frac{t_{\rm obs,dur}/P}{4 \text{ years} / 450 \text{ years}} \right) \left(\frac{R_{\star}/a}{ 1/25,000}\right), 
\end{equation}
where $a/R_{\star} = 25000$ is the fiducial value estimated 
from equation (\ref{duration_app}), $t_{\rm obs, dur}$ is a 
observational period, and $N_{\rm target}$ is the number 
of target stars. The adopted values of $t_{\rm obs, dur}$ and $N_{\rm target}$ are 
the typical values of {\it Kepler}. 
The frequency of planets with $P=450$ yrs would be less than 1, so  
we may see  $n_{\rm tra}$ as the optimistic upper limit of the expected value. 
This current value of $n_{\rm tra}=0.045$ is small, but not too unlikely. 
 Apart from the tiny ring-like feature, the overall shape of the signal is clearly due 
to the transiting event, and it is very difficult to explain the feature from the stellar activities. 

We would like to comment on the reliability of $P\simeq 450$ years. 
The key parameters are $\rho_{\star}$ and 
the eccentricity in Eq (\ref{linear}). For example, if the system 
is a giant star rather than an M dwarf, the density and the period 
become smaller.  In this sense, to specify the correct stellar density, we would need a follow-up observation.
Moreover, the eccentricity can also change the estimated period in Eq (\ref{linear}). If $e = 0.6$, the period can be changed by the factor of  $(1/8.0)$--$8.0$, and if $e = 0.9$, the factor of change is within $(1/82.82)$--$82.82$ (or 5 years $<P<$ 34,000 years). Thus, the planet with a relatively short period and a large eccentricity can also explain the data. Although the period is uncertain, 
the different period does not change the fitting results, so we adopt $P= 450$ years for the fiducial value for the time being. 

 For fitting, we adopt $P= 450$ years,
and $q_{1}$ and $q_{2}$ from the official catalog of {\it Kepler}. 
In summary, there are nine free parameters, 
$t_{0},R_{\rm p}/R_{\star}, b, a/R_{\star}$, and $c_{i}$ ($i=$0--4)
for the model without the ring, and five additional parameters 
$\theta, \phi, r_{\rm in/p}, r_{\rm out/in}$ and $T$ for the model with ring. 
We set the initial values of $c_{i}$ ($i=$0--4) to those obtained 
from a polynomial curve fitting for the out-of-transit data. 

\begin{figure}[htpb]
\centering
    \includegraphics[width=.50\hsize]{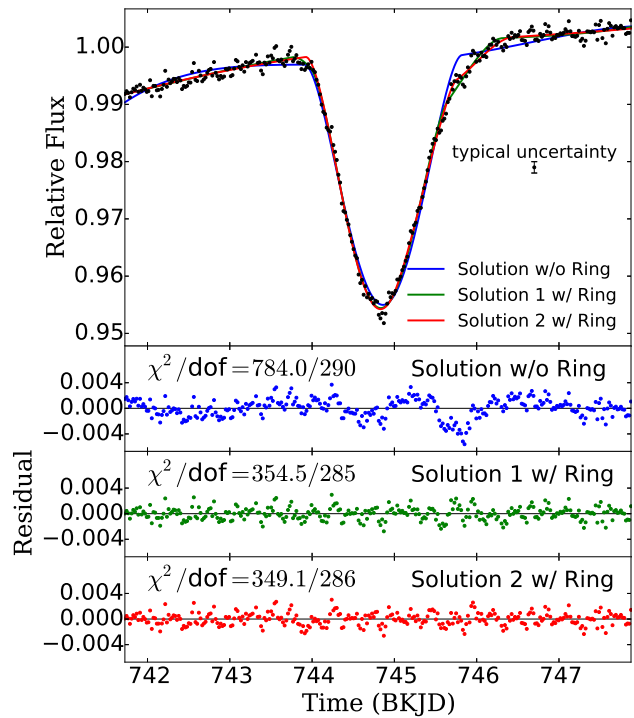}
    \caption{\small Comparison of the light curve of KIC 10403228 (black data) 
    with the models with ring (red line \& green line) and without ring (blue line). 
    We found different configurations for the model with ring.
    While the model without ring cannot explain the data around 
    745.8 day (BKJD), the ringed planet model fits the data well. 
    Residuals and $\chi^{2}$ of fit are shown in each panel. }
    \label{best}
\end{figure}

\begin{figure}[htpb]
  \centering
	\includegraphics[width=.49\linewidth]{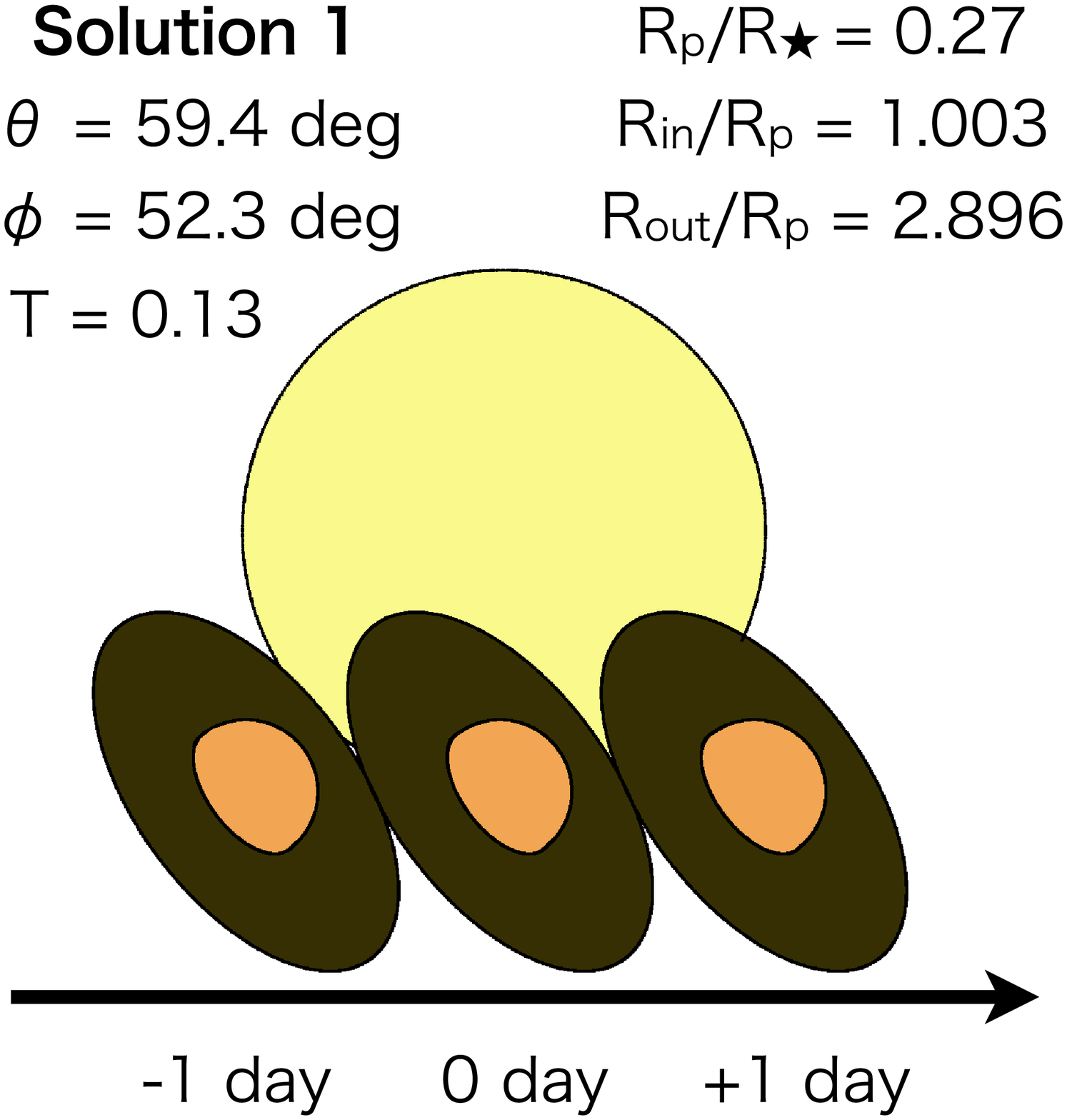} 
	\includegraphics[width=.49\linewidth]{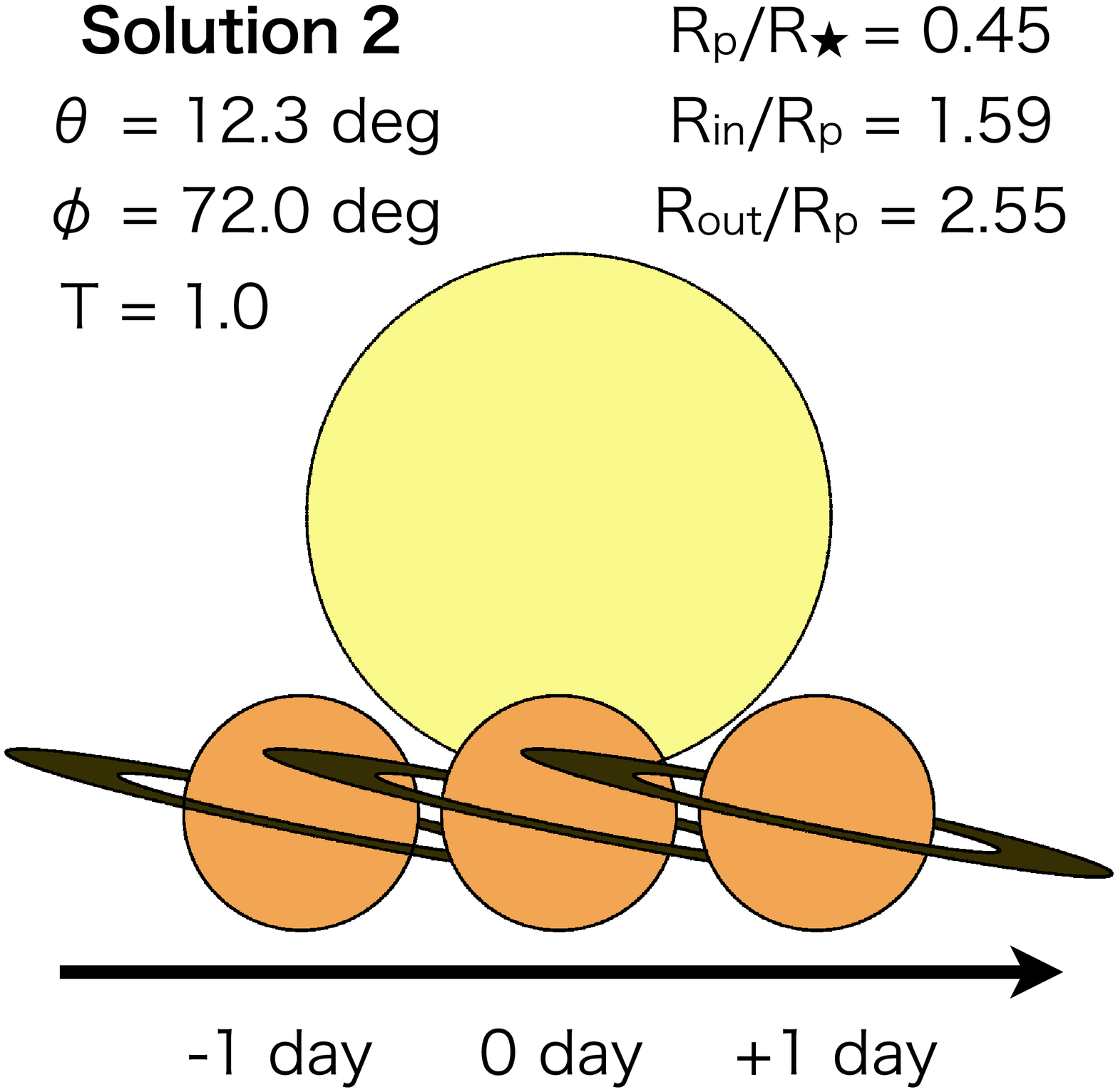} 
    \caption{\small Schematic illustration of the ringed planet models 
    for the two best-fit solutions. Positions of the planet are shown at $-1, 0, +1$ day 
    from the central time of the transit of the planet. Note that 
    the occultation of the star due to the ring continues 
    even after the transit of the planet is completed. }

  \label{ring_geo_fit}
\end{figure}

First, we fit the planet alone model to the data. The blue 
line in Figure \ref{best} is the best-fit model without the ring. 
The best-fit parameters are listed in Table \ref{best_table}. 
The residuals from the fit clearly have some systematic features, 
and the planet alone model fails to fully explain the light curve, 
in particular, around 745.8 day (BKJD) in Figure \ref{best}. 
Therefore, we attempt to interpret the data with the ringed planet model. 
After trying a lot of initial values for fitting, we finally find two solutions, 
which give at least local minima of $\chi^{2}$ in Equation (\ref{chi_def}). 
Figure \ref{best} shows those two solutions in the red and green lines. The 
best-fit parameters are shown in Table \ref{best_table}. 
The geometrical configurations for both solutions are shown 
in Figure \ref{ring_geo_fit}. Clearly,  
models with the ring significantly improve the fit. 

In Table \ref{best_table}, values of $R_{\rm p}$, 
$R_{\rm in}$, and $R_{\rm out}$ are calculated on the assumption of  
$R_{\star} = 0.33 \pm 0.05\,R_{\odot}$ \citep{2015ApJ...815..127W}. 
It turns out that the resulting ratio of ring and planet radii is similar to 
that of Saturn: $R_{\rm in} \simeq 1.5 R_{\rm p}$ and 
$R_{\rm out} \simeq 2.0 R_{\rm p}$. 

%\begin{figure}[htpb]
%\centering
   % %\includegraphics[width=.65\hsize]{ring_geo_zu_1.eps}
    %\caption{\small Comparing the data of KIC 10403228 (black circles) 
    %with the models with the ring (red line \& green line) and without the ring (blue line). 
    %For the ringed planet model we found different configurations.
   % While the model without the ring cannot explain the egress of the transit, 
    %the ringed planet model fit the data well. 
    %Residuals of fitting are shown in bottom. }
    %label{best}
%\end{figure}


\begin{landscape}
\begin{table}[htpb]
\caption{Best fit parameters for the transit of KIC 10403228 }
\begin{center}
\begin{minipage}{\textwidth}
\scalebox{0.8}{

\begin{tabular}{c c c c}
\hline \hline
 & Single planet & Ringed planet (solution 1)  & Ringed planet (solution 2) \\ 
\hline
\hline \hline Fixed parameters & & &  \\ \hline \hline
$P$ (years) & $450 $ & $450  $ & $450 $\\ 
$q_{1}$ & $0.6737 $ & $0.6737  $ & $0.6737 $\\ 
$q_{2}$ & $0.0767 $ & $0.0767 $ & $0.0767 $\\ 
\hline \hline Variables & & &  \\ \hline \hline

%\hline \hline Planet parameters & & & \\ \hline \hline
$t_{0}$ (day)& $0.065 \pm 0.0021$ & $0.017 \pm 0.016$ & $0.0386 \pm 0.0052$\\ 
$R_{p}/R_{\star} $ & $0.46 \pm 9.09$ & $0.27 \pm 0.33$ & $0.45 \pm 0.05$\\ 
$b(= a \cos i /R_{\star} )$ & $1.16 \pm 10.6$ & $0.99  \pm 0.44$ & $1.14 \pm 0.06$\\ 
$a/R_{\star} $ & $24275.0 \pm 31617.1$ & $25394.5  \pm 765.0$ & $25743.9 \pm 215.6$\\ 
%\hline \hline Ring parameters & & & \\ \hline \hline
$T$ & \nodata  & $0.13  \pm 0.07$ & $1.0$ (converged to upper bound)\\ 
$\theta$(deg) & \nodata  & $59.4  \pm 45.0$ & $12.3 \pm 4.0$\\ 
$\phi$(deg) & \nodata & $52.3  \pm 24.9$ & $72.0 \pm 4.5$\\ 
$R_{in}/R_{p} $ & \nodata  & $1.003  \pm 2.95$ & $1.59 \pm 0.42$\\ 
$R_{out}/R_{in}$ & \nodata  & $2.89  \pm 8.0$ & $1.61 \pm 0.59$\\ 
%\hline \hline Polynomial parameters  & & & \\ \hline \hline
$c_{0}$ & $0.9972 \pm 0.0002$ & $1.00023  \pm 0.00034$ & $0.9999 \pm 0.0003$\\ 
$c_{1}$ & $0.00066 \pm 0.00008$ & $0.0014  \pm 0.0001$ & $0.0015 \pm 0.0001$\\ 
$c_{2}$ & $0.00067 \pm 0.00011$ & $-0.00047  \pm 0.00015$ & $-0.00030 \pm 0.00013$\\ 
$c_{3}$ & $0.000158 \pm 0.000013$ & $0.000053  \pm 0.000015$ & $0.000038 \pm 0.000015$\\ 
$c_{4}$ & $-0.000072 \pm 0.000011$ & $0.000021  \pm 0.000014$ & $0.000005 \pm 0.000012$\\ 
\hline \hline(Given $R_{\star} = 0.33 \pm 0.05\,R_{\odot}$) & & & \\ \hline \hline
$R_{\rm p} \,\,(R_J\,\,^{a})$ &$1.48  \,\,^{b}  \,$ &  $0.88 \pm 1.07 $&$1.44 \pm 0.27 $ \\
$R_{\rm in}  \,\,(R_J\,\,^{a})$ &\nodata&$0.89 \pm 2.65 $  & $2.29 \pm 0.72 $   \\
$R_{\rm out}  \,\,(R_J\,\,^{a})$&\nodata &$2.56 \pm 1.36 $ & $3.69 \pm 0.23 $  \\
\hline \hline (Statistical values) & & & \\ \hline \hline
$\chi_{\rm red}^{2}$ (=$\chi^{2}$/dof)& $2.70$ (=$784.0$/290)& 1.24 (=354.5/285) &1.23 (=$349.1/286\,\,^{c}$) \\ 
\hline 
\end{tabular} 
 \label{best_table} 
\footnotetext[1]{\small \,$R_J$ is the radius of Jupiter. }
\footnotetext[2]{\small \,$R_{\rm p}/R_{\star}=0.46$ and $R_{\star} = 0.33 R_{\odot}$ are assumed. }
\footnotetext[3]{\small \,$T$ is converged to the upper bound, so dof = 286 = 285 + 1. }

}
  \end{minipage}

 \end{center}  
 \end{table} 

 \end{landscape}

\newpage
 
%%%%%%%%%%%%%%%%%%%%%%%%%%%%%%%%%%%%%%%%%%%%%%%%%%%%%%%%%%%%%%%%%%%%%%

%1pcの距離からみた地球の公転軌道(1 AU)の角直径は2秒角であるため、KIC10403228までの距離を数百pcとすると、KIC10403228と伴星が数100AU離れていて、トランジットは起こさないだろう
%\subsection{Implication of the fitted model for KIC 10403228}
%\subsubsection{Radiative equilibrium temperature}
We comment on the implication of the fitted model for KIC 10403228 in the following. 
The radiative equilibrium temperature of the ring particle is given by
\begin{equation}
 \;\;\;\;\;\;\;\; T_{\rm eq} \simeq 15.1 \,\text{K} \left(\frac{25000}{a/R_{\star}}\right)^{0.5} 
 \left(\frac{T_{\star}}{3386\,\text{K}}\right)  \left(\frac{ 1-A}{1-0.5} \right)^{0.25},
\end{equation} 
where we fiducially adopt the Bond albedo of the ring particle $A$ of 0.5. 
The stellar effective temperature $T_\star = 3386$ K of KIC 10403228 is 
taken from \cite{2015ApJ...815..127W}. Since the equilibrium 
temperature expected from the model is much lower than 
the temperature $170$ K at the snow line \citep{1981PThPS..70...35H}, 
icy particles around the planet can survive 
against the radiation of the host star. 

%\subsubsection{Tidal equilibrium; implication for planetary spin }
The best-fit values of $\theta=59.4^{\circ}$ for solution 1 implies
a significantly tilted ring with respect to the orbital plane, and $\theta=12.3^{\circ}$ 
for solution 2 implies a slightly tilted ring. We examine the stability 
of those tilted rings on the basis of a simple tidal theory.
Under the assumption that the ring axis is aligned 
with the planetary spin, the damping timescale of the ring axis  
is equal to that for the orbital and equatorial planes 
of the planet to be coplanar. This time-scale is given by 
a tidal theory \citep[e.g.][]{2015A&A...583A..50S}:
\begin{align}
 \;\;\;\;\;\;\;\; \;\;\; \tau_{\rm tidal} &= \frac{ P_{\rm orb} Q }{ 9 \pi k_{2}} 
 \frac{ \rho_{p}}{\rho_{s}} \left ( \frac{ a}{R_{\star}} \right)^{3} 
 \simeq \frac{G P_{\rm orb}^{3} Q}{27 \pi^{2} k_{2}} \rho_{p} \nonumber \\
& = 6.94 \times 10^{16}\, \text{yr} \left( \frac{ P_{\rm orb}} {450\,
\text{years}} \right)^{3} \left( \frac{2.3 \times 10 ^{-4}}{k_{2}/Q} \right)
\left( \frac{\rho_{p}}{0.70\, \text{ g\,cm$^{-3}$ } }\right), \label{time_spin}
\end{align}
where $P_{\rm orb}$ is an orbital period, $Q$ is a dissipation 
factor, and $k_{2}$ is the second Love number. 
If we adopt $k_{2}/Q = 2.3 \times 10^{-4}$ \citep{2012ApJ...752...14L} 
and $\rho_{p} = 0.70$ gcm$^{-3}$ \citep{2000asqu.book.....C} 
of Saturn, the damping timescale is sufficiently long. 
Thus, the best-fit configurations are consistent 
with the spin damping theory even under the assumption 
that the equatorial plane of the planet is coplanar with 
the ring plane. Thus, the tilted rings 
of our best-fits also imply the non-vanishing obliquity 
of the planet. 

The ringed planet model is consistent with the data. However, the V-shape of the transit (Figure \ref{best}) is also a typical feature 
of eclipsing binaries, and the estimated period $\sim 450$ yrs may be too long to be detected in four years of {\it Kepler}'s observation (Equation (\ref{eq:n_tra})). 
Therefore, we discuss other scenarios without a planetary ring. For this purpose, in the following, we present two possible hypotheses, which can also explain the data; a binary with a circumstellar disk and a hierarchical triple.

\subsection{Interpretation with a circumstellar disk}
%In the previous section, we interpret the data with a ringed planet model. 
%So far, we assume that the main transit signal comes from a planet.  
In this section, we pursue the possibility that 
the current transit is caused by an eclipsing-binary with a circumstellar disk rather than a planetary ring. Actually, the fitting result in the previous section is also applicable to this binary scenario, so we may compare the plausibilities of the eclipsing-binary and planet scenarios to test the circumstellar disk model. For this specific purpose, we use the public code VESPA (Validation of Exoplanet Signals using a Probabilistic Algorithm) \citep{2012ApJ...761....6M,2015ascl.soft03011M}. With VESPA, we compare the likelihoods of the following four scenarios; ``HEBs (Hierarchical Eclipsing Binaries)", ``EBs (Eclipsing Binaries)", ``BEBs (Background Eclipsing Binaries)", and ``Planets" (Transiting Planets) adopting a variety of different periods. 

We adopt $JHK$-magnitudes from 2MASS ($\text{J-mag} = 13.429 \pm 0.028$, $\text{H-mag} = 12.793 \pm 0.03$, and $\text{K-mag} = 12.518 \pm 0.027$), (RA, Dec) = $(19^{\rm h}\, 24^{\rm m}\, 54.413^{\rm s},\; +47^{\circ}\, 32'\, 57.5'')$, maxrad = 3.0 arcsec (angular radius of the simulated region), $\text{Kepmag} = 16.064$, and $R_{\rm p}/R_{\star}=0.3$. In reality, those observed colors might be contaminated by the nearby star discussed in Section 5.3, but we assume that the contamination is sufficiently small in the 
present analysis. Given these inputs, VESPA calculates the star populations and the probability distribution of transit shape parameters for the above four scenarios. For our adopted set of input parameters, VESPA identifies the primary star as an M dwarf consistent with the classification of \cite{dressing2013occurrence}. 
We repeat the simulation ten times with different initial random numbers 
according to the prescription of VESPA. 

\begin{figure}[htpb]
  \centering
  %\begin{tabular}[b]{@{}p{0.45\textwidth}@{}}
    \includegraphics[width=.5\linewidth]{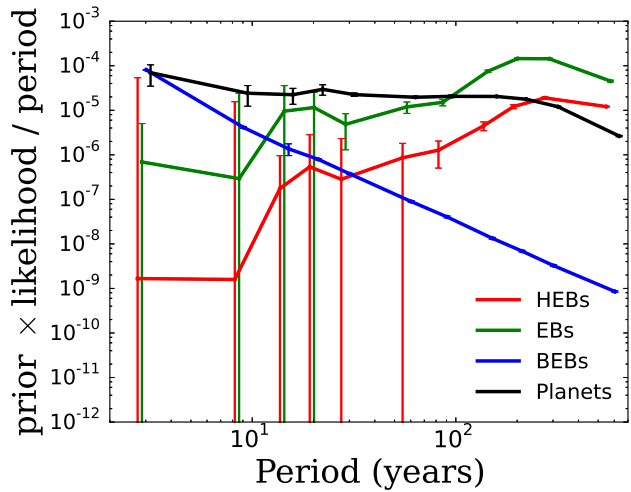}
  \caption{\small   Relative probability of different models against adopted periods. 
  The values in the vertical axis include the prior and  likelihood factors, and 
  the observational probability. Each point is slightly offset for visibility.  }
  \label{evid}
\end{figure}

Figure \ref{evid} shows the relative probability of each scenario for different assumed periods. We define the relative probability as the product of the ``prior" and ``likelihood" computed by VESPA, multiplied by $1000\,\mathrm{days}/P$. The last factor $1000 \text{ days} /P$ corrects for the probability that a long-period transit is observed in a given observing duration much shorter than the orbital period, which is not taken into account in the ``prior" of VESPA. The plot shows the medians and the standard deviations of the probabilities computed from 10 sets of simulations. While the binary scenarios are more likely than the Planets scenario for the shortest and the longest periods investigated here, the Planets scenario is the most preferred in the intermediate region (10 years $\lesssim P \lesssim$ 100 years). The result suggests that the planetary interpretation of the light curve is not so unlikely compared to the binary scenario, although there is a fair amount of probability that this is a false positive. Another important implication of Figure \ref{evid} is that the likelihood of orbital periods in the Planets scenario is much broader than what we intuitively thought before, and not sharply peaked around 450 years. %{\bf This implies that the occurrence rate of the current transit simply inferred from the assumption with $P=450$ yrs in Section 6.1 would be underestimated in a sense that the scenario with the shorter period is more likely than the scenario with $P = 450$ yrs. }

While Figure \ref{evid} represents our final result from VESPA, we point out two additional factors that may be of importance for more detailed arguments.

First, the period distribution and the overall fraction of long-period planets and binaries have not been taken into account. Occurrence rate of giant planets around M dwarfs is given by \cite{clanton2016synthesizing}. They estimated the frequency of the planets with $10^{2} M_{\oplus}<M_{\rm p} <10^{3} M_{\oplus}$ to be $0.039 ^{+0.042}_{-0.025}$ for $10^{3} \text{ days}<P < 10^{4} \text{ days}$ and $0.013  ^{+0.025}_{-0.010}$ for $10^{4} \text{ days}<P < 10^{5} \text{ days}$. On the other hand, \cite{janson2012astralux} estimated the multiplicity distribution of the binaries in $3$--$227$ AU and found the overall occurrence rate $0.27\pm 0.03$ peaked around 10 AU. These results imply that planets around M dwarfs is rarer than its stellar companion by one or two orders of magnitude. This difference in the overall frequency may further increase the relative plausibility of the EBs scenario compared to the Planets scenario.

Second, what also matters in reality is the frequency of planetary rings and circumstellar disks that produce the observed anomaly in addition to the transit signal. It is, however, far beyond our current knowledge to estimate these factors rigorously. 

Given these difficulties, follow-up spectroscopy or high-resolution imaging would be more feasible to distinguish the EBs and Planets scenarios.

\subsection{Interpretation with a hierarchical triple \label{chap:kep}}
%In the different context of a planetary ring or a circumstellar disk, a hierarchical triple 
%can also produce an asymmetric and long transit-like signal as seen in KIC 10403228. 
An eclipse due to a close binary (rather than a single star/planet) on a wide orbit around the primary M star is yet another possibility to explain the asymmetric and long transit-like signal observed for KIC 10403228. This is because the orbital motion of the occulting binary can produce the acceleration that modifies the in-eclipse velocity of the occulting object(s) relative to the primary. To test this possibility, we consider a hierarchical triple system consisting of a short-period binary (``inner" binary) orbiting around and eclipsing the primary M star on a wide orbit (``outer" binary). In the following, we only take into account the luminosity of the occulted star and ignore the flux from the smaller binary. %and focus on a specific configuration where only one component of the smaller binary eclipses the primary star. 
We also assume the orbits of both inner and outer binaries are Keplerian, and use the subscripts ``in" and ``out" to denote their parameters. A mass ratio of the inner binary is fixed to 1 for simplicity.  In this model, the motion of the two components of the inner binary is specified by $t_{\rm 0, in}$ (inferior conjunction of the inner binary), $P_{\rm in}$, $a_{\rm out}/a_{\rm in}$, $i_{\rm in}$, $\Omega_{\rm in}$ (longitude of ascending node relative to that of the outer binary), in addition to the parameters for a single-planet model (now with the subscripts ``out"). We fit all of these parameters except that the stellar density $\rho_{\star} = 13.0\text{gcm}^{-3}$, $q_{1} = 0.6737$, $q_{2} = 0.767$, the time offset $T_{\rm 0}=744.773$ days in Eq (\ref{model_fit}), and the same 
baseline as obtained in Table \ref{best_table} (solution 2) are fixed.
The mass ratio of the outer binary is related to $P_{\rm in}$, $P_{\rm out}$, and $a_{\rm out}/a_{\rm in}$ as 
\begin{equation}
q \equiv \frac{M_{\rm in}}{M_{\rm out}} = \frac{1}{(a_{\rm out}/a_{\rm in})^{3} (P_{\rm in}/P_{\rm out})^{2} - 1}, 
\end{equation}
where $M_{\rm in}$ is a total mass of the inner binary, and $M_{\rm out}$ is the mass of the primary star. 

\begin{figure}[htpb]
  \centering
  %\begin{tabular}[b]{@{}p{0.45\textwidth}@{}}
    \includegraphics[width=.5\linewidth]{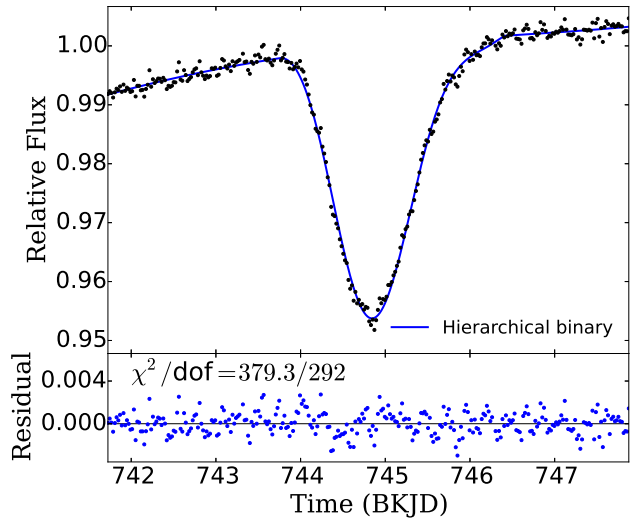}
  \caption{\small Light curve of KIC 10403229 and fitting curve with a model of a hierarchical triple. }
  \label{triple}
\end{figure}

Figure \ref{triple} shows one of the best-fitting models with $P_{\rm out} = 1396.615$ days, 
$a_{\rm out}/a_{\rm in} = 54.53$, $\Omega_{\rm in} = -0.00945$, $P_{\rm out} = 10.96$ days, $b_{\rm out} = 2.024$, $\cos i_{2} = 0.103$, $t_{\rm 0, out} = 1.965\times 10^{-3}$ days, and $t_{\rm 0, in} =1.965\times 10^{-5}$ days. In this solution, we find $q=0.0956$, which leads to $M_{\rm in} = 30 M_{\rm J}$. We also obtain $\chi^{2} / \text{dof} = 379.3/292$, which is comparable to the ringed-planet model. In this solution, the observed $\sim2\,\mathrm{days}$ duration is reproduced despite that the value of $P_{\rm out}$ is much shorter than required for the planetary scenario. This is made possible because the orbital motion of the inner binary cancels out the high orbital velocity of the outer binary. In addition to this particular solution, we find various other solutions with similar $\chi^2$ values for a wide range of $P_{\rm out}$.  In general, the solutions with longer $P_{\rm out}$ are found to correspond to smaller $q$; for example, we find $q\simeq 0.02$ for $P_{\rm out} \simeq 30 \text{\,yrs}$, and $q\simeq 0.003$ for $P_{\rm out} = 300 \text{\,yrs}$.  For $M_{\star} = 0.3 M_{\odot}$, these mass ratios $q=(0.1, 0.02, 0.003)$ translate into $M_{\rm in} = (30 M_{\rm J}, 6 M_{\rm J}, M_{\rm J})$. Thus, in this scenario, the system can be composed of three low-mass stars or a star with a binary planet.  

The advantage of this scenario is that the observed long transit can be reproduced with much smaller $P_{\rm out}$ than in the ringed-planet model, which leads to far higher transit/eclipse probability. On the other hand, it is also true that the parameters need to be finely tuned to cancel the two orbital motions. While the degree of required fine tuning is crucial in comparing the evidence of this hypothesis with the planetary or stellar ring models, the evaluation of this factor is not trivial given the large parameter space. In addition, there still remain uncertainties in frequency of the hypothetical hierarchical triple (three low-mass stars or a star with a binary planet). Given these complexities, it is difficult to conclude whether or not this scenario is favored compared to the above two. Again, the follow-up observation will be effective for the further study. 

\subsection{Possibilities other than a ringed object and a hierarchical triple}
So far, we present the three leading scenarios ``planetary ring scenario", ``circumstellar disk scenario", and ``hierarchical triple scenario". There still remain other possibilities that may potentially account for the light curve of KIC 10403228. In this section, we examine these possibilities and show that they are unlikely to explain the data. Throughout this section, we basically assume that the transit is caused by a planet, but the results in this section are also applicable for the stellar eclipse. 

\subsubsection{Oblate planet}
A significant oblateness of a single planet may mimic a ring-like anomaly 
during a transit. Indeed our model reduces to an oblate planet if we set 
$R_{\rm p} = 0$, $R_{\rm in} = 0$, and $T=1.0$ with 
an appropriate choice of $\theta$ and $\phi$. 
We attempt the fit of this oblate planet model to the light curve, 
and obtain the best-fit with  $\chi^{2} / {\rm dof} = 492.4/288$. 
This value is much larger than the best-fit value $\chi^{2}/ {\rm dof} = 349.1/286$ 
with the model with a ring. Furthermore, 
the best-fit oblate planet model requires the projected 
ellipticity of $f=(a-b)/a= 0.79$, 
where $a$ is the major axis, and $b$ is the minor axis. 
This solution is an unstable configuration; 
the rotating object will break up due to the centrifugal force 
when $a\geq 1.5b$ (Equation (2.14) in \cite{2009pfer.book.....M}). 
Thus, we conclude that the oblateness of the planet is 
unlikely to explain the observed anomaly.

\subsubsection{Additional transit due to exomoon}
In Section 6.3, we only consider an additional motion of an occulting object due to an accompanying object. However, a transit of the accompanying object (e.g. exomoon) itself is yet another possibility for the peculiar light curve of KIC 10043228. As shown below, this possibility is ruled out by the shape of the anomaly.

\begin{figure}[htpb]
  \centering
  %\begin{tabular}[b]{@{}p{0.45\textwidth}@{}}
    \includegraphics[width=.5\linewidth]{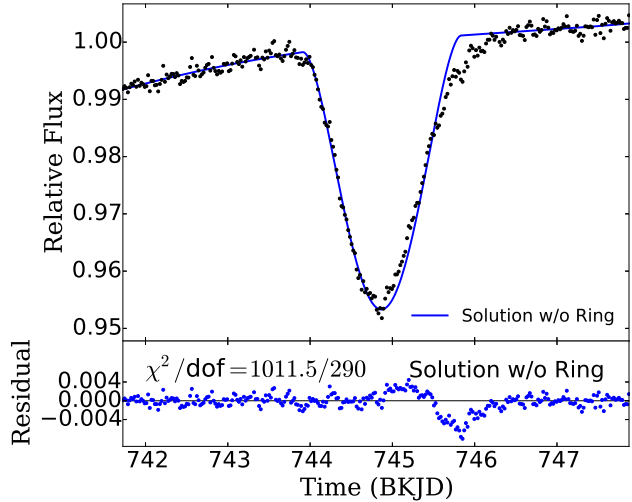}
  \caption{\small Planet-alone fit to only the first half of the transit light curve. The baseline polynomial is the same as that in solution 1 in Table \ref{best_table}. The anomaly consists of a rise followed by a dip. 
   }
  \label{Moon}
\end{figure}

As shown in Figure \ref{best}, the anomaly in the light curve is significant only in the latter half. Motivated by this fact, we fit the light curve using the planet-alone model, masking the latter half of the transit and adopting the same baseline as obtained in Table \ref{best_table} (solution 2); the difference between this model and the observed light curve would represent the anomalous contributions from anything other than the main transiting planet. The result in Figure \ref{Moon} clearly shows that the anomaly consists of a short rise in the flux followed by a more significant dip. Such a feature is clearly inconsistent with the transit of an exomoon.
%For the further discussion, we would need the precise and complex 
%modeling of the light curve, and we leave it to the future works. }

%Throughout this discussion, we suppose that the exomoon 
%orbits around the planet with $M_{\rm Jup}$ and $R_{\rm Jup}$. 
%First of all, we evaluate the relative velocity of the exomoon. 
%The orbital velocity of the exomoon is written by 
%\begin{equation}
%v_{\rm rel,moon} = \sqrt{G M_{\rm Jup}/a} = 42.57 \sqrt{R_{\rm Jup}/a}  \text{ km/s}
%\end{equation}
%The range of $v_{\rm rel,moon}$ is determined by 
%the range of $a$. For the stability of the orbit of the exomoon,  the value of $a$ should be within the Roche radius and the Hill radius. 
%While the Roche radius is an order of $R_{\rm p}$, the Hill radius is given by 
%\begin{equation}
%R_{\rm Hill } = a (M_{\rm Jup}/3\times 0.3M_{\odot})^{1/3} = R_{\star}(a/R_{\star})(M_{\rm Jup}/3M_{\star})^{1/3} \simeq 10000R_{\rm Jup} ,
%\end{equation}
%where, in the last expression, we suppose $M_{\star}= 0.3 M_{\odot}$, $R_{\star} = 0.3R_{\odot}$, and 
%$a/R_{\star} = 30000$. Then, we obtain $R_{\rm Jup} < a < 10,000 R_{\rm Jup}$, which 
%leads to $0.42  \text{ km/s} < v_{\rm rel,moon}  < 42.57  \text{ km/s}$. 
%On the other hand, an orbital velocity of the planet is given by 
%\begin{equation}
%v_{\rm pla} =   \sqrt{G M_{\rm \star}/a_{\rm p}} = 2.415 \text{ km/s} 
%\end{equation}
%Comparing $v_{\rm pla}$ and $v_{\rm rel,moon}$ gives the estimation of the 
%relative motion of the exomoon

\subsubsection{Anomalies specific to in-transit data}
There exist anomalies specific to in-transit data; spot crossing and gravity darkening. 
If the planet crosses spots on the stellar surface, 
the light curve is deformed \cite[e.g.][]{2011ApJ...733..127S}. 
In general, however, spots are dark, so spot-crossing causes a bump in 
the light curve. The observed anomaly in the bottom 
panel of Figure \ref{Moon} is inconsistent with a single bump, 
so the spot is unlikely to cause the anomaly. 
Gravity darkening makes the light curve asymmetric 
\citep[e.g.][]{2011ApJS..197...10B, 2015ApJ...805...28M}. 
In section 5.3, we identify the target star as the slow-rotating star, 
and the gravity darkening is negligible. 
In conclusion, these mechanisms are unlikely to explain 
the ring-like signal in the light curve.

\subsubsection{Stellar noise}
The ring-like structure in the light curve shows up only for a short duration. 
Thus, the short-term stellar noise might mimic the ring-like anomaly just by chance. 
To discuss this possibility, we investigate the statistical property of the stellar activity of 
KIC 10403228. 
Specifically, we consider how frequently one encounters stellar noises 
comparable to the anomalous in-transit residuals. As will be shown, we find 
it difficult to reproduce the feature with stellar activities of KIC 10403228. 
In principle we could check to see if the similar feature arises in stars other than KIC 10403228 more generally, but it is a separate question and does not answer if the signal for the particular star is due to that stellar activity.  Therefore we analyze the light curve of KIC 10403228 alone in this section.

To focus on the short-term noises, we remove the long-term variations by dividing the light curves into short segments and fitting each of them with polynomials. The more specific procedure is as follows. We exclude in-transit data as well as data around gaps in the light curve. From the remaining data, we pick up a segment of $6.18$-day long light curve centered around a randomly chosen time and fit it with a quartic polynomial to remove the variation within the segment. In principle, one could use different functions (e.g. a spline function) or different time-window for detrending, but in any case the final results are insensitive to these choices. For consistency, we adopt the same baseline and time-window as those used in Section 6.1. 

\begin{figure}[htpb]
  \centering
  \includegraphics[width=0.5 \linewidth]{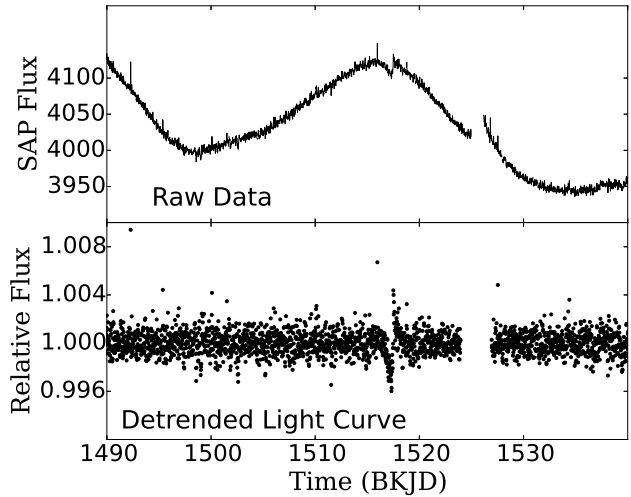} 
  \caption{\small An example of a detrended light curve of KIC 10403228. 
  The flare-like events are visible at several epochs. To create this curve, 
  we generate 1000 light curves as described in the main text.
   }
      \label{hist_stellar}
\end{figure}

We iterate ``picking up a segment" and ``detrending" procedures 1000 times and obtain  
1000 segments of detrended  light curves, whose centers 
are randomly distributed over the whole observing duration. 
We note that the total number of points in the detrended segments 
is $1000\times 300=3.0\times 10^{5}$, 
which is sufficiently large to sample all the original data points ($N=10,000$). 
By averaging the 1000 detrended light curves 
at each time, we obtain one light curve. 
This averaging operation suppresses the dependence on the choice of the 
central time of each segment. 
Figure \ref{hist_stellar} shows the resulting detrended light curve (bottom) along 
with the light curve before detrending (top). 

\begin{figure}[htpb]
\includegraphics[width=.49\linewidth]{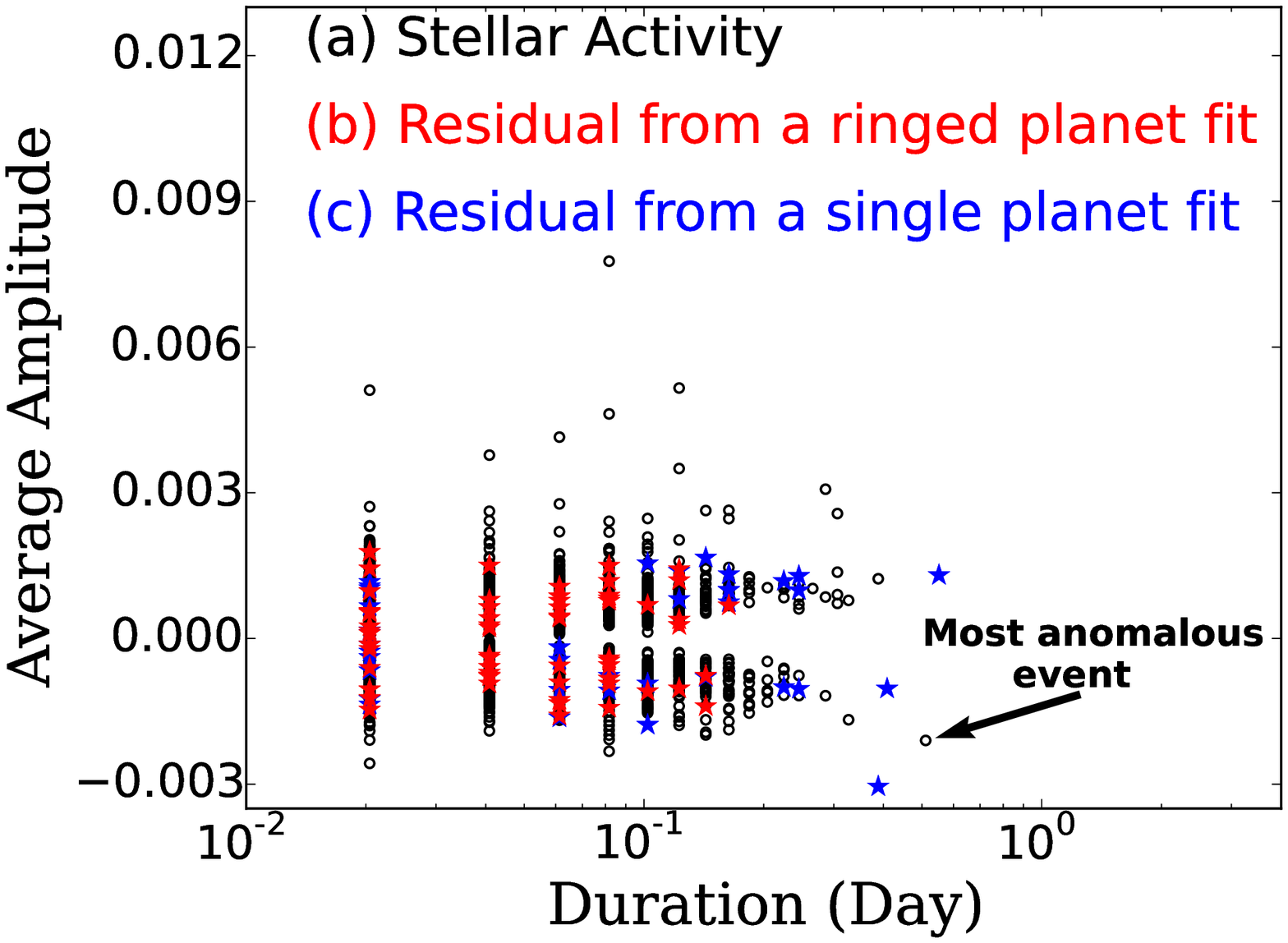} 
\includegraphics[width=.49\linewidth]{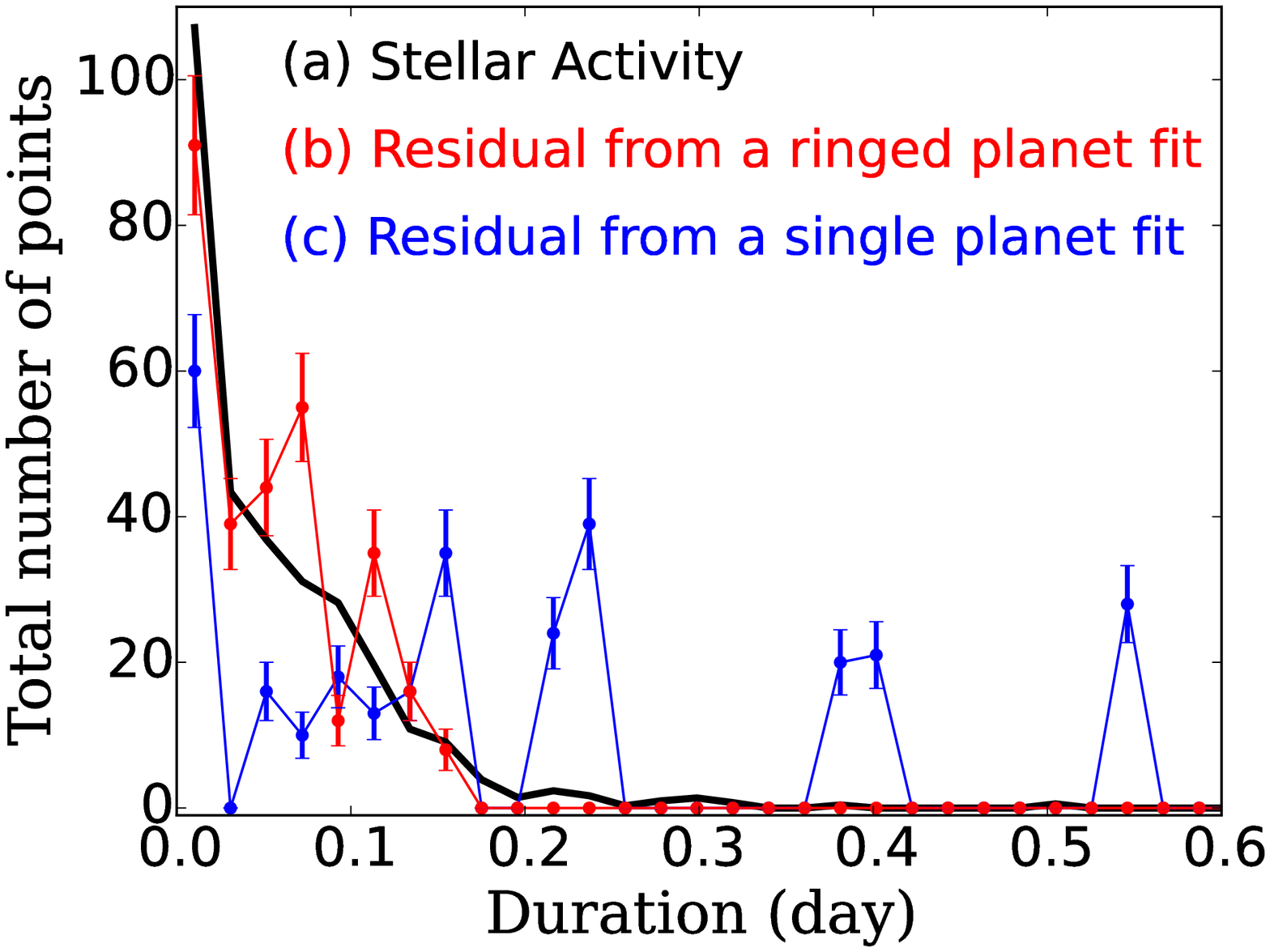} 
   \caption{\small Statistical property of stellar activities of KIC10403228 and 
   the residuals in Figure \ref{best}. (Left)
  Duration and average amplitude of continuously brighting 
  $(F(t) > 1)$ and fading $(F(t) < 1)$ events are plotted. The 
  black points are plotted from the stellar activities in Figure \ref{hist_stellar}, 
  and red points and blue points are plotted from the residuals of a ringed planet 
  and a single planet model fitting in Figure \ref{best} respectively. (Right) Histogram of 
  duration. The color has the same meaning as in the left panel.  }
  %\caption{\small Property of stellar activities of KIC10403228. (left)
  %Durations and amplitudes of continuously brighting 
  %$(F(t) > 1)$ and fading $(F(t) < 1)$ events are plotted. The 
  %black points are from the light curve out-transit, and red points 
  %from the residuals of the planet model fitting in Figure \ref{best}. (right) The frequency of the 
  %duration obtained by integrating the frequency in the amplitude-axis in the left figure. 
  %The red and blue line are made from the residuals in Figure \ref{best}. }
  \label{duration}
\end{figure}

Now we move on to the comparison of the statistical property 
of stellar activities and the residuals of fit in Figure \ref{best}. 
Let us define $F_{\rm data}(t)$ as the flux ratio of the detrended 
light curve with respect to the mean. To investigate 
the short-term correlation of stellar activities, 
we divide the light curves into continuously brightening events $(F_{\rm data}(t)> 1)$ 
and fading events $(F_{\rm data}(t) < 1)$.  Then, 
we compute the duration and amplitude (average of the deviation from the 
mean $|F_{\rm data}(t) - 1|$) for each event. 
For comparison, we also calculate the duration and average relative flux 
for events in residuals in Figure \ref{best}. 
The left panel in Figure \ref{duration} is the scatter plot 
of the duration and average relative flux of events for three groups; 
\begin{itemize}
\item[(a)] all events out of the transit (the black data in Figure \ref{hist_stellar}). 
\item[(b)] residuals of the ringed-planet fit (the red line in Figure \ref{best}). 
\item[(c)] residuals of the single-planet fit (the blue line in Figure \ref{best}).
\end{itemize}
The right panel in Figure \ref{duration} shows 
the distributions of durations for the three groups. 
In each duration bin, the vertical axis shows 
the total number of points in all events with that duration. 
The distribution of (a) is normalized to give the 
same total number of events as (b) and (c). 
The quoted error-bars are simply computed from
Poisson statistics of the number of each event. 
Figure \ref{duration} shows 
that the distribution (b) is closer to (a) than (c). Thus, the ringed planet model 
is better than the planet model in terms of property of the 
correlated noise.  

\begin{figure}[htpb]
    \centering\includegraphics[width=.49\linewidth]{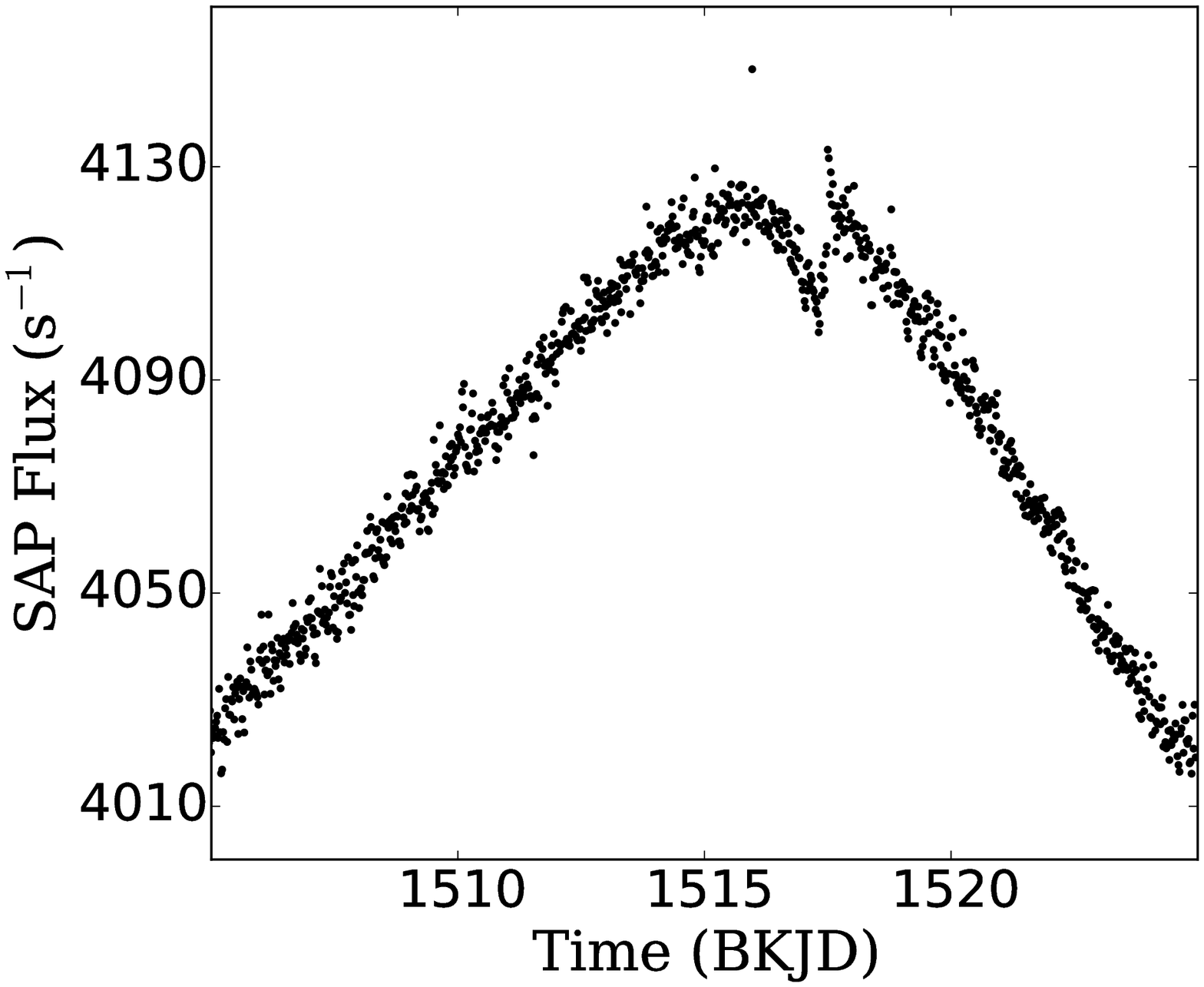} 
    \centering\includegraphics[width=.49\linewidth]{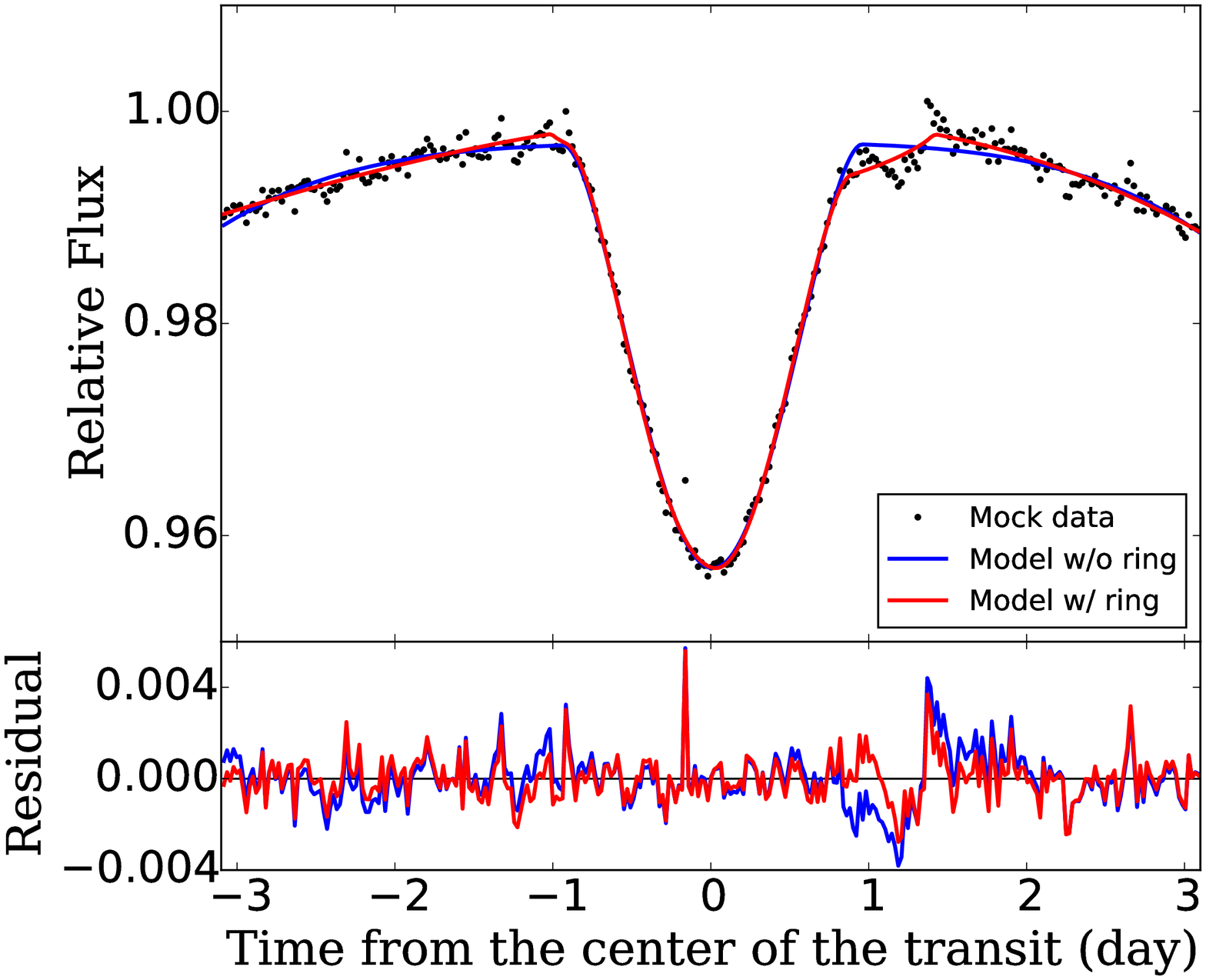} 
  \caption{\small Testing the model of a transit of a planet with the stellar activities. 
  (left) The most significant anomaly in KIC 10403228 as 
  indicated with an arrow in the left panel of Figure \ref{duration}. 
  (right) The light curve of the transit of the 
  planet embedded into the event in the left figure. The data are 
  fitted with the ringed planet model (red) and the planet alone model (blue).
  The results show $\Delta \chi^{2} \simeq 157.9$, which is smaller than the real 
  value $\Delta \chi^{2}= 434.8$ in 
  Section 6.1. }
      \label{test_stellar}

\end{figure}

So far, we have shown that the ring-like anomaly cannot be explained statistically. 
We further consider whether the stellar noise can mimic the light-curve shape itself. 
We examine this hypothesis by focusing on the most significant fading 
event in the out-of-transit data; see the left panel of Figure \ref{duration}. 
The light curve of this event is shown in Figure \ref{test_stellar}. 
We would like to see if the combination of the planet 
model and this event can reproduce the ringed-planet like feature. 
To do this, we appropriately embed the transit 
of the planet into the light curve around the fading event. 
Here, the parameters of the planet are the same as in Table \ref{best_table}. 
Then we fit the two models with and without a ring to those 
data, as shown in Figure \ref{test_stellar} (b). As a result, we find the difference in 
$\chi^{2}$ of the two models to be
%$\chi^{2}_{\rm pla}(=502.6, \text{dof } =290) 
%- \chi^{2} _{\rm ring}(=344.7, \text{dof } =285) 
%=\Delta \chi^{2} \simeq 157.9$. Considering  
%$\chi^{2}_{\rm pla}(=784.0, \text{dof } =290) 
%- \chi^{2} _{\rm ring}(=349.1, \text{dof }=286) \Delta \chi^{2} \simeq  434.9$ in Section 4.5, we 
$157.9$, which is smaller than $434.9$ obtained in Section 6.1 for solution 2. 
Thus, we conclude that it is difficult to reproduce the ring candidate by 
combining the stellar activities and the transit of the planet. 

\subsubsection{Combination of the above mechanisms}
In principle, a combination of the mechanisms discussed above could be invoked to reproduce the observed anomaly. In Figure \ref{Moon}, for example, the bump and dip in the residual might be explained separately by a spot crossing and an exomoon. However, such a probability is a priori very low, and so we do not discuss those possibilities any further.

\section{Conclusion and Future prospects}
In this paper, we present a methodology to detect exoplanetary rings and apply it to the 89 {\it long-period} transiting planet candidates in the {\it Kepler} sample for the first time. After fitting a single planet model to light curves of target objects, we classify them into four groups depending on the observed anomalies and model predictions.  Assuming grazing geometry and a titled ring, we obtain upper limits on $R_{\rm out}/R_{\rm p}$ for 12 planet candidates, and find $R_{\rm out}/R_{\rm p}<1.5$ for six of them.  While we select five preliminary ringed planet candidates using the results of classification, four of them turn out to be false positives, but KIC 10403228 still remains as a possible ringed-planet system.
 
  We fit our ringed planet model to the light curve of KIC 10403228, and we obtain two consistent solutions with the tilted ring. However, the V-shape of the current transit is a typical feature of an eclipsing binary, and the estimated orbital period $P$=450 years on the assumption of a circular orbit may be too long for the transit to be detected. Therefore, we also consider other two possibilities accounting for the data. One model assumes that the transit is caused by an eclipsing binary, and the ring-like feature is caused by a circumstellar disk rather than a planetary ring. For comparison, using the public code VESPA, we calculate the plausibility of this scenario and the planet scenario, and find that we cannot exclude both possibilities at the current stage. The other model we consider assumes the observed eclipse is caused by two objects orbiting around each other (hierarchical triple configuration), where the orbital motion of the smaller binary produces the long and asymmetric eclipse as observed for KIC 10403228. Assuming this model, we find various solutions for a wide range of orbital periods down to $P\simeq 1400\,\mathrm{days}$, although it requires more or less fine-tuned configurations. 
In addition to the above scenarios, we also discuss other possibilities, and 
find that none of them are likely to explain the data. 
In conclusion, there remain the three leading scenarios accounting for the data: ``planetary ring scenario," ``circumstellar disk scenario," and ``hierarchical triple scenario."
A follow-up observation would play an important role in the further study.

The current research can be improved in several different ways.  We can
enlarge the sample of target objects towards those with shorter orbital
periods. The interpretation of KIC 10403228 is fundamentally limited by
the fact that it exhibits the only one transit. Obviously the
credibility significantly increases if a system exhibits a robust
ring-like anomaly repeatedly in the transits at different epochs. 
 Moreover, difference in transit shapes at different epochs would enable us to discriminate between ``disk scenario" and ``hierarchical triple scenario."
In addition, our current methodology puts equal weights on the data points over the
entire transit duration. Since the signature of a ring is particularly
strong around the ingress and egress, more useful information on
$R_{\rm out}/R_{\rm p}$ would be obtained with more focused analysis of
the features around those epochs.  We plan to improve our methodology,
and attempt to apply it to a broader sample of transiting planets in due
course. We do hope that we will be able to affirmatively answer a
fundamental question ``Are planetary rings common in the Galaxy?".

%%%%%%%%%%%%%%%%%%%%%%%%%%%%%%%%%%%%%%%%%%%%%%%%%%%%%%%%%%%%%%%%%%%%%%

%%%%%%%%%%%%%%%%%%%%%%%%%%%%%%%%%%%%%%%%%%%%%%%%%%%%%%%%%%%%%%%%%%%%%%
\section*{Acknowledgements }
We are grateful to the {\it Kepler} team for making the 
revolutionary data publicly available.
We thank Tim Morton for helpful conversation, and anonymous referees for a careful reading of the manuscript and 
constructive comments. 
M.A. is supported by the Advanced Leading Graduate Course for Photon
Science (ALPS).  K.M. is supported by the Leading Graduate Course for
Frontiers of Mathematical Sciences and Physics (FMSP).  This work is
supported by JSPS Grant-in-Aids for Scientific Research No.\,26-7182
(K.M.), No.\,25800106 (H.K.) and No. 24340035 (Y.S.) as well as 
by JSPS Core-to-Core Program ``International Network of Planetary Sciences".
This work was performed in part under contract with the Jet Propulsion Laboratory (JPL) funded by NASA through the Sagan Fellowship Program executed by the NASA Exoplanet Science Institute.

%%%%%%%%%%%%%%%%%%%%%%%%%%%%%%%%%%%%%%%%%%%%%%%%%%%%%%%%%%%%%%%%%%%%%%
%\appendix

\appendix

\section{NUMERICAL INTEGRATION IN EQUATION (5)} 
We present a formulation for fast and accurate 
numerical  integration of Equation (\ref{genkou}). In addition to $(x,y)$ coordinates defined in Section 2, we also introduce the cylindrical 
coordinates $(r,\theta)$, whose origin is at the 
center of the star. The ranges of $(r,\theta)$ 
integration  are $ 0 < r < R_{\star}$ and 
$0\le \theta < 2 \pi$. We integrate Equation 
(\ref{genkou}) by dividing the total range of 
integration into several pieces as follows: 

\begin{align} 
&\int  I(x,y) D(x,y) dS = \int^{R_{\star}} _{0} \int^{2\pi} _{0} I( \sqrt{1-(r/R_{\star})^{2}} ) D(r, \theta) r dr d\theta \nonumber \\
&= \sum_{i} \sum _{l}   D_{i,l}\int _{r_{i}} ^{r_{i+1}} \int _{\theta_{i,l }(r)}^{\theta_{i,l+1}(r)} I( \sqrt{1-(r/R_{\star})^{2}} )r dr d\theta \nonumber \\
&= \sum_{i} \sum _{l}   D_{i,l}\int _{r_{i}} ^{r_{i+1}} ( \theta_{i,l+1}(r) -\theta_{i,l}(r))  I( \sqrt{1-(r/R_{\star})^{2}} )r dr. 
\end{align}

\begin{figure}[htpb]
\centering    \includegraphics[width=.90\linewidth]{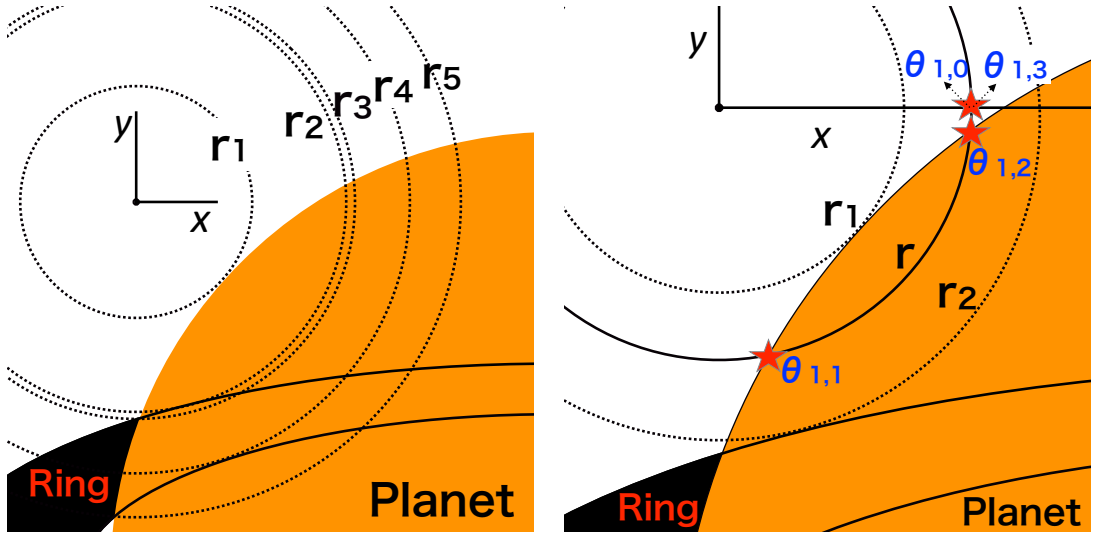} 
  \caption{\small Illustrations of $r_{\rm i}$ (left) and $\theta_{i,j}(r)$ (right). 
  A ringed planet is depicted with colors. In the figure,  the 
  boundary lines of rings behind the planet are expressed for clarity.  } 
      \label{ri_thetai}
\end{figure}{
The intervals of integration are specified by $r_{\rm i}$ and $\theta_{i,j}(r)$. 
We will define them in the following, and the corresponding schematic illustration is 
depicted in Figure \ref{ri_thetai}. }

The number of the intersection points 
between a circle with the radius $r$ and 
the ringed planet depends on the value of 
$r$; there exists boundary values $r$ 
for the number of intersection points. We define $r_{i}$ as 
the $i$-th boundary value, and we arrange a set of $r_{i}$ 
in ascending order.  If we have elements 
$ r_{i} >R_{\star}$, we insert $R_{\star}$
into the set of $r_{i}$, and exclude elements 
that satisfy $r_{i} >R_{\star}$. 

Next, let us 
suppose $r_{i} < r < r_{i+1}$, where the number of intersections 
remains the same. In this range, we define 
$\theta_{i,j}(r)$ to be the $j$-th value of $\theta$ 
of the intersection points between a ringed planet and 
a circle with the radius r. A set of $\theta_{i,j}(r)$ 
is also rearranged in ascending order, and we add 
$0, 2\pi$ before and behind the set of 
$\theta_{i,j}(r)$. We define $D_{i,l}$ to be 
the values of $D(r,\theta,t)$ for 
$\theta_{i, l}(r) < \theta < \theta_{i, l+1}(r)$ 
and $ r_{i} < r < r_{ i +1}$. We will derive the 
equations for $r_{i}$ and $\theta_{i,j}(r)$ in the rest of appendix. 

\subsection{Derivation of $r_{i}$}
Conditions for possible values of $r_{i}$ are divided into the following three cases: 
 \begin{itemize}
 \item[(a)] Intersections of the edge of the planet (circle) and the edge of the ring (ellipse).
 \item[(b)] Extreme points of the distance function from the center of the star to the edge of the planet (circle).
 \item[(c)] Extreme points of the distance function from the center of the star to the edge of the ring (ellipse).
\end{itemize}
The number of $r_{i}$ is at most eight for (a), two for (b), and two for (c). (a) and (b) are reduced to quadratic equations, which can be easily solved. The last case can be reduced to quartic equations. Here, we derive the quartic equations using the method of Lagrange multiplier. Let the length of the major axis be $2R$ and that of the minor axis be $2R(1-f)$, where $f$ is the oblateness. We set the center of the ellipse to be at $(x_{\rm p}, y_{\rm p})$. For $(x,y)$ on the edge of the ellipse, we define the following function:
 \begin{equation}
A(x,y,\lambda) = x^{2} +y^{2} + \lambda \left[ \left(\frac{x-x_{\rm p}}{R}  \right)^{2} + \left(\frac{y-y_{\rm p}}{R (1-f)} \right) ^{2} -1 \right]. f
 \end{equation}
 From the condition, we need 
 \begin{align}
 \frac{\partial A}{\partial x} &= 2 x + \frac{2 (x-x_{p}) \lambda} { R^{2}}  =0 \label{x_siki} \\
 \frac{ \partial A}{\partial y} &= 2 y + \frac {2 (y-y_{p}) \lambda}{ (1-f)^{2} R^{2} } =0 \label{y_siki} \\
 \frac{\partial A } {\partial  \lambda} &= \left(\frac{x}{R}  \right)^{2} + \left(\frac{y}{R-Rf} \right) ^{2} -1 = 0 
 \end{align}
We reduce the above three equations to the following:
 \begin{align}
%\scriptsize
 & \frac{ \lambda ^{4}} {(1-f)^{4}R^{8}}+ \frac{2 \lambda ^{3}}{ ( 1-f)^{2} R^{4}} \left[ \frac{1}{R^{2}} + \frac{1}{ (1-f)^{2} R^{2}} \right]\nonumber \\&+ \lambda^{2} \left[\frac{1}{R^{4} } +\frac{4}{(1-f)^{2} R^{4}} + \frac{ 1 }{(1-f)^{4} R^{4}} - \frac{ x_{\rm p} ^{2}}{(1-f)^{4} R^{6}} -\frac{ y_{\rm p} ^{2}}{(1-f)^{2} R^{6}} \right]\nonumber\\ & + \lambda \left[\frac{2}{R^{2}} + \frac{ 2 } {(1-f)^{2} R^{2}} -  \frac{ x_{\rm p}^{2} + y_{\rm p}^{2}}{(1-f)^{2}R^{4} } \right]+ 1 - \frac{x_{\rm p}^{2}}{R^{2}} - \frac{y_{\rm p}^{2}}{(1-f)^{2}R^{2}} =0. 
 \end{align}
 
 In general, quartic equations are analytically solved, but we compute the solutions for the equation using a root-finding algorithm, because of complexity of the analytic solution. $x$ and $y$ are calculated from derived $\lambda$ as follows:
 \begin{equation}
 x = \frac{ -x_{\rm p}}{1 + (\lambda/R^{2})} +x_{\rm p}, \; y=  \frac{ -y_{\rm p}}{1 + (\lambda/((1-f)R)^{2})} + y_{\rm p} \label{bubu}.
 \end{equation}
The number of solutions for $(x,y)$ is at most four. We exclude the solutions including complex numbers and/or $(x,y)$ not on the ellipse. Equation (\ref{bubu}) gives the singular solutions when
 \begin{equation}
 \lambda = - R^{2},\; -(1-f)^{2} R^{2} .
 \end{equation}
 Inserting the above values into Equation (\ref{x_siki}) or (\ref{y_siki}), we find $x_{\rm p} =0$ or $y_{\rm p}=0$. In this case, we cannot use Equation (\ref{bubu}), but the conditions are reduced to the quadratic equations, which can be easily solved.
 \subsection{Derivation of $\theta_{i,l}(r)$}
To derive $\theta_{i,l}(r)$, we calculate the intersections of a circle, centered at $(0,0)$, with the radius $r$, and a transiting object, composed of the circle (planet) and two ellipses (rings). The center of the ring system is $(x_{\rm p}, y_{\rm p})$. The intersections of two circles are easily computed and the number of the intersection points is two at most. Here, we derive the equations for intersection points of a circle and an ellipse. Let the radius of the circle be $r$. We select the same ellipse as before. For simplicity, we introduce the following parameters:
  \begin{align}
 &A =  1 - (1-f)^{2},\; B = 2 x_{\rm p} (1-f)^{2}, \nonumber\\ &C = (1-f)^{2} R^{2} - r^{2} - (1-f)^{2} x _{\rm p}^{2} - y_{\rm p}^{2}, \; D = -2 y_{\rm p}.
 \end{align}
Then, an equation for $x$, the $x$-coordinate of intersections, is given by:
 \begin{equation}
 A^{2} x ^{4} + 2 A B x^{3} + ( 2 A C + B^{2} + D ^{2} ) x^{2} +  2 B C x + C^{2} - D^{2} r^{2} = 0, \label{last}
 \end{equation} 
 
Equation (\ref{last}) is a quartic equation, which is analytically solved. We solve this equation with the root-finding method in the same way as before. The number of the solutions for this equation is four at most. In total, there are up to 10 possible solutions for $\theta_{i,l}(r)$.

\subsection{Precision and computational time}

To test the precision and the computational time in our scheme, we simulate a 
transit of a Saturn-like planet with $R_{\rm p}/R_{\star} = 0.083667,\;R_{\rm in/p} = 1.5, 
\; R_{\rm out/p} = 2.0, \; \theta = \pi/3, \;\phi = \pi/3, \; T = 1.0$. 
We take $P=10759.3$ days, $a/R_{\star} = 2049.89$, $b=0.5$, $q_{1} = 0.49$, and $q_{2} = 0.34$ for orbital parameters and stellar parameters. 
For comparison, we prepare another 
integration scheme, which adopts pixel-by-pixel integration around the planetary center \citep[e.g.][]{2009ApJ...690....1O}.

First, we check the precision of the integration of our proposed method by comparing the 
precision of the pixel-by-pixel integration methods with $5000\times 5000$ pixels. 
As a result, two methods are in agreement to the extent of $10^{-7}$. Thus, our 
proposed method achieves the numerical error less than $10^{-7}$, which is 
much smaller than the typical noise in the {\it Kepler} data $10^{-4}$. 

Second, we check the computational time of our template. Our proposed method typically takes $3.0$ ms for calculating one point and 200 s in fitting in Section 6.1. 
For comparison, we also check the computational time of the planetary transit using PyTransit package \citep{2015MNRAS.450.3233P}, and we find that it takes 0.3 ms to  compute all the 300 data points and 0.3 s in fitting in Section 6.1.

Finally, we compare our method with the pixel-by-pixel integration. If we set the pixel sizes to satisfy the same computational time as that of our method, the precision of the integration becomes $10^{-5}$ in the fiducial configurations. This precision depends on the specific configuration; it becomes $10^{-4}$ for if we adopt ``$R_{\rm p} = 0.17$ and $b=0.8$" 
and $3 \times 10^{-6}$ for ``$R_{\rm p} = 0.042$ and $b=0.3$". 
In summary, when we need a high-precision model, one should use our proposed method, and, if not, one may use the pixel-by-pixel integration to save the amount of calculation. 

Incidentally, in a practical case of fitting with the Levenberg-Marquardt algorithm, 
our method is useful in a sense that gives the smooth value of $\chi^{2}$. This is because the smoothness is needed to calculate the differential values 
for $\chi^{2}$ in LM method.

\section{Method of target classification in Section \ref{s:sec3}}
\subsection{Concept}
As we demonstrated in the main text, signatures of a ringed planet
can be detected by searching for any deviation
from the model light curve assuming a ringless planet.
The deviation is, however, often very tiny and comparable to the noise level,
and so careful quantitative arguments are required to discuss the presence 
or absence of the ring in a given light curve.
In the following, we present a procedure to evaluate the detectability of 
a ring based on the comparison between the residual from the ``planet-alone"
model fit and the noise level in the light curve.

Let us denote one light curve including a transit 
by $I_i$ $(i=0, 1, \cdots, N_{\rm data}$), 
where $N_{\rm data}$ is the number of data points. 
We also define $\delta_{i}$ as the residual of fitting $I_i$ 
with the planet-alone model.
As a quantitative measure of this residual signal $\delta_i$ relative to the noise level, 
we introduce the following signal-to-noise ratio:
\begin{equation}
S/N = \frac{\sum_{i} \delta_{i}^{2}}{\sigma^{2}} = \frac{\sum_{i} \delta_{i}^{2}} {N_{\rm data}} \; \frac{N_{\rm data}} {\sigma^{2}} = \Delta^{2} \frac{1}{(\sigma/\sqrt{N_{\rm data}})^{2}}\;\;\;\;\;\; \left( \Delta ^{2} \equiv  \frac{\sum_{i} \delta_{i}^{2}} {N_{\rm data}}\right).  \label{s_n_def}
\end{equation}
In the last equality, we further define $\Delta^{2}$ as the variance of the residual time series, 
and $\sigma$ is evaluated as the standard deviation of the 
out-of-transit light curve.
We use the subscript ``obs" to specify the above quantities obtained
by fitting the planet-alone model to the real observed data: 
$\delta_{i,\,\mathrm{obs}}$, $S/N_{\rm obs}$, and $\Delta^{2}_{\rm obs}$. 

On the other hand, we can also compute the corresponding values 
of $\delta_i$, $S/N$, and $\Delta$, by fitting the {\it simulated} light curve
of a ringed planet with the planet-alone model.
We denote these values as $\delta^{2} _{i,\,\mathrm{sim}}(p)$, 
$S/N_{\rm sim}(p)$, and $\Delta^{2}_{\rm sim}(p)$,
where $p$ represents the set of parameters of the ringed-planet model.
If these values are sufficiently large compared to the noise variance
(see $\Delta_{\rm thr}^2$ below), the signal of the ringed planet 
is distinguishable from the noise.
In addition, comparing these theoretically expected residual levels
with observed ones, we can relate the observed residuals to 
the parameters of the ringed model, even in the absence of clear anomalies.

To simplify the following arguments, 
we mainly use $\Delta ^{2}$ instead of $S/N$ 
to evaluate the significance of the anomaly (see also Section \ref{ssec:delta_sim}
for detailed reason).
Practically, conversion from one to the other is rather simple, as the conversion
factor $\sigma/\sqrt{N_{\rm data}}$
is well determined from the observed data alone;
given a transit light curve, the transit duration $T_{\rm dur}$ 
and the bin size $t_{\rm bin}$ give the number of data points 
$N_{\rm data} = T_{\rm dur}/t_{\rm bin}$, and
the standard deviation $\sigma$ can also be inferred from the out-of-transit flux.

\begin{figure}[htpb]
\centering
%\begin{tabular}[b]{@{}p{0.45\textwidth}@{}}
\includegraphics[width=.49\linewidth]{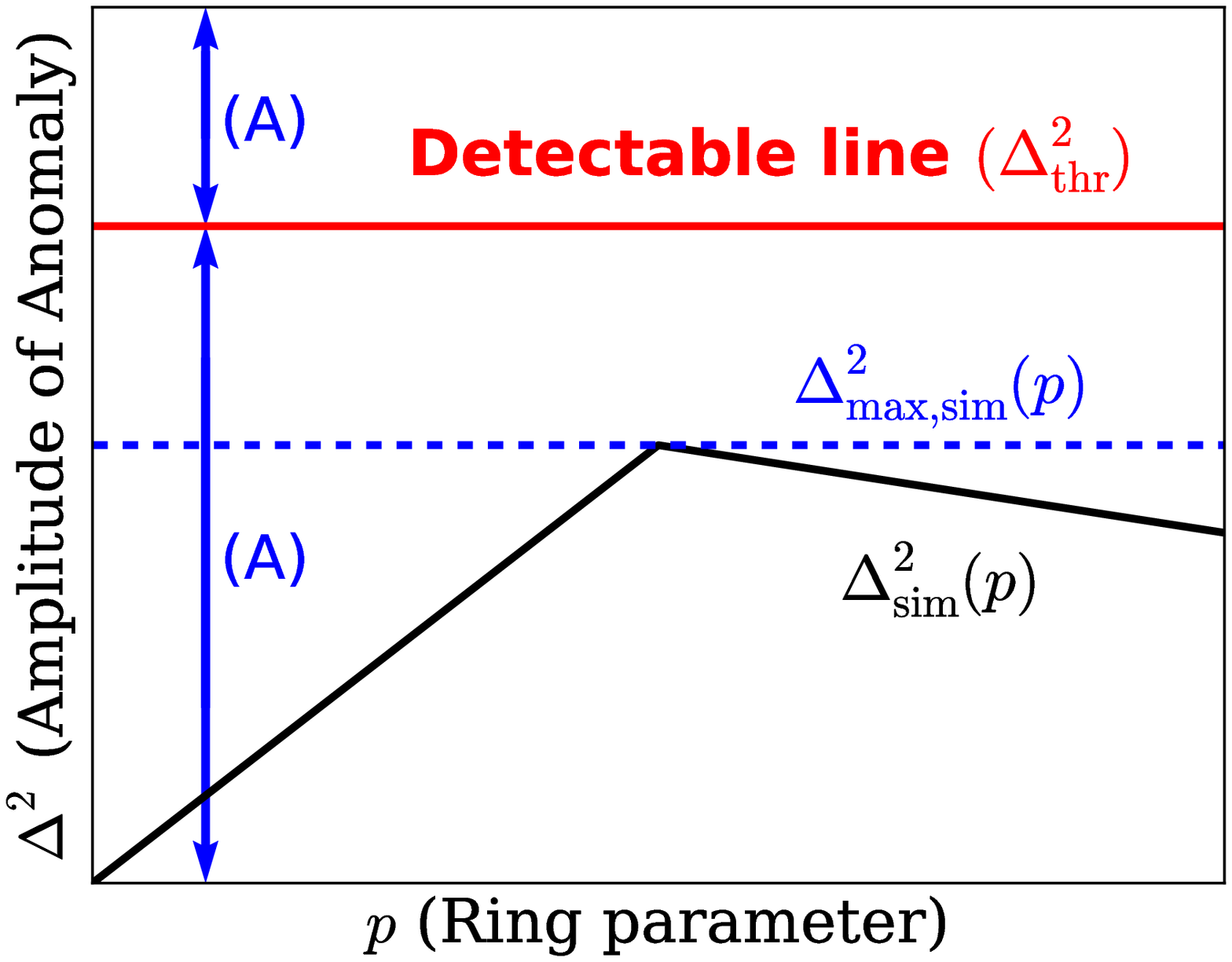} 
%\centering\small (a) $R_{p}/R_{\star}$ is multiplied by $\alpha$.
%\end{tabular}%
%\quad
%\begin{tabular}[b]{@{}p{0.45\textwidth}@{}}
\includegraphics[width=.49\linewidth]{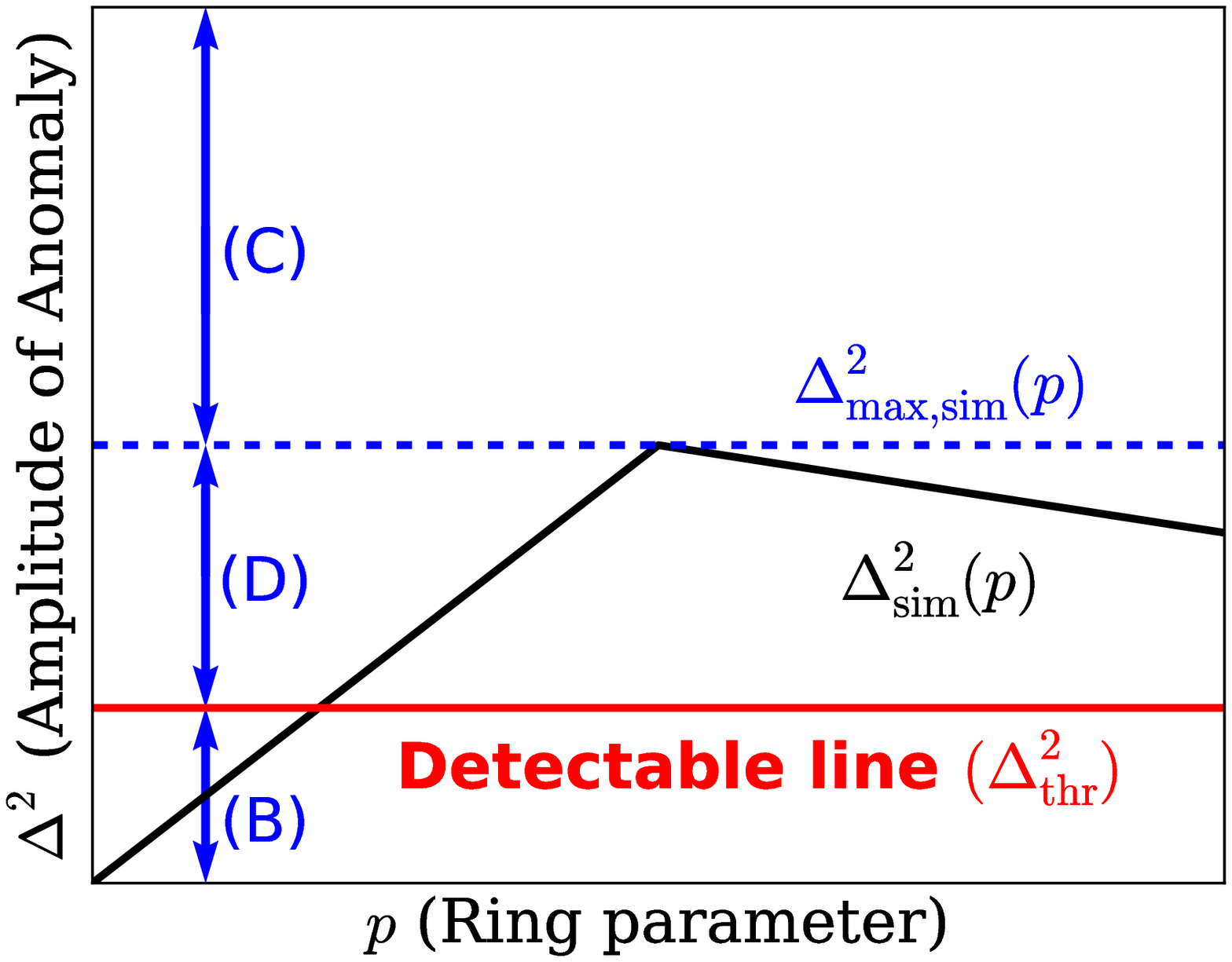} 
%\centering\small (b)$r_{\rm out/in}$ is multiplied by $\alpha$.
%\end{tabular}
\caption{Classification using $\Delta^{2}_{\rm thr}, \Delta^{2}_{\rm obs}$, and $ \Delta_{\rm sim}^{2}(p)$. Candidates are classified into (A)$\sim$(D) depending on the value of $\Delta^{2}_{\rm obs}$. 
(Left) $\Delta^{2}_{\rm max,sim} <  \Delta^{2}_{\rm thr}$. 
The detectability of rings is small. 
(Right) $\Delta^{2}_{\rm thr} <  \Delta^{2}_{\rm max,sim} $. The detectability of rings is large. 
 }
\label{s_n_obs}
\end{figure}

For a given region of parameter space $p$, 
$\Delta^{2}_{\rm sim}(p)$ has the maximum value $\Delta^{2}_{\rm max,\,sim}$. 
If $\Delta^{2}_{\rm max,\,sim}$ is smaller than 
some threshold value $\Delta^{2}_{\rm thr}$ determined by the noise level
in the light curve, 
the ringed-planets with the corresponding value of $p$,
even if they exist, cannot be detected in the system.
Then, the comparison of $\Delta ^{2}_{\rm obs}$, 
$\Delta^{2}_{\rm max,\,sim}$, and $\Delta^{2}_{\rm thr}$ 
allows for classification into four categories schematically 
illustrated in Figure \ref{s_n_obs}: 
\begin{itemize}
\item[(A)]: $\Delta^{2}_{\rm max,\,sim}  <\Delta^{2}_{\rm thr} $ \\ 
The expected signal from the ring is so small compared to the noise level
that we cannot discuss its detectability.
\item[(B)]: $ \Delta^{2}_{\rm obs}  <\Delta^{2}_{\rm thr} <\Delta^{2}_{\rm max,\,sim} $ \\
Although the rings with $\Delta^{2}_{\rm thr} < \Delta_{\rm sim}^{2}(p)$ could have been detected, no significant anomaly is observed ($\Delta_{\rm obs}<\Delta_{\rm thr}$) in reality.
Thus, the parameter region that gives $\Delta^{2}_{\rm thr}
< \Delta_{\rm sim}^{2}(p)$ is excluded. 
\item[(C)]: $ \Delta^{2}_{\rm thr} <\Delta^{2}_{\rm max,\,sim}  < \Delta^{2}_{\rm obs}  $ \\
A significant anomaly is detected, but its amplitude 
is too large to be explained by the ringed-planet model with the given range of $p$.
\item[(D)]: $ \Delta^{2}_{\rm thr}<\Delta^{2}_{\rm obs}   <\Delta^{2}_{\rm max,\,sim}   $ \\
A significant anomaly is detected, and its amplitude is compatible
with the ring model. In this case, we may find the ring parameters consistent with the observed
anomaly.
\end{itemize}

The value of $\Delta^{2}_{\rm thr} $ is arbitrary. 
In this paper, we choose $\Delta^{2}_{\rm thr}$ so that it corresponds to $S/N = 10$ in 
Equation (\ref{s_n_def}):
\begin{equation}
\Delta^{2}_{\rm thr} %=  \frac{\sigma^{2}}{N_{\rm data}} \times S/N 
= \frac{10 \,\sigma^{2}}{N_{\rm data}} \label{s_n_10},
\end{equation}
where $\sigma$ and $N_{\rm data}$ are calculated from 
the observed data. 
The methods to calculate the other variances, $\Delta^{2}_{\rm obs}$, 
$\Delta^{2}_{\rm sim}(p)$, and $\Delta^{2}_{\rm max,\,sim}$
will be presented in the following subections.

Before proceeding further, let us consider the orbital period dependence of 
$N_{\rm data}=T_{\rm dur}/t_{\rm bin}$ in Equation (\ref{s_n_def}). 
From Kepler's third law, $T_{\rm dur}\propto P\, ( R_{\star}/ a) \propto P^{1/3}$. 
For the short-period planets, $t_{\rm bin}\propto P$ because the 
number of folded transits is proportional to $1/P$. 
Thus, the number of the data $T_{\rm dur}/t_{\rm bin}$
is proportional to $P^{-2/3}$. This means that the detectability of rings ($S/N$)
is higher for the shorter-period planets 
for a given value of $\Delta^{2}$. 
This explains the strong constraints on the ring parameters 
obtained by \cite{2015ApJ...814...81H} for hot Jupiters. 

\subsection{Calculation of $\Delta^{2}_{\rm obs}$} 
\subsubsection{Definition}
The residual $\delta_{i,\,\mathrm{obs}}$ 
is obtained by fitting the planet-alone model to the data. 
If the ring does not exist, the value of $S/N_{\rm obs}$ in Equation (\ref{s_n_def}), 
which is formally equivalent to the chi-squared, is expected to be close to
the degree of freedom $DOF_{\rm obs}$. 
In contrary, if the ring does not exist, $S/N_{\rm sim}(p)$ is equal to zero. 
This mean that $S/N_{\rm obs}- S/N_{\rm sim}(p) \simeq DOF_{\rm obs}$ in the limit 
of the non-ring system. 
Thus, for comparison of $\Delta^{2}_{\rm sim}(p)$ and $\Delta^{2}_{\rm obs}$,  
the value of ($S/N-DOF_{\rm obs}$) serves as a good estimator of the observed anomaly 
rather than $S/N$. We thus slightly modify Equation (\ref{s_n_def})
to define $\Delta^{2}_{\rm obs}$ so that it
corresponds to ($S/N -DOF_{\rm obs}$):
\begin{align}
\Delta^{2}_{\rm obs} &=\left(\chi^{2} - DOF_{\rm obs}\right)  
\frac{1}{(\sigma/\sqrt{N_{\rm data}})^{2}},  
\end{align}\\
where
\begin{align}
\chi^{2}  &=  \sum_{i} \left(\frac{\delta_{i,\,\mathrm{obs}}}{\sigma}\right)^{2}. 
\label{chi_def} 
\end{align}
The residual $\delta_{i,\, {\rm obs}}$ is defined for the best-fit planet-alone model
obtained by minimizing $\chi^{2}$ as described in Section \ref{sssec:fit_detail} below.
The value of $\chi^2$ is computed using the data just around the transit
(within $0.6\,T_{\rm dur}$ from the transit center)
so that the value is not strongly affected by the out-of-transit data.
We assume $DOF_{\rm obs} = N_{\rm data} - N_{\rm para} -1$, where 
$N_{\rm para}$ is the number of fitted parameters.

\subsubsection{Detail of fitting\label{sssec:fit_detail}}
In fitting, we minimize $\chi^{2}$ using the Levenberg-Marquardt algorithm 
by implementing cmpfit \citep{2009ASPC..411..251M}. 
The adopted model $M(t)$ is composed of 
a fourth-order polynomial and a transit model $F(t)$: 
\begin{equation}
M(t) = F(t) [c_{0} +  c_{1} (t - T_{0})  +  c_{2} (t-T_{0})^{2} 
+ c_{3}(t-T_{0})^{3} + c_{4}(t-T_{0})^{4} ], \label{model_fit}
\end{equation}
where $c_{i}$ are coefficients of polynomials, and $T_{0}$ is a time offset. 
The polynomials are used to remove the long-term flux variations in the light curve. 
The transit model $F(t)$ is implemented 
by the PyTransit package \citep{2015MNRAS.450.3233P}.  
PyTransit generates the light curves based on 
the model of \cite{2002ApJ...580L.171M} with the 
quadratic limb darkening law. 

The above model $M(t)$ includes 12 parameters, 
$t_{0},R_{\rm p}/R_{\star}, b, a/R_{\star}, P,  q_{1}, q_{2}$, and $c_{i}\, (i=0 \sim 4)$. 
For KOIs, the initial values of $a/R_{\star}, R_{\rm p}/R_{\star},$ and $b$ 
for fitting are taken from the KOI catalog. 
The initial values of the limb darkening 
parameters are taken from the {\it Kepler} Input Catalog. 
For a single transit event, where we cannot 
estimate the orbital period from the transit interval, 
we choose $P$, instead of $a/R_\star$, as a fitting parameter
and estimate $a/R_{\star}$ from $P$ using Kepler's third law
and the mean stellar density given in the catalog.

In fitting, we remove outliers iteratively to correctly evaluate $\chi^{2}$. 
We first fit all the data with the model $M(t)$, and flag the 
points that deviate more than $5\sigma$ from the best model. 
We then refit only the non-flagged data using the same model, 
and update the flags of {\it all} the original data points,
including the ones classified as outliers before,
on the basis of the new best model and the same $5\sigma$ criterion. 
We iterate this procedure until the flagged 
data are converged. While this process gives a more robust evaluation
of $\chi^2$, it may also erase the signature of the ringed planet;
thus we visually check all the light curves in any case
not to miss the real ringed planets. 

The noise variance $\sigma^{2}$ is estimated for each transit light curve 
by fitting the out-of-transit light curve with a fourth-order polynomial, 
and calculating the variance of the residuals.
Flare-like events are excluded from the 
estimation of the noise variance. 

\subsection{Calculation of $\Delta^{2}_{\rm sim}(p)$ and $\Delta^{2}_{\rm sim, max}$\label{ssec:delta_sim}}  
Since the parameter space $p$ for a ringed planet is very vast, 
we wish to reduce the volume we need to search with simulations as much as possible.
First we show that $\Delta^2_{\rm sim}(p)$ does not depend on $P$ and $a/R_\star$ with 
other parameters fixed including limb darkening parameters $q_{1}, q_{2}$, 
the transit impact parameter $b$, planet-to-star radius ratio $R_{\rm p}/R_\star$, 
inner and outer ring radii relative to the planetary radius $r_{\rm in/p}$ and $r_{\rm out/p}$,
, the direction of the ring $(\theta, \phi)$, and a shading parameter $T$.
This property becomes apparent by rewriting $\Delta^2_{\rm sim}(p)$ into the following 
integral form approximately, assuming that the sampling rate ($t_{\rm bin}$)
is sufficiently small compared to the duration $T_{\rm dur}$:
\begin{align}
\Delta_{\rm  sim}^{2}(p)
\simeq \frac{ \int_{-T_{\rm dur}/2}^{T_{\rm dur}/2} \delta _{\rm sim}^{2}(t,  p) dt}
{\int_{-T_{\rm dur}/2}^{T_{\rm dur}/2} dt } 
= \frac{ \int_{-1/2}^{1/2} \delta _{\rm sim}^{2}(T_{\rm dur}t',  p) dt'}
{\int_{-1/2}^{1/2} dt'}
= \int_{-1/2}^{1/2} \bar{\delta}_{\rm sim}^{2}(t',  p) dt', 
\label{eq:delta_sim_bar}
\end{align}
where
\begin{align}\bar{\delta}_{\rm sim}(t, p) \equiv \delta_{\rm sim}(T_{\rm dur} t,p) \label{app_int}
\end{align} 
and the origin of time is shifted to the transit center.
Assuming that the values of $q_{1}$, $q_{2}$, $b$, $R_{\rm p}/R_\star$, $r_{\rm out/p}$, $r_{\rm in/p}$,
$\theta$, $\phi$, and $T$ are fixed, 
$\bar{\delta}(t, p)$ defined above does not depend on $T_{\rm dur}$ explicitly.
Therefore, $\Delta_{\rm  sim}^{2}(p)$ given by Equation (\ref{eq:delta_sim_bar}) 
does not depend on the time scale of the transit $T_{\rm dur}$,
which is determined by $P$ and $a/R_\star$, and 
we do not need to simulate the dependence of $\Delta_{\rm sim}^2(p)$ on these two parmeters.

To constrain the parameter space further, we use the observed transit depth. 
Here we also assume that the values of $q_{1}$, $q_{2}$, $b$, $T$, 
$r_{\rm in/p}$, and the ring direction are fixed
and that $R_{\rm p}/R_\star$ and $r_{\rm out/p}$ are the only free parameters.
Then, the constraint on the observed transit depth leaves only one degree
of freedom, specified by contours in the $R_{\rm p}/R_\star$-$r_{\rm out/p}$ plane;
henceforth we rewrite $\Delta_{\rm sim}^{2}(p)$ as $\Delta^{2}_{\rm sim}(r_{\rm out/p})$
to explicitly show this dependence.

To compute the relation $\Delta^{2}_{\rm sim}(r_{\rm out/p})$ for a given transit depth, 
we first calculate the value of $\Delta^2_{\rm sim}$ and the transit depths 
for a sufficient number of points
in the $(r_{\rm out/p}, R_{\rm p}/R_\star)$ plane. The necessary number of points depends on 
the fiducial model, and, in our simulation, 
we prepare about two hundred points for each model in Table \ref{sec3_ring}. 
%Then, from these grid points, we extract a sequence of $(r_{\rm out/p}, R_{\rm p}/R_\star)$
%that yield the observed transit depth 
%using the interpolation method implemented in the python module ``scipy.interpolate.interp1d."
%The detail procedure is as follows.
For any $r_{\rm out/p}$, the observed transit depth 
uniquely translates into $R_{\rm p}/R_{\star}$ by the interpolation 
in the $R_{\rm p}/R_{\star}$-transit depth plane, because 
the transit depth is a monotonically increasing function of the 
$R_{\rm p}/R_{\star}$. Thus, the given value of $r_{\rm out/p}$ 
is uniquely related to $\Delta^{2}_{\rm sim}$ given the 
transit depth. By repeating this procedure for many different values of $r_{\rm out/p}$, 
we can compute the relation $\Delta^{2}_{\rm sim}(r_{\rm out/p})$. 
We note that once a sufficient number of interpolated lines are prepared, 
one transit depth determines the relation
$\Delta^{2}_{\rm sim}(r_{\rm out/p})$ without additional 
calculation.

\begin{figure*}[htbp]
\begin{center}
\includegraphics[width=0.48 \linewidth]{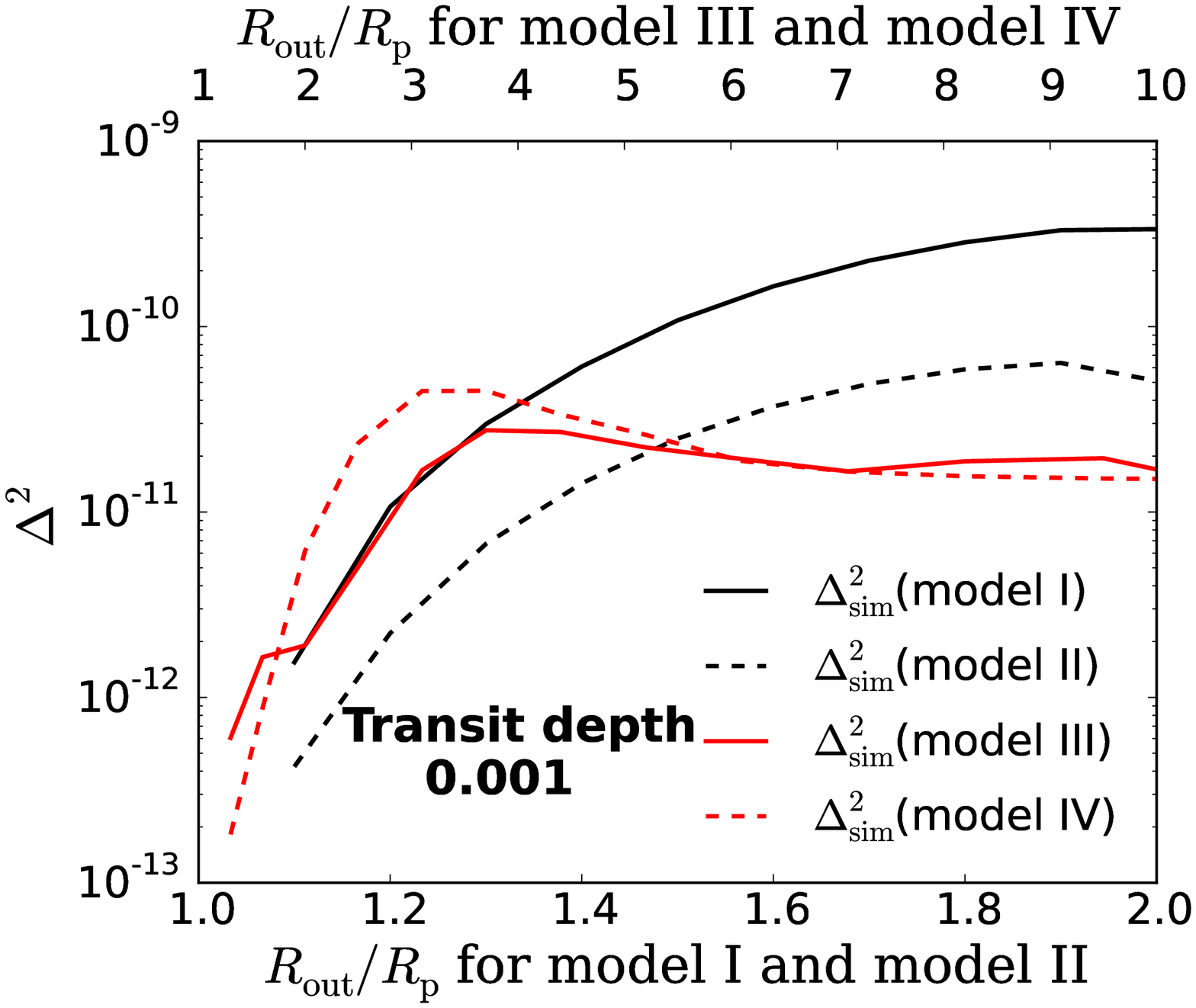}\hfill%
\includegraphics[width=0.48 \linewidth]{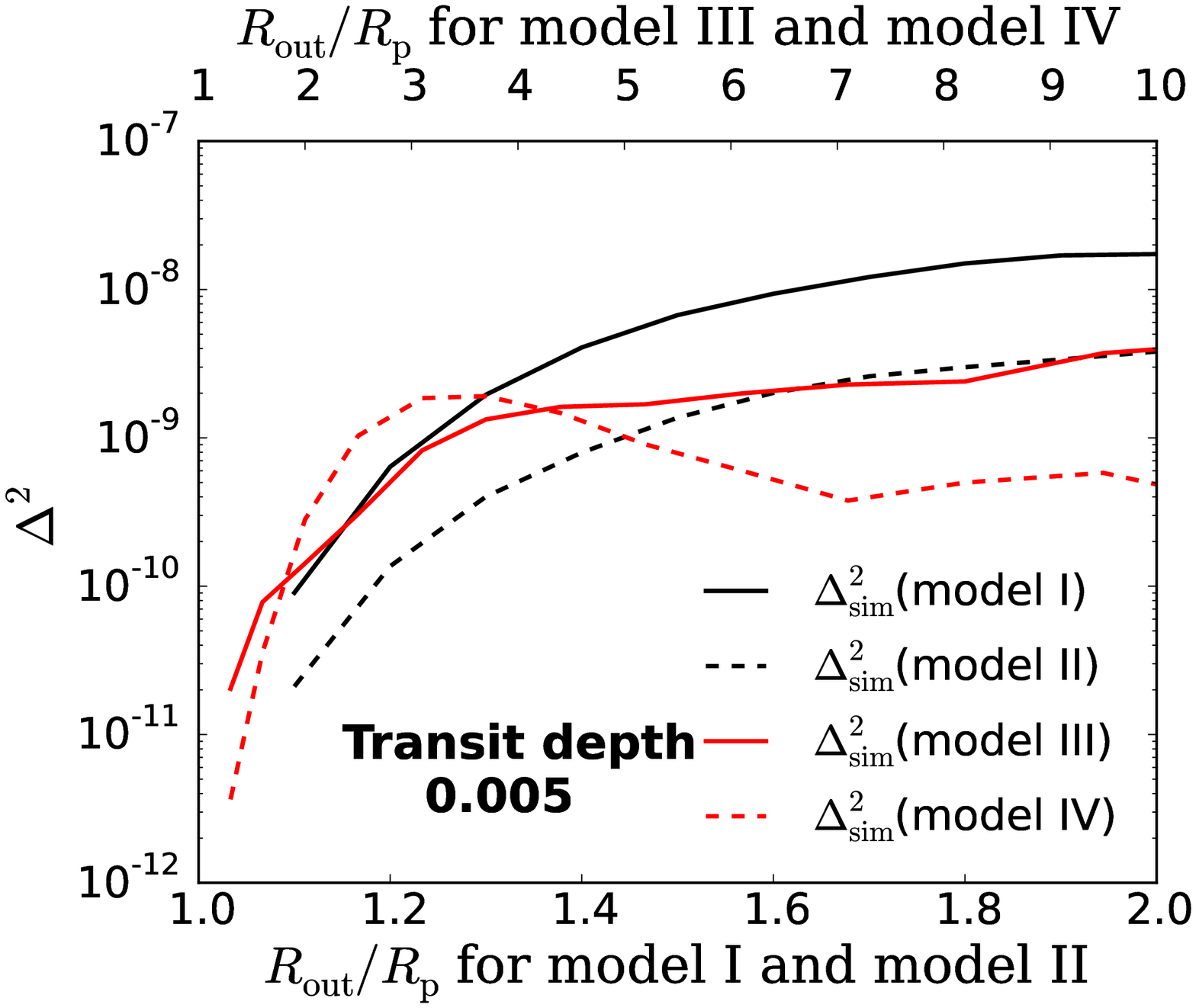}
\vspace{4pt}

\includegraphics[width=0.48 \linewidth]{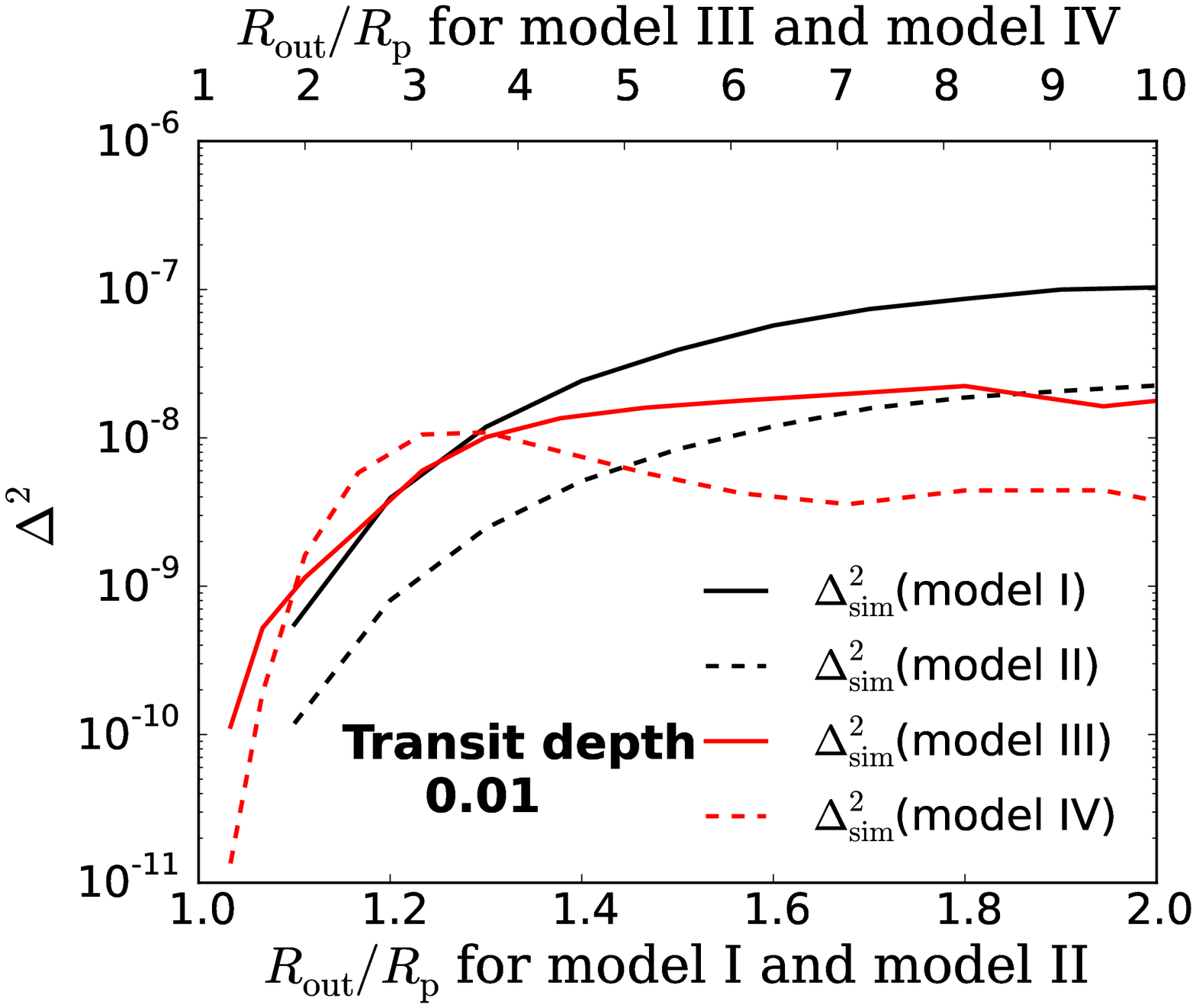}\hfill%
\includegraphics[width=0.48 \linewidth]{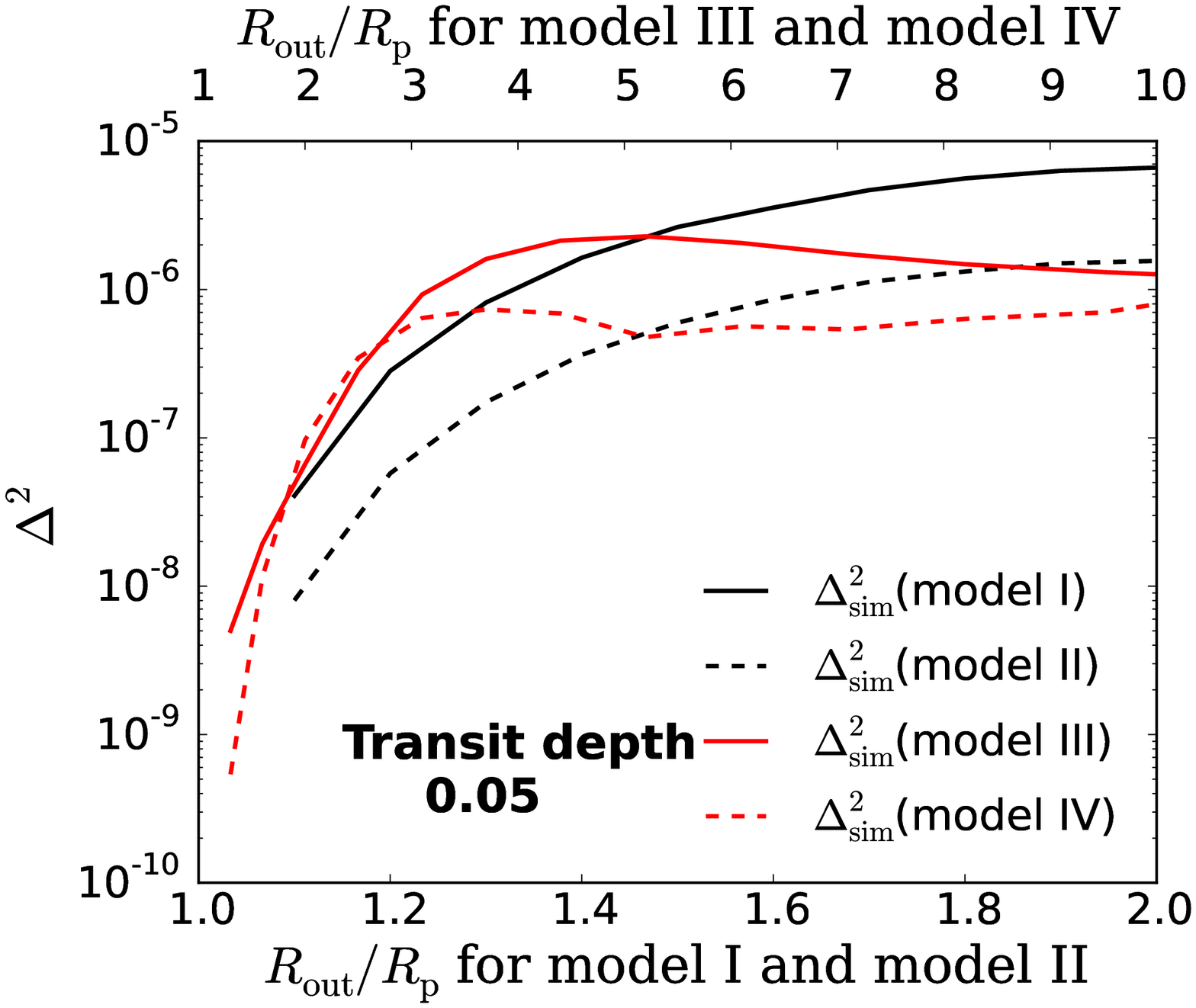}
\hspace{0.2pt}

\caption{The value of $\Delta^{2}_{\rm sim}(p)$ as a function of $r_{\rm out/p}$ 
for four different transit depths. 
The four lines in each panel correspond to four different sets of parameters 
(model I $\sim$ model I\hspace{-.1em}V) summarized in Table \ref{sec3_ring}. \label{fig:flux}}
\end{center}
\end{figure*}

Figure \ref{fig:flux} shows $\Delta^{2}_{\rm sim}(r_{\rm out/p})$ curves
created in this way, for
$4\times4 = 16$ different sets of impact parameters, ring directions, and transit depths.
The four sets of $p$ adopted here (model I $\sim$ model I\hspace{-.1em}V) 
are summarized in Table \ref{sec3_ring}, and four transit depths 
are chosen to be 0.001, 0.005, 0.01, and 0.05. 
We fix $T = 1$ and $r_{\rm in/p} = 1$ in all of these simulations. 

Here we simulate $\Delta^{2}_{\rm sim}(r_{\rm out/p})$ only for 
$1 \leq r_{\rm out/p} \leq r_{\rm eq}$,
where $r_{\rm eq}$ is the value of $r_{\rm out/p}$ for which 
the minor axis of the sky-projected outer ring is equal to the planetary radius,
computed for each model.
This is because the value of $\Delta^{2}_{\rm sim}(r_{\rm out/p})$ shows
no $R_{\rm p}/R_\star$ dependence beyond $r_{\rm eq}$, 
when $T = 1$ and $r_{\rm in/p} = 1$ are adopted;
if this is the case, the planetary disk is within the outer disk
and the transit depth is solely determined by the latter.

In this paper, we only use the observed constraint on the transit depth.
However, this is just for simplicity and we can certainly  
take into account the constraints on other parameters including
$b$, $q_{1}$, and $q_{2}$ from the 
morphology of the observed transit light curve 
(e.g. egress and ingress durations). 
Such constraints further restrict the ring models that could be 
consistent with the observed light curve and thus help more elaborate  
discussions on the ring parameters, which we leave to future works. 

\section{Derivation of the upper limit of $r_{\rm out/p}$: case of KOI-1466.01}
If a system is classified into group (B), the ring models with 
$\Delta^{2}_{\rm thr} < \Delta_{\rm sim}^{2}(r_{\rm out/p})$
are excluded.
The upper limits of $r_{\rm out/p}$ thus obtained are summarized 
in Sections 4 and 5.
Here we describe how the limit is derived using the relation 
$\Delta^2_{\rm sim}(r_{\rm out/p})$, taking KOI-1466.01 for example.
 
\begin{figure}[htpb]
  \centering
  \includegraphics[width=0.5 \linewidth]{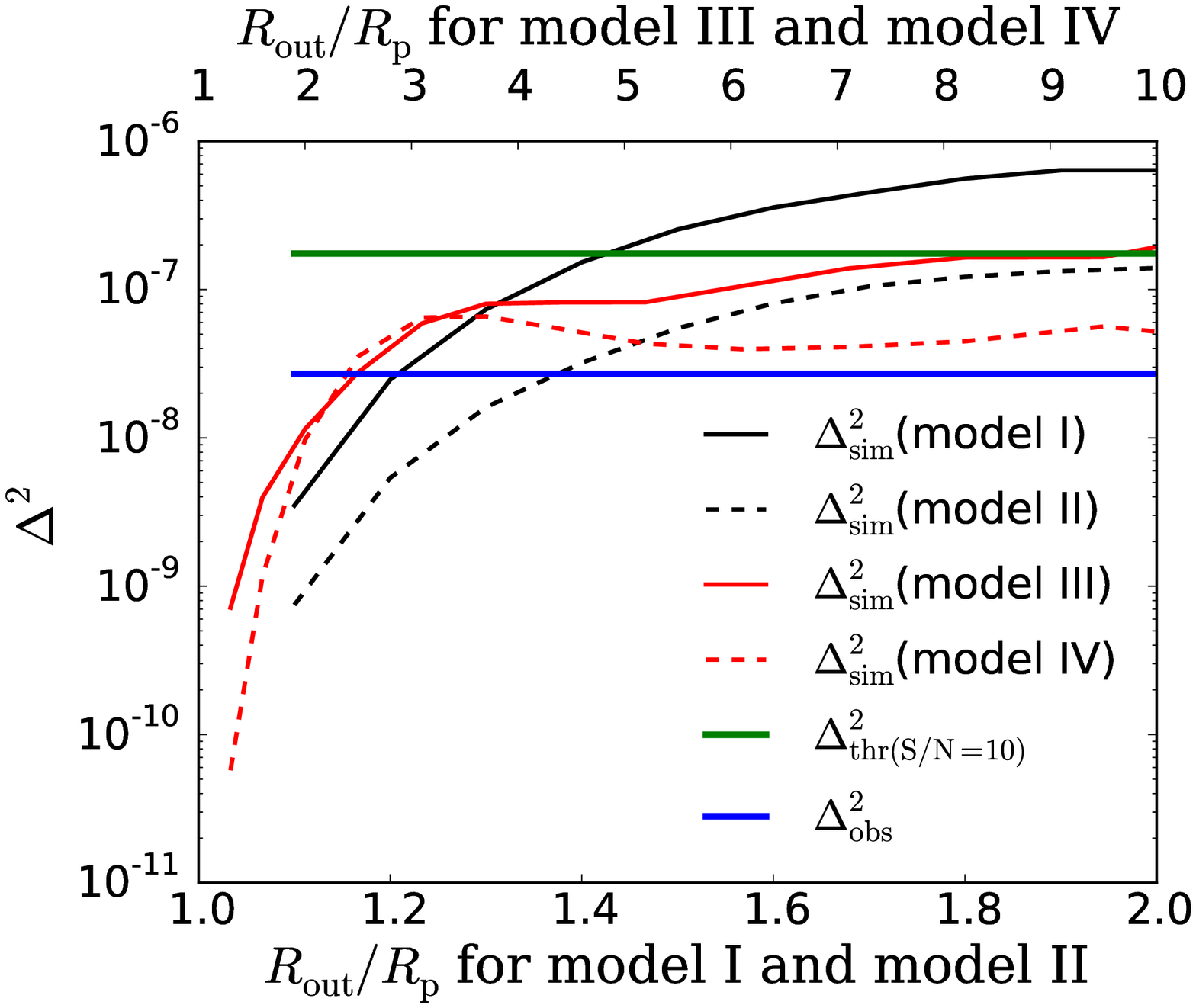} 
  \caption{Derivation of the upper limit of $r_{\rm out/p}$ in the case of 
  KOI-1466.01. The observed signature of the ring in the blue line is small 
  compared with the detectability of the ring shown in the green line. 
  The red and black lines are the theoretically expected values of $\Delta^2$   
  for models I-IV in Table \ref{sec3_ring}.
  Assuming model I (black solid line), for example, 
  the region with $1.5 < r_{\rm out/p}$ (i.e. a part of the line
  above the green line) is inconsistent
  with the non-detection of the anomaly and thus excluded.  
  }
      \label{can_1466}
\end{figure}

The black and red lines in
Figure \ref{can_1466}
are theoretically expected signals from the ringed planets 
(i.e., $\Delta^2_{\rm sim}(r_{\rm out/p})$)
for model I $\sim$ model I\hspace{-.1em}V
and for the transit depth of 0.0202 inferred from the observed data.
The green line shows the threshold value of $\Delta^2_{\rm thr}$ that satisfies $S/N =10$, 
and the blue line shows the observed residual level $\Delta^2_{\rm obs}$
obtained by fitting the planet-alone model to the data.
Here $\Delta^2_{\rm obs}<\Delta^2_{\rm thr}$, which means that
no significant deviation from the planet-alone model is detected.
In this case, we can in turn exclude the models above the green line,
because any anomaly above this level should have been detected if present.
In the case of the black solid line (model I), for example,
the ring with $r_{\rm out/p}>1.5$ would have produced the anomaly with $S/N>10$, 
which is not detected in reality. Thus, we can set the upper limit of 
$r_{\rm out/p}<1.5$ for model I. Note that the upper limits depend on the adopted
parameter set; this situation is clearly illustrated in Figure \ref{can_1466}, 
where similar limits cannot be derived for the other models.

%\input{app_4.tex}

%%%%%%%%%%%%%%%%%%%%%%%%%%%%%%%%%%%%%%%%%%%%%%%%%%%%%%%%%%%%%%%%%%%%%%
%\bibliography{IOPEXPORT_BIB}
%\end{document}
%%%%%%%%%%%%%%%%%%%%%%%%%%%%%%%%%%%%%%%%%%%%%%%%%%%%%%%%%%%%%%%%%%%%%%
%%%%%%%%%%%%%%%%%%%%%%%%%%%%%%%%%%%%%%%%%%%%%%%%%%%%%%%%%%%%%%%%%%%%%%
\end{document}